\documentclass[a4paper,11pt]{article} 
\usepackage[margin=0.75in]{geometry} 
\usepackage{amssymb}
\usepackage{amsmath,epsfig, epsf, epic, graphicx} 
\usepackage{fancybox, lscape, subfigure} 
\usepackage{bm}
\usepackage{natbib}
\usepackage[margin=2cm]{caption}
\usepackage{setspace}
\usepackage{hyperref}
\usepackage{cleveref}

\allowdisplaybreaks

\usepackage{floatrow}

\title{Heterogeneous causal effects of neighborhood policing in New York City with staggered adoption of the policy}
\author{Joseph Antonelli and Brenden Beck}

\begin{document}
\date{}
\maketitle

\begin{abstract}
    In New York City, neighborhood policing was adopted at the police precinct level over the years 2015-2018, and it is of interest to both (1) evaluate the impact of the policy, and (2) understand what types of communities are most impacted by the policy, raising questions of heterogeneous treatment effects. We develop novel statistical approaches that are robust to unmeasured confounding bias to study the causal effect of policies implemented at the community level. We find that neighborhood policing decreases discretionary arrests in certain areas of the city, but has little effect on crime or racial disparities in arrest rates. 
\end{abstract}

\section{Introduction}

 In this paper, we consider the problem of estimating the effect of neighborhood policing on arrest rates in New York City (NYC) over the years 2006-2018. Neighborhood policing is a policy implemented at the police precinct level in which the New York Police Department (NYPD) restructured its precincts, hired specialized community engagement officers, and gave all patrol officers time away from responding to 911 calls to devote to preemptive problem solving. The policy is meant to encourage officers to build community relationships and has them patrol a small sector with the twin goals of reducing crime and promoting trust between the police and residents \citep{bratton2015nypd}. The City began implementation in May 2015 with two of the 76 police precincts, and all precincts had adopted neighborhood policing by October 2018. In the statistics and economics literature this type of policy implementation is referred to as ``staggered adoption" \citep{athey2018design, shaikh2019randomization, ben2019synthetic}. Our overarching goal is two-fold: 1) to understand the causal effect of this policy on crime and arrest levels, and 2) to estimate how the effect varies over time and how communities with different characteristics might respond differently.

\paragraph{Review of neighborhood policing}

Neighborhood policing updates a policing approach common since the 1980s: community policing. Community policing de-emphasizes traditional police actions like arrests and prioritizes problem solving with community members, though research on its consequences is mixed. One reason for the unclear findings is that ``many of the [past] studies were characterized by weak evaluation designs” \citep[p. 7]{national2018proactive}. In \cite{gill2014community}, a meta-analysis found that many studies compared only one treated and one control neighborhood. When looking at studies with rigorous pre- and post-intervention measures, they found that community policing did not have a significant effect on crime, and concluded there is ``a need for further research around community policing" \citep[p.399]{gill2014community}. Another review of studies found that high levels of police-community contact had the promise to reduce crime, but such findings did not hold up when examining only RCTs \citep{sherman2003policing}. 

Despite these null findings of community policing's effect on crime, the NYPD made crime reduction the first goal of its new community-oriented policy \citep{oneill2018commissioner}. When announcing the policy, Mayor Bill de Blasio suggested several mechanisms that might link the policy to decreased crime, including  the ``constant vigilance" of neighborhood policing deterring crime, officers being more inclined to help would-be offenders desist from crime, and improved relationships between police and community members leading to more cooperation in crime solving \citep{deblasio2015annoucement}.

Previous research has focused on crime rates, but there might be additional, unintended consequences of neighborhood policing. The new practice might decrease arrests and the racial disparities of arrests by reorienting the goals of officers away from aggressive enforcement. A large body of evidence finds Black people are more likely to be arrested than white people, even controlling for differences in offending, so any racial equity impacts of neighborhood policing would be important \citep{lytle2014effects, kochel2011effect}. Alternatively, the increased police-community contact the policy generates could expose police to more potentially criminalizable behavior and therefore increase arrests and racial disparities. Furthermore, if police are deployed in higher numbers to predominantly black or Latino neighborhoods and/or if officers' bias makes them more likely to arrest black or Latino residents, neighborhood policing might increase the racial disparities in low-level, discretionary arrests through this increased contact.

\paragraph{Statistical literature on causal inference and panel data}

There is an extensive literature on the estimation of causal effects from observational, time-series data. Interrupted time series (ITS) methods are one of the most common methods for estimating causal effects in such situations and have been used for decades in economics and epidemiology, e.g. \citep{campbell1979quasi, gillings1981analysis}. See \cite{bernal2017interrupted} for both a review of ITS and practical details on their implementation. Intuitively, ITS methods fit a time-series model that allows the outcome to depend on time and an indicator of whether the policy has been initiated. Under certain assumptions, such as no unmeasured time-varying confounding, the effect of the policy can be obtained from the parameters of the model. Similar ideas have been extended to more complex problems, such as estimating spillover effects in time series studies \citep{bojinov2019causal} or estimating heterogeneous treatment effects from Bayesian state-space models \citep{li2018estimating}. 

Another commonly used method for identifying causal effects from observational time series data is the difference in differences (DiD) design \citep{ashenfelter1978estimating, angrist2008mostly}. Traditionally these methods have been employed in settings with \textit{two} groups and two time points. In the first time point, neither group has received treatment, while in the second time point only one group has received treatment. These designs rely on the parallel trends assumption that states that the counterfactual outcome under no treatment has the same time trend in both groups.  For a review of DiD methods and their extensions to more complex settings, see \cite{lechner2011estimation} and references within. Of most relevance to the problem at hand are methods that account for the staggered adoption of treatment over time, which has seen a spike in interest within DiD methodology \citep{athey2018design, goodman2018difference, callaway2019difference}. These methods focus on first defining relevant causal estimands in the presence of multiple time periods where treatment is initiated. Correspondences between these estimators and traditional DiD estimators are highlighted, and multiple different inferential strategies are used to acquire uncertainty measures for treatment effects. \cite{athey2018design} take a design-based approach to causal inference where uncertainty stems from the treatment assignment. \cite{callaway2019difference} highlight new causal estimands unique to the multiple time point setting, and derive a novel bootstrap approach to inference that is able to account for correlation and clustering within their data. 

Synthetic controls present a different approach to causal inference in this setting \citep{abadie2010synthetic, abadie2015comparative}. These were initially developed for the setting where only one unit receives treatment, and their potential outcome under control is estimated using a weighted average of control units with weights estimated using data prior to treatment initiation. This method has been extended to multiple treated units with staggered adoption by using the synthetic control method separately on each treated unit \citep{dube2015pooling, donohue2019right}. This approach has been shown to not be optimal if interest lies in average treatment effects, and was extended to the staggered adoption regime in \citep{ben2019synthetic}. Synthetic controls, along with other estimators in the panel data setting, were placed in a broader framework of matrix completion methods by \cite{athey2018matrix}. This treats the matrix of potential outcomes over time as a partially observed matrix and uses matrix completion methods to impute the missing values of the potential outcomes. Further, synthetic controls and DiD estimators have been combined to provide doubly robust estimates such that only one of the synthetic control weights or fixed effects regression model needs to be correctly specified in order to obtain consistent estimates of treatment effects \citep{arkhangelsky2019synthetic}.  

More recently, time series methods have been used to estimate causal effects by forecasting what would happen in the absence of treatment \citep{brodersen2015inferring, papadogeorgou2018causal, miratrix2019simulating}. The original approach in \cite{brodersen2015inferring}, is developed for the setting when one time series is measured over many time points both in the pre- and post-treatment time periods. The pre-treatment data is used to estimate a Bayesian state-space model that is then used to predict the counterfactual outcome under control during the post-treatment period. 

\paragraph{Review of our contribution}

We develop a framework for causal inference with multiple time series in the presence of staggered adoption that allows for estimation of causal quantities and heterogeneous treatment effects that vary over time and across precinct characteristics that is robust to unmeasured confounding bias. We use multivariate Bayesian time series models that allow for a high-dimensional set of observed time series to produce posterior predictive distributions of subject-specific treatment effects, which account for temporal and spatial correlation in the data. We couple the posterior predictive distribution with regression models to find conditional average treatment effect functions in a straightforward manner. Our paper extends the existing literature in a number of ways. Synthetic controls require there to be units without the treatment, while in our study every precinct adopts the policy by the end of the study. Many of the estimators proposed in the literature with staggered treatment are targeting average treatment effects of a policy, while our goal is to estimate conditional treatment effects that are functions of observed precinct characteristics. Interrupted time series type approaches estimate heterogeneous treatment effects in time-series settings \citep{li2018estimating}, but rely on an assumption of no unmeasured confounding, while we show our approach is robust to certain types of confounding bias from unmeasured covariates. Another key challenge we address is that our data are highly correlated across space, while existing approaches do not account for this spatial dependence in the observations. We show using pre-treatment data in NYC that our approach to inference is able to provide valid inferences for treatment effects in this setting, and then apply our approach to answer important, unanswered questions of the effects of neighborhood policing on crime and arrests. Lastly, we provide an R package to implement the proposed methodology that is available at \url{https://github.com/jantonelli111/HeterogeneousTEpanel}

\section{Crimes, arrests, and policing in New York City}

We gathered data on crimes, arrests, and community characteristics from three sources. Data on crimes reported to the police come from the NYPD's Historic Complaint Database, while arrest data are from the NYPD's Arrest Database. Crime and arrest data are publicly available on New York City's Open Data Portal. Data on precincts' demographic, economic, and housing characteristics come from the Census Bureau's American Community Survey (ACS) five-year estimates. Data were spatially linked and acquired at the precinct level. Address-level crime and arrest data were placed into precincts using a precinct shapefile map provided by the City and Stata's geoinpoly command \citep{picard2015geoinpoly}. ACS data at the census tract-level, a smaller geography than the precinct, were placed into the precinct that hosted its centroid. In total, the data contains 156 months of data for 76 precincts.

\begin{figure}[!t]
		\centering
		\includegraphics[width=0.8\linewidth]{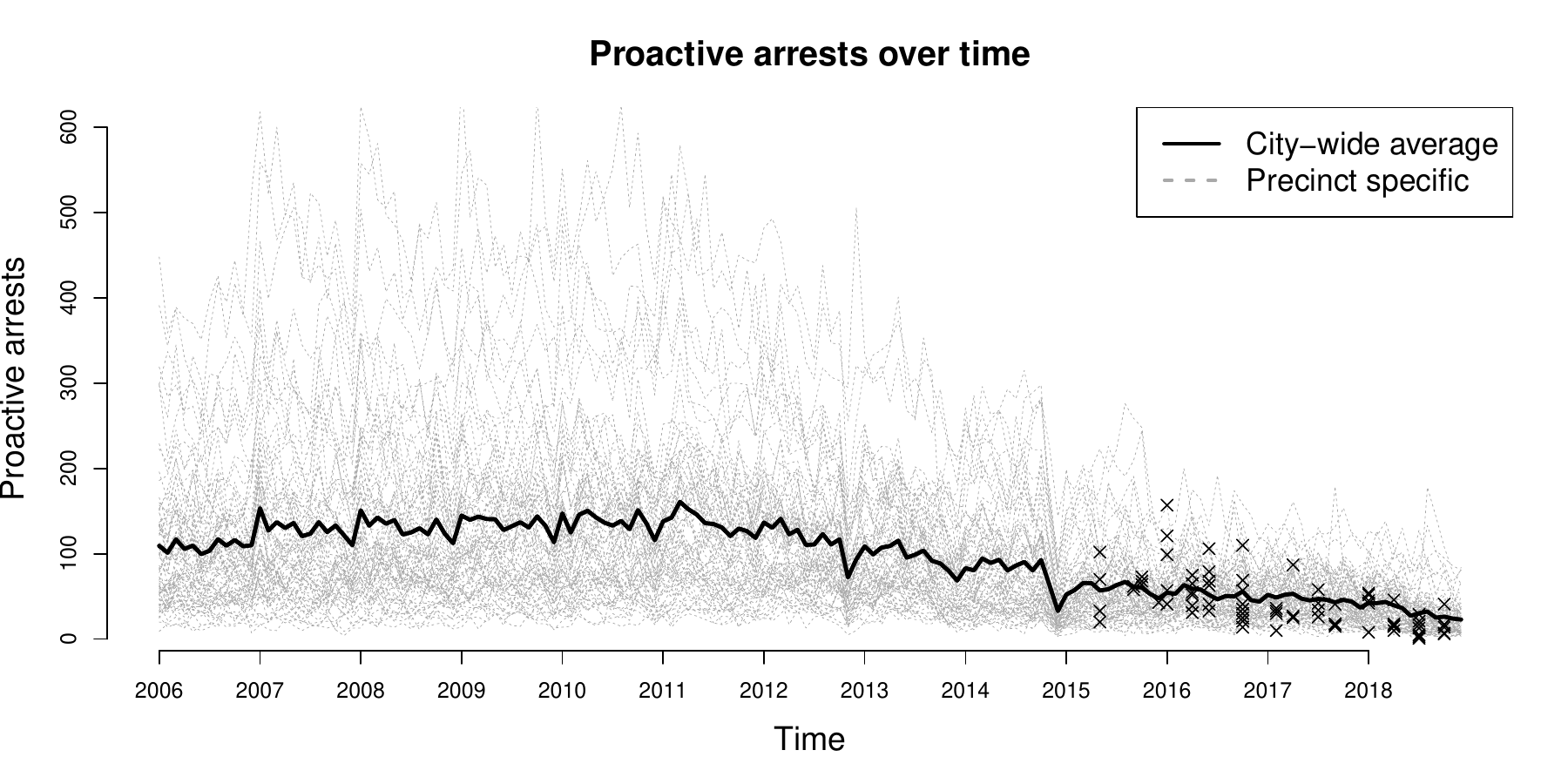}
		\caption{Preliminary look at the NYC policing data showing the time series for proactive arrests for each precinct and the city average. The points marked by an x are the times at which each precinct adopted neighborhood policing.}
		\label{fig:prelim}
\end{figure}

We are interested in studying how the adoption of neighborhood policing affected (1) overall crime, (2) arrest levels, and (3) racial disparity in enforcement. To study the effect on overall crime, we use the number of \emph{violent crimes}, defined as the number of murders, manslaughters, robberies, and felony assaults reported to the police in each precinct. We use \emph{violent crime} rather than total crime because violent crimes are the most likely to be reported to police. Misdemeanor crimes are less frequently reported and are therefore more reliant on police action to be recorded, reflecting police enforcement priorities more than actual crime levels. To analyze arrest levels, we focus on both the number of \emph{misdemeanor arrests} and \emph{proactive arrests}. \emph{Misdemeanor arrests} is a count of arrests for 133 misdemeanor crimes, the most common of which were marijuana possession, misdemeanor assault, theft of services (transit fare evasion), possession of stolen property, and trespassing. \emph{Proactive arrests} are a subset of misdemeanor arrests that reflect the 55 crimes most often identified by police activity rather than victim complaints. We focus on these two arrest types, rather than an aggregate measure of all arrests, because their discretionary nature makes them the most likely to fluctuate as a result of policy changes. An illustration of proactive arrests over time can be found in Figure \ref{fig:prelim}. We see that proactive arrests were generally increasing in the early years of the study, though they declined in recent years. Finally, we are interested in the impact on racial disparities in arrest rates. To measure this, we define a measure of \emph{racial disparity in proactive arrests} as the difference in the number of proactive arrests a precinct makes of black people and the number of proactive arrests of white people in the precinct.

We obtained the implementation date of neighborhood policing for each precinct from the NYPD's Commissioner Report in 2018 \citep{oneill2018commissioner}. Treatment adoption times are shown in Figure \ref{fig:prelim}, which highlights how precincts steadily began implementation in 2015 and shows there was no period of time when the majority of precincts began neighborhood policing. To understand heterogeneity of the effects of neighborhood policing, we measure a vector of precinct-specific demographic, economic, and housing variables that have been shown to relate to crime and arrest rates. Understanding how the effect of neighborhood policing varies by these characteristics will inform the types of communities for which the policy is most useful. 

\section{Potential outcomes, estimands, and identification}
\label{sec:estimands}

Let $A$ and $Y$ denote the treatment and outcome of interest, respectively, from a population of $n$ units. We observe each of these $n$ units at $T$ time periods and therefore our data consists of $A_{it}$ and $Y_{it}$ for $i=1, \dots, n$ units, and $t=1, \dots, T$ time periods. Each unit in our population eventually adopts treatment and we let $\boldsymbol{T}_0 = (T_{10}, T_{20}, \dots, T_{n0})$ be the vector of initiation times for each unit. We will be working under the framework that once a unit initiates treatment, it cannot revert back to the control condition, i.e $A_{it} = 1 \ \forall \ t \geq T_{i0}$. Lastly, we denote the unit-specific vector of covariates by $\boldsymbol{X}_{i}$.  

\subsection{Potential outcomes and estimands}

We denote potential outcomes by $Y_{it}(t_0)$, where $t_0$ is the time at which treatment is initiated.  This represents the potential outcome we would observe for subject $i$ at time $t$ had they initiated treatment at time $t_0$. Similar to \cite{athey2018design} we let $Y_{it}(\infty)$ denote the potential outcome for a subject if they never receive treatment. To link potential outcomes to the observed data we make a standard consistency assumption that $Y_{it}(T_{i0}) = Y_{it}$. This consistency assumption implicitly assumes there is no interference between units.  This assumption states that the treatment status of one unit can not affect the outcomes of other units. As we discuss in greater detail in Section \ref{sec:AnalysisAssumptions}, this is reasonable in the policing data as it is unlikely for neighborhood policing in one precinct to affect crime or arrest rates in neighboring precincts. We also alleviate this assumption in Appendix B and find that results in the NYC policing example remain unchanged. Lastly, we assume that there are no anticipatory effects, i.e. that $Y_{it}(t_0) = Y_{it}(\infty)$ for all $i$ and $t < t_0$. This states that units do not respond to treatment before it gets initiated. This could be violated if units are aware of the impending treatment change, and subsequently change their behavior due to the upcoming change in policy. 

We first define unit-level treatment effects at specific time points, and then extend them to sample-level treatment effects, as well as estimands that target heterogeneity of the treatment effect. Let us first define the \textit{unit-level causal effect} for unit $i$ at time $t$ if they initiated treatment at time $t_0$ as
\( \displaystyle
    \Delta_{i, t, t_0} = Y_{it}(t_0) - Y_{it}(\infty).
\)
This contrast compares what would have happened if a unit initiated treatment at time $t_0$ versus what would have happened if they never initiated the treatment. While unit-level treatment effects are of interest themselves, in many settings average treatment effects are of more interest as they highlight the impact of a policy over an entire region, such as New York City. We define the time-specific sample average treatment effect as
\begin{align}
    \Delta(q) = \frac{1}{n}\sum_{i=1}^n \Delta_{i, T_{i0} + q, T_{i0}} =  \frac{1}{n} \sum_{i=1}^n \Big[ Y_{i, T_{i0} + q}(T_{i0}) - Y_{i, T_{i0} + q}(\infty) \Big]. \label{eqn:DeltaQ}
\end{align}
This is the average impact of the treatment $q$ time periods after treatment initiation over all units. For different values of $q$, $\Delta(q)$ illuminates how the treatment effect varies over time after treatment initiation. In the policing example, this represents the effectiveness of neighborhood policing $q$ time points after the observed initiation times, which measures the overall impact that the policy had on crime and arrest levels.

\subsection{Identification of population estimands}
\label{sec:identification}

The estimand $\Delta(q)$ is a sample-specific estimand and is therefore not strictly identifiable from the observed data in the sense that it can be written as a function of the observed data distribution in finite samples \citep{balzer2016targeted}. Nonetheless, we can examine a population-level counterpart to $\Delta(q)$, and study the assumptions under which it is identified from the observed data. This will provide an improved understanding of the assumptions required for estimation.

Define a population analog of $\Delta(q)$ as 
\( \displaystyle
E(Y_{t_0 + q}(t_0) - Y_{t_0 + q}(\infty) \vert T_0 = t_0).
\)
Similarly to $\Delta(q)$, this estimand looks at the average effect of the policy $q$ time points after the realized start time, which is denoted by $t_0$ here. The first term is immediately identifiable by the consistency assumption, which implies that $E(Y_{t_0 + q}(t_0) \vert T_0 = t_0) = E(Y_{t_0 + q} \vert T_0 = t_0)$, which is a function of the observed data distribution only. We show in Appendix A that under the consistency and no anticipatory effects assumptions, we can write the second term, $E(Y_{t_0 + q}(\infty) \vert T_0 = t_0).$, as a function of 
\begin{align}
    P(Y_{t_0+1}(\infty) \vert Y_{t_0}(\infty), T_0 = t_0), \dots,
    P(Y_{t_0+q}(\infty) \vert Y_{t_0 + q - 1}(\infty), T_0 = t_0), \label{eqn:Stationarity1}
\end{align}
where $P(Y_{t_0+1}(\infty) \vert Y_{t_0}(\infty), T_0 = t_0)$ is the distribution of $Y_{t_0+1}(\infty)$ given $Y_{t_0}(\infty)$ and $T_0 = t_0$.  These are not identifiable from the observed data without additional assumptions because they are functions of unobserved counterfactual values. We prove in Appendix A that these terms, and hence the causal effect, are identified under the following assumption:
\begin{align}
    \textbf{Assumption 1: } P(Y_{t_0+l}(\infty) \vert Y_{t_0 + l - 1}(\infty), T_0 = t_0) &= P(Y_{t_0+m}(\infty) \vert Y_{t_0 + m - 1}(\infty), T_0 = t_0) \nonumber \\ \forall \ l,m &\in \{-t_0 + 2, -t_0 + 3, \dots, q-1, q \}. \label{eqn:Stationarity2}
\end{align}
This states that the distribution of the potential outcome time series given the past value of the time series is stationary up to $q$ time periods post-treatment initiation. Note this assumption is only on the potential outcome in the absence of treatment and does not assume anything about the potential outcome under treatment. This differs from certain estimators, such as DiD estimators, that make assumptions about potential outcomes across treatment and control groups. For a more in-depth comparison of this stationarity assumption with existing assumptions used in the literature, see Appendix C where we additionally highlight a situation when stationarity holds, but existing assumptions are violated. 

A crucial point regarding our strategy to identifying causal effects is that we only require assumption 1 and \textit{do not} rely on an assumption that there are no unmeasured confounders, a common assumption in causal inference with observational data. This means that there can be unmeasured variables $U_i(t)$ that affect both the time of treatment initiation and the outcome, yet we are still able to identify and estimate causal effects. The only assumption we rely on is stationarity as defined in (\ref{eqn:Stationarity2}). Note that assumption 1 can be loosened to condition on time-varying covariates or additional lags beyond the single time point lag that is used currently, which we describe further in Appendix A. In Appendix C, we highlight scenarios where there exists a time-varying or seasonal unmeasured confounder and our approach is still able to obtain unbiased estimates of causal effects. It is important to note that our approach is not robust to all types of unmeasured confounding, as some may violate the stationarity assumption. If the effect of the unmeasured variable on the outcome changes after treatment initiation, or if treatment initiation affects the distribution of the unmeasured variable, then stationarity would be violated. A key feature of this assumption is that one can assess whether stationarity holds in the pre-treatment time periods to assess the plausibility of the assumption. If stationarity does not hold in the pre-treatment period, we would not believe that it holds in the post-treatment period. However, if it holds in the pre-treatment period, this would provide increased confidence in the stationarity assumption. We assess the ability of our approach to estimate treatment effects in the pre-treatment period for the motivating NYC study in Section \ref{sec:sim}, where we find that our approach obtains accurate estimates of treatment effects with valid measures of uncertainty.

\subsection{Heterogeneous treatment effects}

In the policing example, we are interested in understanding how the treatment effect varies as a function of precinct level characteristics, such as the racial or socioeconomic distribution in a precinct. We can study how $\Delta_{i, t, t_0}$ varies with $\boldsymbol{X}_i$ using
\( \displaystyle
    \Delta_{i, t, t_0} = f(\boldsymbol{X}_{i}, t - t_0),
\)
which conveys how treatment effects are expected to change as a function of precinct level characteristics. The $f(\cdot)$ function can be a complex, nonlinear function of covariates, but many times it is of interest to study heterogeneity by a specific covariate. Let $\boldsymbol{X}_{-j}$ be the matrix of observed covariates in our data excluding covariate $j$, and let $x_j$ and $x_j'$ denote two distinct values for covariate $j$. We define
\begin{align}
    \overline{\Psi}(j, q) = \frac{1}{n} \sum_{l=0}^q \sum_{i=1}^n \Big\{ f([\boldsymbol{X}_{-j}, x_j], l) - f([\boldsymbol{X}_{-j}, x_j'], l) \Big\}, \nonumber
\end{align}
which highlights the difference in causal effects after $q$ time points for units with the same $\boldsymbol{X}_{-j}$ but different values for covariate $j$. We focus on setting $x_j'$ and $x_j$ to the $25^{th}$, and $75^{th}$ quantile of $\boldsymbol{X}_j$, though other values would work analogously. In some instances, however, this may not be a realistic comparison. One example is that it is unlikely that a precinct could have increasing unemployment levels without also having increasing poverty levels. In situations such as these, we can compare $f(\boldsymbol{X}_{i}, t - t_0)$ for distinct values of the entire covariate vector that correspond to feasible levels of the covariates. In Section \ref{sec:application} we investigate the effects of neighborhood policing on distinct types of precincts that are found via a clustering algorithm.

\section{Estimation and inference}
\label{sec:estimation}

Based on assumption 1 and the identifiability results in Section \ref{sec:identification}, we require a model that predicts future values of $Y_{it}(\infty)$ given previous time periods. Therefore, we develop a Bayesian multivariate time series model that accounts for spatial correlation across precincts. Our goal is to find the posterior distribution of the potential outcome time series in the absence of the policy. Letting $\boldsymbol{\widetilde{Y}}(\infty)$ represent all unknown values we are interested in predicting, interest lies in $P(\boldsymbol{\widetilde{Y}}(\infty) \vert \boldsymbol{Y}, \boldsymbol{X})$, the posterior predictive distribution of these predictions given the observed data. Both temporal and spatial correlation in the data must be accounted for properly if we want our estimators to have good inferential properties, such as frequentist interval coverage. We describe one such model here, and explore an additional vector autoregressive model in Appendix E, but the ideas that follow will hold for any model for $P(\boldsymbol{\widetilde{Y}}(\infty) \vert \boldsymbol{Y}, \boldsymbol{X})$. 

\subsection{Bayesian multivariate structural time series model}

To account for both spatial and temporal dependencies, we specify a Bayesian structural time series model of the form
\begin{align}
\boldsymbol{Y}_t &= \boldsymbol{\mu}_t + \boldsymbol{\epsilon}_t, & \boldsymbol{\epsilon}_t &\sim \mathcal{N}(\boldsymbol{0}, \boldsymbol{\Sigma}) \label{eqn:MainModel} \\
\boldsymbol{\mu}_t &= \boldsymbol{\mu}_{t-1} + \boldsymbol{\delta}_{t-1} + \boldsymbol{\eta}_t^{\mu}, & \boldsymbol{\eta}_t^{\mu} &\sim \mathcal{N}(\boldsymbol{0},  \boldsymbol{D}_{\mu}) \nonumber \\
\boldsymbol{\delta}_t &= \boldsymbol{\delta}_{t-1} + \boldsymbol{\eta}_t^{\delta}, & \boldsymbol{\eta}_t^{\delta} &\sim \mathcal{N}(\boldsymbol{0}, \boldsymbol{D}_{\delta}), \nonumber
\end{align}
where $\boldsymbol{D}_{\mu}$ is a diagonal matrix with elements given by $\sigma_{\mu,i}^2$ for $i = 1, \dots, n$. $\boldsymbol{D}_{\delta}$ is defined analogously, but with variance parameters given by $\sigma_{\delta,i}^2$ for $i = 1, \dots, n$. Independent inverse-gamma prior distributions are assigned for all variance parameters $(\sigma_{\mu,i}^2, \sigma_{\delta,i}^2)$ for $i=1, \dots, n$. This is one example of a more general class of structural time series models, and it is straightforward to add additional complexities such as terms capturing seasonality. For a more general discussion of these models, see \cite{scott2014predicting}. The trend terms $\boldsymbol{\mu}_t$ capture the underlying trend of the multivariate time series at time $t$, while $\boldsymbol{\delta}_t$ represents the slope of the trend at time $t$. Each of these follow random walks that induce dependence over time, the extent of which is governed by $\boldsymbol{D}_{\mu}$ and $\boldsymbol{D}_{\delta}$. Lastly, the error term $\boldsymbol{\epsilon}_t$ allows for spatial dependence across units in the study at a particular time period through the covariance matrix $\boldsymbol{\Sigma}$. 

\subsection{Reducing parameter space of $\Sigma$}

In high-dimensional time series settings, we do not have sufficient data to estimate all $n(n-1)/2$
parameters of the covariance matrix $\boldsymbol{\Sigma}$. Existing dimension reduction approaches include imposing sparsity on the inverse of the covariance matrix, assuming it has a low-rank structure \citep{fox2015bayesian}, or decomposing the covariance matrix into the product of upper triangular matrices \citep{george2008stochastic}. We utilize the first of these approaches, but will use geographic information in the data to inform the sparsity. Our approach finds an initial estimator of model \ref{eqn:MainModel} using smoothing approaches such as natural cubic splines or smoothing splines for each unit separately, then uses the model residuals and an optimization algorithm to acquire an estimate of $\boldsymbol{\Sigma}$. From the initial fitted model, we acquire predicted values $\widehat{\boldsymbol{Y}}_t$ for all $t < t_{min}$, where $t_{min} = \text{min}(T_{10}, \dots, T_{n0})$ is the earliest time point that treatment is initiated, and calculate
$$\widehat{\boldsymbol{S}} = \frac{1}{t_{min} - K - 1}\sum_{t=1}^{t_{min}} (\boldsymbol{Y}_t - \widehat{\boldsymbol{Y}}_t) (\boldsymbol{Y}_t - \widehat{\boldsymbol{Y}}_t)',$$
 where $K$ is the degrees of freedom used for the individual models for each unit. For high-dimensional data sets such as the NYC policing data, this estimator will be very unstable, so we regularize the estimated covariance matrix by imposing sparsity on $\boldsymbol{\Sigma}^{-1}$. 
 It is known in Gaussian graphical models that if the $(i,j)$ element of $\boldsymbol{\Sigma}^{-1} \equiv \boldsymbol{\Omega}$ is zero, then $Y_{it}(\infty)$ and $Y_{jt}(\infty)$ are conditionally independent given the remaining observations. In our policing example, a reasonable assumption is that precincts that are more than one neighbor apart are conditionally independent. We enforce this in the estimation of $\boldsymbol{\Sigma}$ by solving the following constrained optimization:
\begin{align*} 
\widehat{\boldsymbol{\Omega}} = \underset{\Omega}{\operatorname{argmin}} \text{ tr}(\boldsymbol{\Omega \widehat{S}}) - \log \det \boldsymbol{\Omega}, \quad \quad
\text{such that } \boldsymbol{\Omega} \in \mathcal{Q},
\end{align*}
where $\mathcal{Q}$ is the space of all positive semi-definite matrices whose $(i,j)$ element is zero for any $i$ and $j$ that are not neighbors. This finds the value of $\boldsymbol{\Omega}$ that is closest to $\boldsymbol{\widehat{S}}^{-1}$ while enforcing the desired sparsity. We first estimate $\widehat{\boldsymbol{\Sigma}}$ and run the remaining procedure conditional on this estimate. Note that this is not a fully Bayesian procedure and will ignore uncertainty due to estimation of $\widehat{\boldsymbol{\Sigma}}$. We empirically evaluate the quality of model \ref{eqn:MainModel} in Section \ref{sec:sim} and find that it performs very well on the New York City policing data and leads to credible intervals with nominal coverage rates, despite conditioning on $\widehat{\boldsymbol{\Sigma}}$.  

\subsection{Posterior distribution of treatment effects}

Sampling from the posterior distribution of all parameters in model \ref{eqn:MainModel} allows us to obtain the posterior predictive distribution of future time points. This characterizes our uncertainty around what would have happened in the absence of treatment for all units in the sample. The unit-level treatment effects of interest are $\Delta_{i,t,T_{i0}} = Y_{it}(T_{i0}) - Y_{it}(\infty)$ for $t \geq T_{i0}$. The first of these two values is observed and known as it is simply the observed outcome after treatment is initiated, while the second of these two is the unknown quantity for which we have a posterior distribution. We automatically obtain the posterior distribution of the unit-level treatment effects, denoted by $$P(\boldsymbol{\Delta} \vert \boldsymbol{Y}, \boldsymbol{X}) = P(\boldsymbol{Y}_{obs} - \boldsymbol{\widetilde{Y}(\infty)} \vert \boldsymbol{Y}, \boldsymbol{X}),$$ 
where $\boldsymbol{Y}_{obs}$ is the corresponding vector of observed outcomes after treatment initiation. Now that we have posterior distributions for $\Delta_{i,t,T_{i0}}$ for all $i$, and all $t \geq T_{i0}$, we can proceed with obtaining estimates and credible intervals for the estimands of interest from Section \ref{sec:estimands}. $\Delta(q)$ is obtained directly by averaging the relevant unit-level treatment effects over the correct time periods. Inference is straightforward using the posterior distribution of these quantities.

For treatment effect heterogeneity, we focus on inference for $f(\cdot)$. To estimate $f(\cdot)$, we first draw a sample from the posterior distribution of unit-level treatment effects, $\boldsymbol{\Delta}^{(b)}$. We then regress these values on $\boldsymbol{X}_i$, the observed characteristics for each unit, as well as $t-T_{i0}$. This can be done using linear models, nonlinear models, or more complicated machine learning approaches. We can repeat this process for $b=1, \dots, B$ posterior draws, each time keeping track of estimates of treatment effect heterogeneity that we are interested in, such as $ \overline{\Psi}(j,q)$, and inference proceeds from the posterior distribution of these quantities. We can additionally improve estimation of $\Delta(q)$ by assuming a smooth function of $t - T_{i0}$ in the model for $f(\cdot)$. The predicted values from $f(\cdot)$ can be used to estimate $\Delta(q)$ with $\frac{1}{n} \sum_{i=1}^n f(\boldsymbol{X}_i, q)$, and we show in Appendix H that smoothness can lead to more efficient estimates when the true treatment effect is smooth in time.

\subsection{Distinguishing spatial correlation and interference}

One of the key underlying assumptions necessary for the estimation of causal effects in our setting is the no interference assumption. This states that the potential outcome for a precinct does not depend on the treatment status of other precincts. In spatio-temporal settings it is important to distinguish between spatial dependence across units and spillover of treatment effects into neighboring units. First, we stress that model \ref{eqn:MainModel} is for the potential outcomes in the absence of treatment and does not imply anything regarding the nature of the treatment effect. Importantly, this means that any spatial dependence in model \ref{eqn:MainModel} does not imply spillover of treatment effects, i.e. interference. An example of when interference is present but spatial correlation is absent would be a setting where the \textit{mean} of the potential outcome for unit $i$ depends on the treatment status of neighboring units, but that the correlation across units is still zero. Alternatively, spatial dependence can occur without interference if there is an underlying predictor that has spatial structure that induces dependence of the outcomes. We expect there to be spatial dependence in the NYC data as crime levels tend to have spatial structure. Additionally, in Section \ref{sec:AnalysisAssumptions} and Appendix B, we discuss how the no interference assumption is expected to hold within the context of our study. 

\section{Simulations using observed NYC precinct data}
\label{sec:sim}
Here we present simulation studies using the observed data in New York City. We focus on observed data for misdemeanor outcomes, which is one of the four outcomes we analyze in Section \ref{sec:application}. We ran similar simulations for the other three outcomes to ensure that our method works in all four situations, and those results can be found in Appendix G. These simulations allow us to test a number of features about our approach such as 1) how plausible our identification assumptions are in the NYC data, and 2) how well our model performs in estimating effects for the NYC policing data. We follow a similar approach to \cite{schell2018evaluating} to generating simulated data sets.

We first generate a new time of treatment initiation for each precinct between times 71 and 100, which we can denote by $T_{i0}^*$. Note that $\min_i T_{i0} = 112$, which ensures that we have more than 10 time periods between the actual and simulated start time of treatment. We will only estimate treatment effects 10 time points into the future, which ensures that our simulation results will not be impacted by the introduction of neighborhood policing. We let $T_{i0}^*$ be dependent on the outcome prior to treatment initiation. Specifically, we randomly sample $n$ numbers between 71 and 100 with replacement and sort them in increasing order. We refer to this ordered vector of potential times as $\widetilde{T}_1, \dots, \widetilde{T}_n$ where we have that $\widetilde{T}_1 \leq \widetilde{T}_2 \leq \dots \leq \widetilde{T}_n$. For $j=1, \dots, n$, we assign the $j^{th}$ time $\widetilde{T}_j$ to be the start time for unit $i$ with probability proportional to $\overline{Y}_{i,50} = (1/50)\sum_{t=1}^{50} Y_{it}$. Once a unit has been assigned a start time, they are removed from the pool of units moving forward. This process ensures that units with higher values of the outcome are more likely to initiate treatment earlier, which is a realistic situation in practice. On average, the correlation between $T_{i0}^*$ and $\overline{Y}_{i,50}$ was -0.39 across simulated data sets. Our simulated data set is therefore the observed data from the NYC policing example, but now the start times are given by $T_{i0}^*$. For all time periods after $T_{i0}^*$ for unit $i$, we shift the observed time series, and the amount that we shift the outcome is the magnitude of the causal effect for that unit and time combination. We repeat this process 1000 times and average results over all simulated data sets.

This simulation framework is extremely informative about the performance of our approach on the application of interest, because we only control the magnitude and form of the treatment effect, and we have \textit{no control on the data generating process for the outcome}. This is a far more realistic simulation than simulations based completely on user-specified data generating models and will provide more insight into the performance of our approach for the \textit{data set at hand}. One can think of this simulation as a form of model checking or model validation, where we empirically evaluate the performance of our model on the observed data. If our model performs poorly that would indicate that either stationarity does not hold in the data or our model does not capture all sources of uncertainty. Note that approaches based on a no unmeasured confounding assumption can not evaluate their approach in this same manner as this simulation requires randomly assigning treatment times, which would alter the no unmeasured confounding assumption. 

We consider two distinct estimands for the simulation study: a time specific treatment effect ($\Delta(q)$ for $q=0, \dots, 9$), and $\overline{\Psi}(j, 9)$ for each covariate in our study. To estimate $f(\cdot)$, we specify a linear regression model in the covariates $\boldsymbol{X}_i$ and time since treatment adoption, $t - T_{i0}$. We focus on bias, interval coverage, and efficiency for estimating each estimand. Interval coverage is the proportion of simulations in which the 95\% credible interval covers the true parameter. For brevity, we explore one simulation design here, though additional extensive simulations can be found in Appendices C,E, G and H, and a summary of which can be found in Section \ref{sec:addSim}. All prior distributions and model specifications are as described in Section \ref{sec:estimation}.

\subsection{Results with homogeneous treatment effects}

\begin{figure}[htbp]
		\centering
		\includegraphics[width=0.315\linewidth]{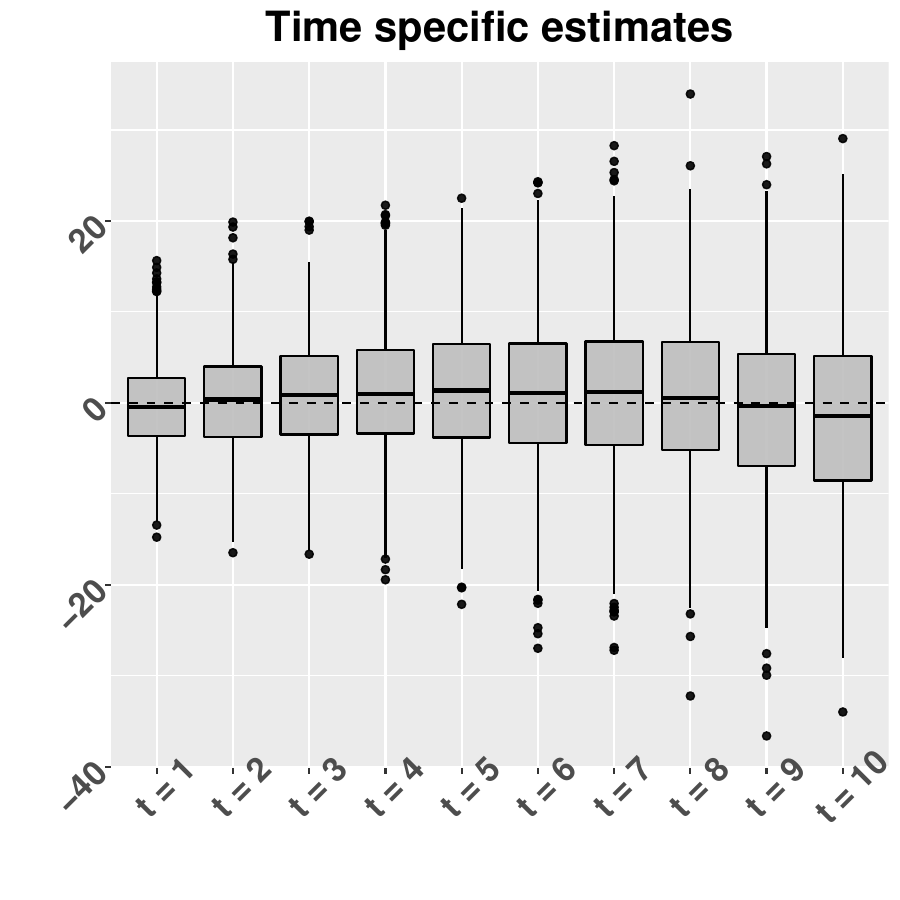}
		\includegraphics[width=0.33\linewidth]{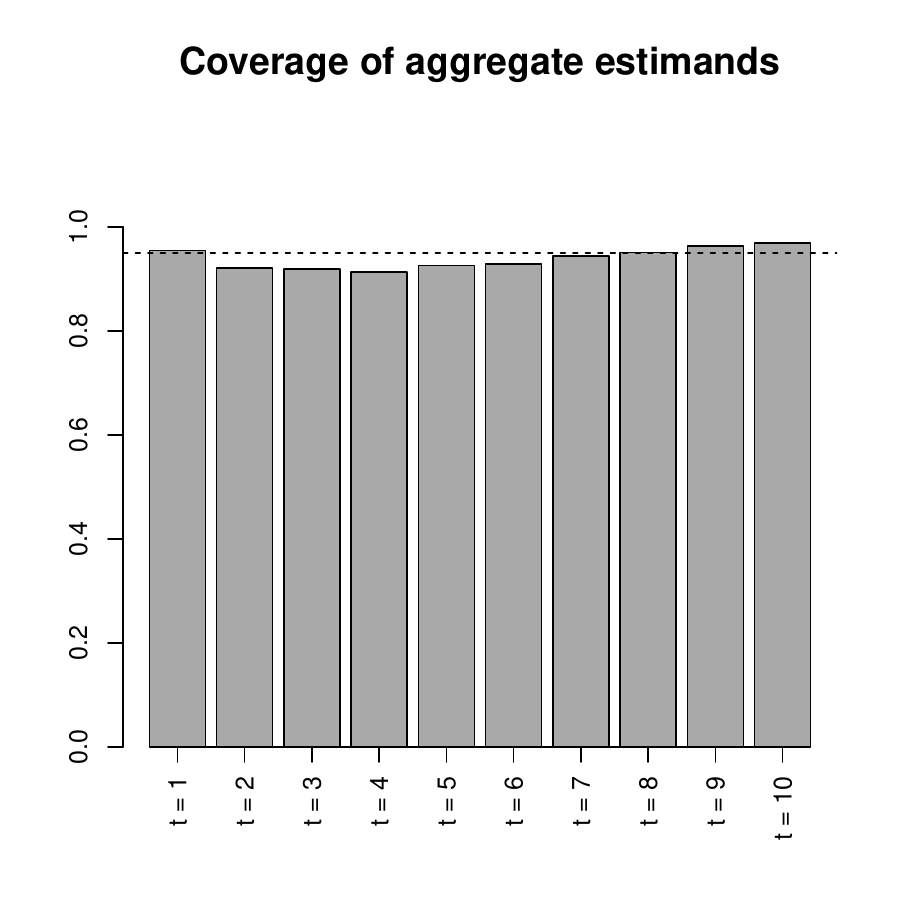}
		\includegraphics[width=0.333\linewidth]{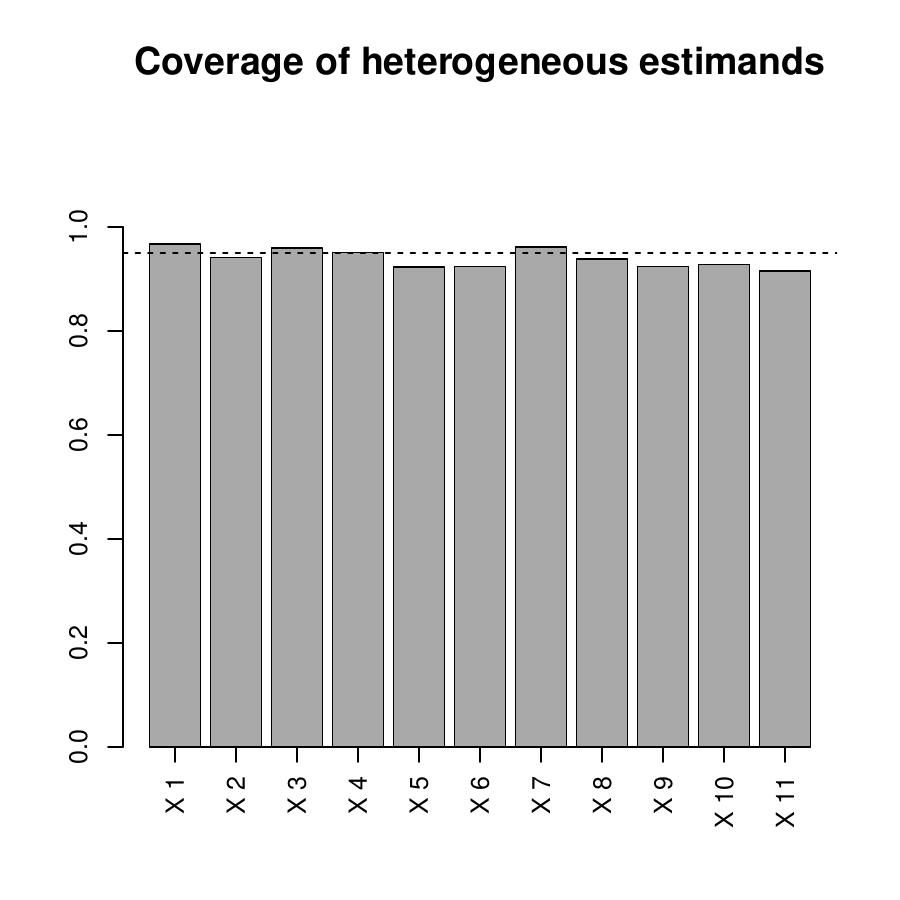}
		\caption{Results from the homogeneous treatment effect simulation study. The left panel shows estimates of $\Delta(q)$ for $q=0, \dots 9$. Estimates are mean shifted so that an unbiased estimator would be centered at zero. The middle panel shows coverage for all marginal estimands, while the right panel shows coverage for heterogeneous estimands.}
		\label{fig:43Sim}
\end{figure}

We first simulate treatment effects such that there is no heterogeneity by covariates $\boldsymbol{X}_i$ and that the unit-specific treatment effects for each precinct across the 10 time points are given by $\boldsymbol{\Delta}_i = 0.1 \overline{Y}_{i,50} + (1, 2, 2, 1, 0.5, 0, 0, 0, 0, 0)$. This ensures that the treatment effect is larger in areas with larger values of the outcome, and that the treatment effect is approximately 10\% of $\overline{Y}_{i,50}$. The range of $\overline{Y}_{i,50}$ is 78.26 to 767.02 with a mean of 256.9, leading to time-specific treatment effects $\Delta(q)$ that are between 25.7 and 27.7. The results from this simulation study can be seen in Figure \ref{fig:43Sim}. The left panel shows a boxplot of estimates for $\Delta(q)$ for ten time points post-treatment, where the estimates are shifted by the true mean so that unbiased estimates would be centered around zero. Estimates are essentially unbiased for all values of $q$ considered. Additionally, in the middle panel, we can see that interval coverages are close to 95\% for all time points considered. A similar story emerges for heterogeneous treatment effects, which can be seen in the right panel of Figure \ref{fig:43Sim}. We are able to obtain interval coverages at or near the nominal rate showing the ability of our approach to account for all sources of uncertainty in the NYC policing data.  

\subsection{Summary of additional simulations}
\label{sec:addSim}
We have run a number of additional simulation studies that can be found in Appendices C,E, G and H that cover a wide range of situations including the presence of an unmeasured variable affecting both treatment times and the outcome, smoothed estimates of $\Delta(q)$, heterogeneous treatment effects, and simulations based on other outcomes in the NYC data. One key takeaway from these simulations is that the presence of an unmeasured covariate affecting treatment times and the outcome does not necessarily bias our results or affect the validity of our inferential procedure. We simulate situations where an unmeasured confounder follows an AR(1) or seasonal process, and we are still able to obtain unbiased results while existing methods are biased. This confirms the theoretical results in Section \ref{sec:estimands} suggesting that our approach is robust to unmeasured confounders as long as stationarity holds. Time-varying, unmeasured confounders could still negatively impact inference if they are non-stationary themselves or if their effect on the outcome changes over time. We also explore scenarios where the true $\Delta(q)$ values are smooth in $q$, and show that if smoothness of $q$ is incorporated into the estimation procedure, then more efficient estimates of marginal treatment effects can be obtained. We further evaluated model performance in a similar manner on the remaining three outcomes of interest in our data analysis: violent crimes, proactive arrests, and the racial disparity in proactive arrests. The results are mostly identical across the four outcomes with some minor differences. For all three outcomes our approach obtains credible interval coverages that are at, or near, the nominal level for both marginal and heterogeneous estimands. We see a small amount of bias in the estimates of $\Delta(q)$ for violent crimes and proactive arrests that could be due to mild amounts of model misspecification or non-stationarity. However, it is not substantial enough to drastically impact coverage rates for either outcome.

Overall, the simulations showed that our approach is well-suited to estimating treatment effects in the policing data in NYC. Additionally, we have seen that our approach is robust to certain types of unmeasured confounding as long as stationarity continues to hold.

\section{The effects of neighborhood policing in NYC}
\label{sec:application}

We estimate the impact of neighborhood policing on four outcomes: The number of proactive arrests, the number of misdemeanor arrests, the number of violent crimes, and the difference in the number of black and white proactive arrests. We choose these outcomes because they are plausibly affected by the initiation of neighborhood policing and are of interest for understanding the impact of the policy throughout New York City. We include eleven covariates as potential effect modifiers: population size, percentage of the population that is Black, percentage of the population that is Latino, percentage of housing units that are vacant, percentage unemployed, percentage living in poverty, percentage young men, percentage foreign-born, percentage of owner-occupied housing units, percentage of people with a bachelor's degree, and the average outcome in the first 50 time periods of the study. 

\subsection{Plausibility of causal assumptions}
\label{sec:AnalysisAssumptions}

Before estimating the effects of neighborhood policing on crime and arrest rates, it is important to discuss the assumptions required to identify them from the data. The three key assumptions are the no interference assumption, the no anticipatory effects assumption, and the stationarity assumption. In evaluating the effects of policing interventions, spillover effects are often of interest. For example, hot spots policing places a large number of officers in a small area with very high crime rates, and such targeted enforcement might push crime to adjacent areas \citep{puelz2019graph, collazos2020hot}. However, in our setting, neighborhood policing is unlikely to have spillover effects that cross precinct lines. Neighborhood policing involves hiring new officers, not to target enforcement, but to engage with community members. The theorized mechanism linking the policy to crime reduction is not the incapacitation of offenders through increased arrests, but the preemptive reduction in crime through improved community trust, which is unlikely to push crime to nearby areas. Neighborhood policing also reduces the time existing patrol officers spend responding to 911 calls. Given that officers typically stay within their precinct---in our data 99\% of arrests in any precinct are made by officers of that precinct---it is unlikely that this additional time devoted to community engagement will have impacts on nearby precincts. Nonetheless, this is a key assumption, and therefore we have provided two additional approaches in Appendix B that alleviate the no interference assumption and allow for spillover of the treatment effect. We find that there is very little evidence of spillover of the treatment effect, which increases our belief in the no interference assumption, as well as the findings presented here. 

The no anticipatory effects assumption ensures that the potential outcome if neighborhood policing is never adopted is the same as the potential outcome if neighborhood policing has not been adopted yet. This would fail if the police officers in a precinct changed their behavior in preparation for the change to neighborhood policing. Given that the new community engagement officers would not be working yet and the traditional officers would not have the additional free time allotted for community engagement, this assumption is expected to hold. Regardless, if this assumption were to be violated, it would likely bias results towards null effects. This is because the pre-treatment data directly before initiation of the policy would reflect the impact of neighborhood policing and our predictions for the post-treatment period in the absence of the policy would be shifted in the direction of the treatment effect, thereby making the estimated effect smaller in magnitude. As a sensitivity analysis we ran all analyses using earlier treatment initiation times and do not find that estimates differ. 

The assumption of stationarity is arguably a strong assumption. Fortunately, this is the one assumption that we were able to partially assess in the simulation study of Section \ref{sec:sim}. While we can never formally test this assumption, because it involves unobserved counterfactuals in the post-treatment period, we are able to assess whether the stationarity assumption holds in earlier time periods. If stationarity were to be violated in the earlier time points, then our simulations based on the observed New York City arrest data would show biased estimates of the treatment effects and coverage rates below 95\%. The fact that our estimates remained relatively unbiased and led to nominal coverage rates gives us increased confidence that this assumption holds in the data example.

\subsection{Marginal effects}

\begin{figure}[!t]
		\centering
		\includegraphics[width=0.24\linewidth]{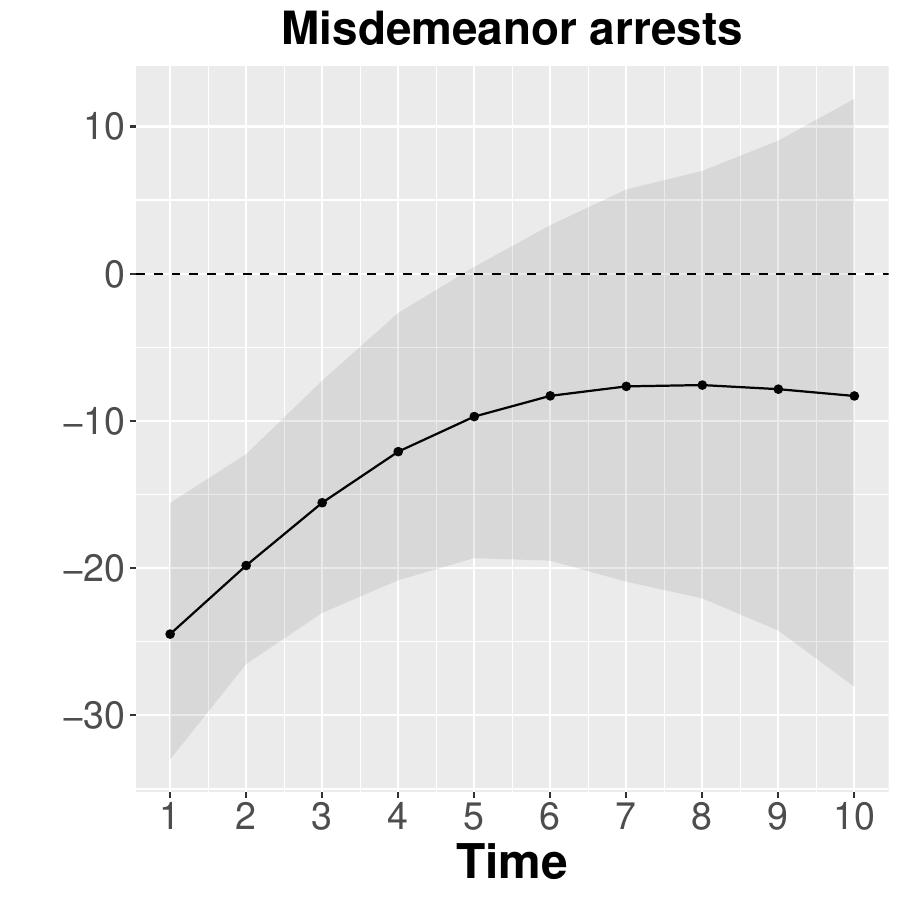}
		\includegraphics[width=0.24\linewidth]{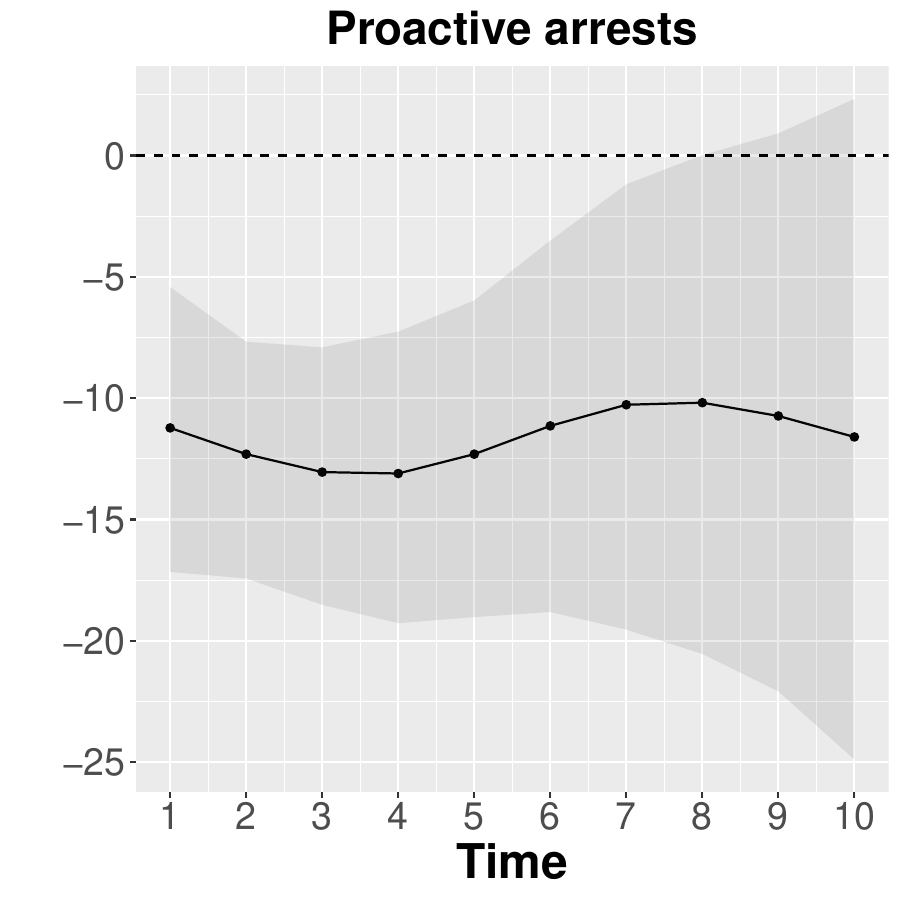}
		\includegraphics[width=0.24\linewidth]{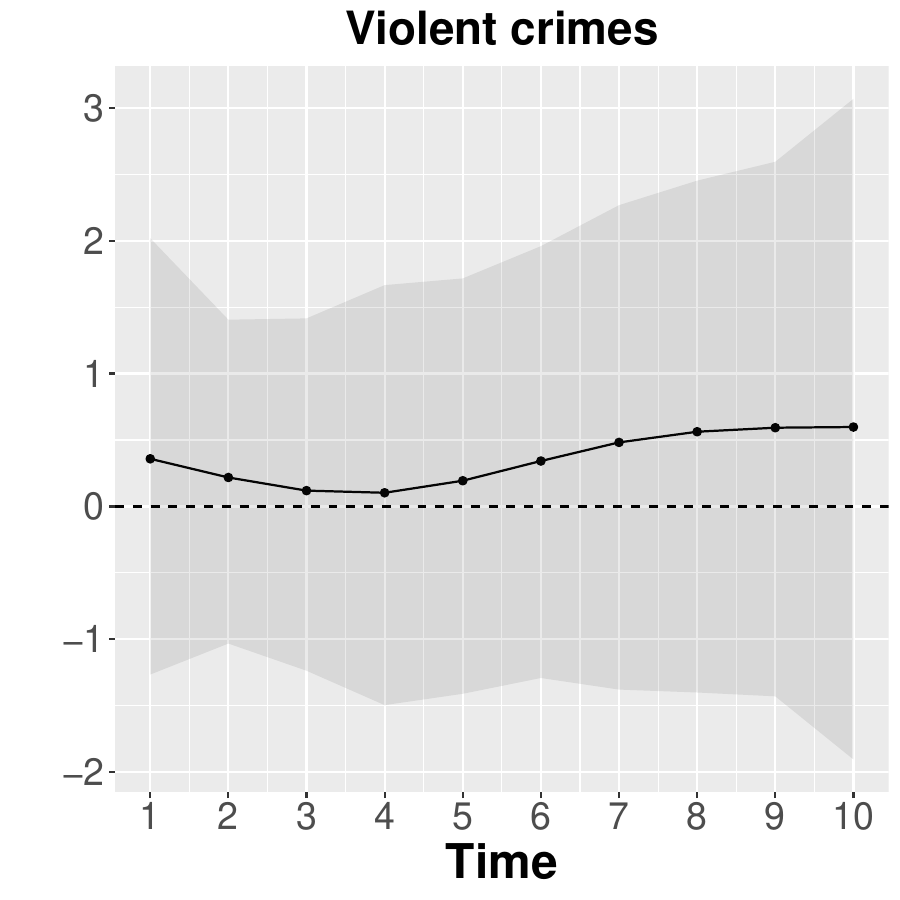}
		\includegraphics[width=0.24\linewidth]{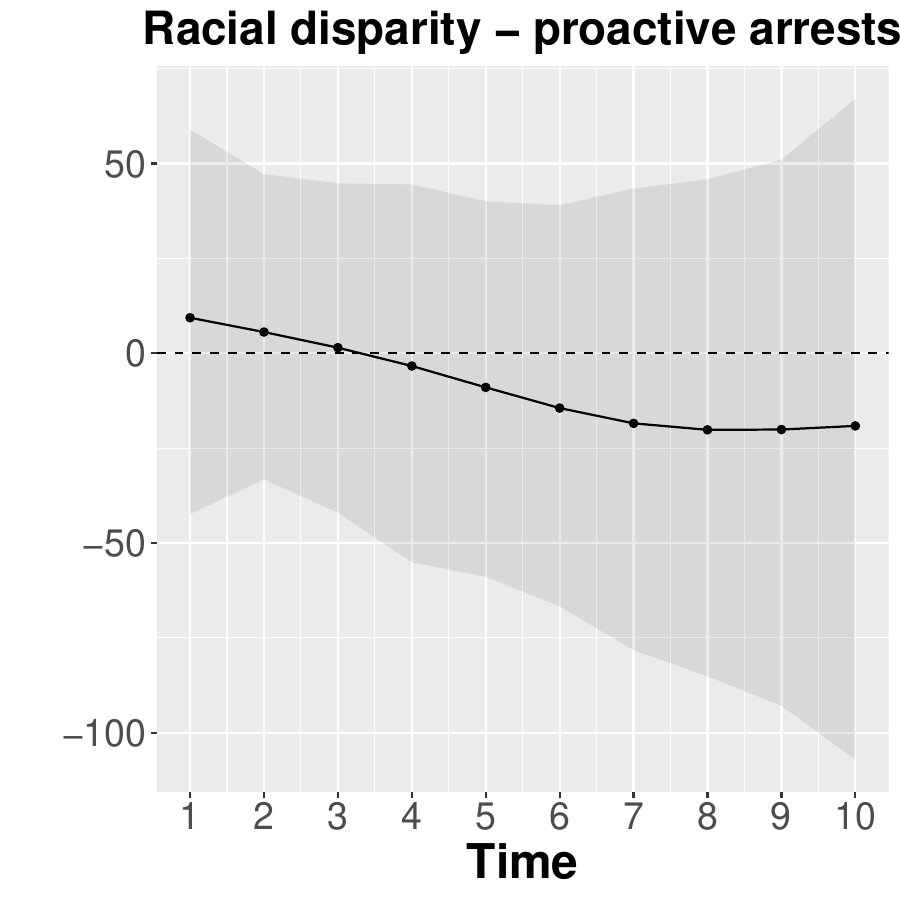}
		\caption{Estimates and 95\% credible intervals for time specific effects $\Delta(q)$ of neighborhood policing on misdemeanor arrests (first panel), proactive arrests (second panel), Violent crimes (third panel), and the difference of black and white proactive arrests (fourth panel).}
		\label{fig:NYCmarginal}
\end{figure}

First we focus on time-specific effects denoted by $\Delta(q)$ for $q=0, \dots 9$ and for each of the four outcomes. Throughout, we assume that $\Delta(q)$ is smooth in $q$ by using 3 degree of freedom splines in $f(\boldsymbol{X}_i, t-t_{i0})$. The estimates and pointwise 95\% credible intervals are depicted in Figure \ref{fig:NYCmarginal}. Estimates are negative for both misdemeanor and proactive arrests indicating that neighborhood policing leads to a reduction in low-level, discretionary arrests. The effect remains relatively constant over time for proactive arrests, and the credible interval only contains zero in the final three time points. The effect of misdemeanor arrests is negative at all time points, but decreases in magnitude at later time points, with the credible interval containing zero beginning at the fifth time point. The average number of misdemeanor and proactive arrests in the month before neighborhood policing implementation was 183 and 55, respectively. This indicates that the estimated reductions of 25 and 11.5 misdemeanor and proactive arrests in the first month of implementation are substantial (13.6\% and 20.9\%) reductions in arrest levels. The estimates of the effects on both violent crime and the difference in black to white arrest levels show essentially no effect of the policy, as the estimates for these two outcomes are very close to zero for all time points considered. These results indicate that the policy does not reduce crime levels, but does reduce arrests for low-level offenses. The increased community contact does not appear to lead to increased arrests. Instead, the policy's de-emphasis of arrests as a primary goal and its increased emphasis on community trust has led to fewer discretionary arrests. As a sensitivity analysis to confirm these results, we utilized both a difference in differences and synthetic control estimator in Appendix G. These rely on different assumptions and different model choices, but the overall findings remain relatively similar. In Appendix E, we evaluate the sensitivity of our results to model specification where we implement a vector autoregressive model for the outcome time series, and again find very similar results. 

\subsection{Heterogeneous effects}

\begin{figure}[htbp]
		\centering
		\includegraphics[width=0.98\linewidth]{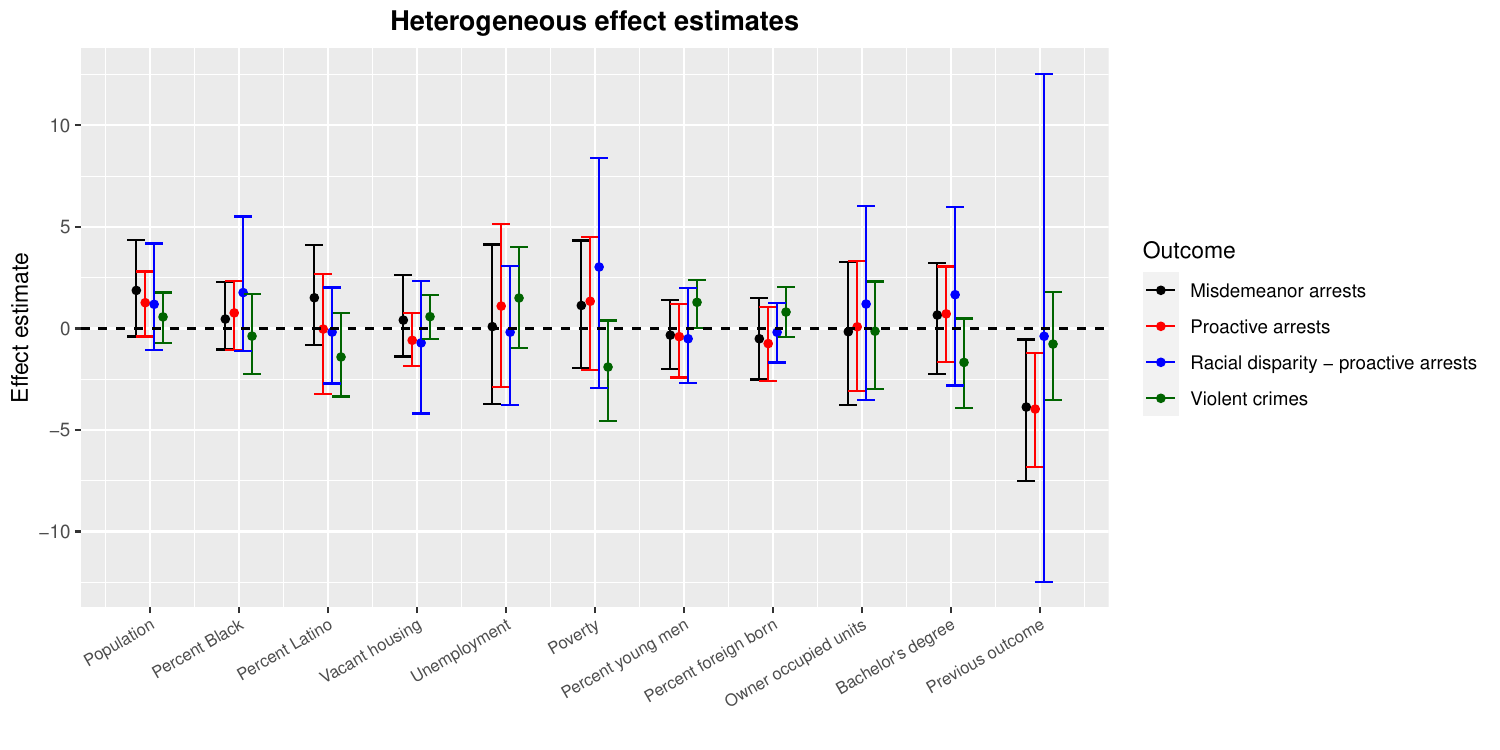}
		\caption{Estimates of coefficients from the heterogeneous treatment effect functions.}
		\label{fig:NYChetero}
\end{figure}

Understanding the impact of the policy on different communities is potentially more useful than marginal effects alone. It is possible that marginal effects show no effect of the policy, but only because certain communities have a positive treatment effect while others have a negative treatment effect. Ignoring such differences could lead to an incomplete assessment of the success (or lack thereof) of the policy. We estimate $\overline{\Psi}(j, 9)$ for each covariate using the $25^{th}$ and $75^{th}$ quantiles of the observed covariate distribution as $x_j$ and $x_j'$. We use a linear regression to model $f(\boldsymbol{X}_i, t-t_0) = \beta_0 + g(t - t_0) + \sum_{j=1}^p \beta_j X_{ij}$. We have tried more flexible approaches, such as random forests and super learners, but did not find substantively different conclusions and therefore we restrict attention to the simpler case here. Figure \ref{fig:NYChetero} shows the coefficient estimates ($\beta_j$) for each of the eleven covariates considered in our study. Precincts with higher prior values of the outcome have more negative treatment effects on misdemeanor arrests and proactive arrests, indicating that the reduction in arrests is more prominent in areas with higher arrest rates. 

\begin{figure}[htbp]
\caption{Illustration of the results of the clustering algorithm on NYC precincts as well as the estimates of the treatment effect on proactive arrests for each of these clusters.}
\label{fig:NYCcluster}
\begin{minipage}{0.48\textwidth}
	\includegraphics[width=0.9\linewidth]{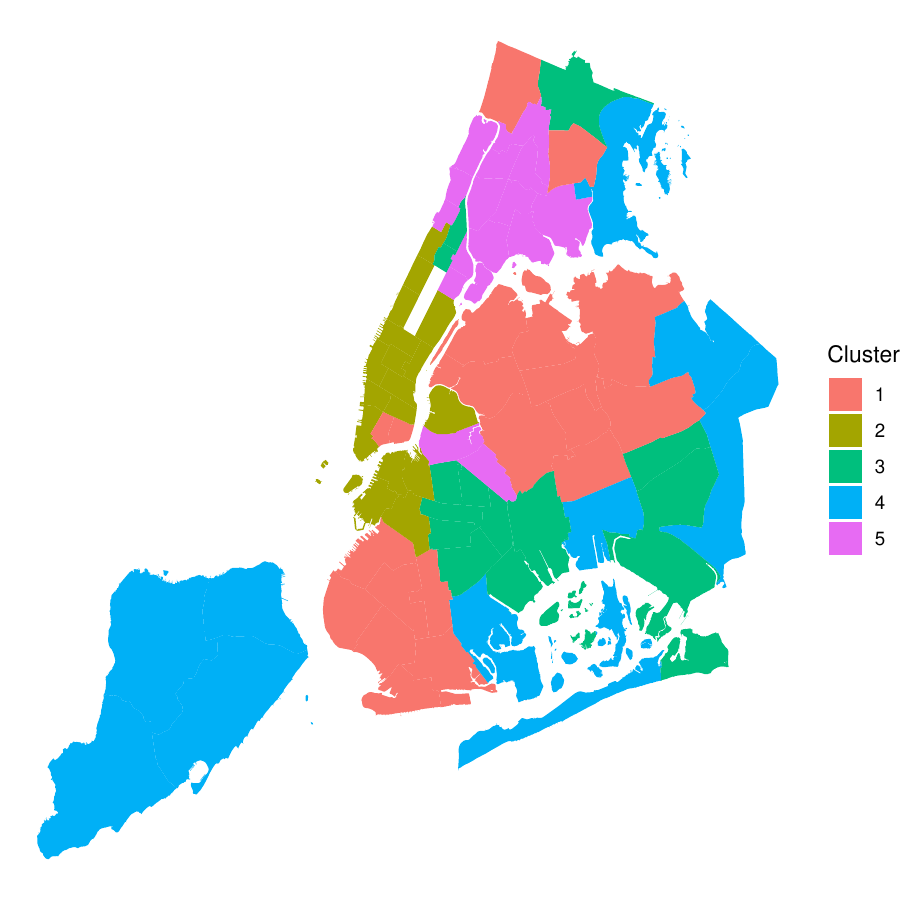}
\end{minipage}
\begin{minipage}{0.3\textwidth}
\begin{tabular}{lr}
  \hline
Cluster & Treatment Effect \\ 
  \hline
  1 & -9.86 (-15.48, -3.85) \\ 
    2 & -2.41 (-9.74, 4.75) \\ 
    3 & -13.44 (-28.96, -0.82)  \\ 
    4 & -7.7 (-16.46, 1.43) \\ 
    5 & -28.79 (-46.51, -10.91) \\ 
   \hline
\end{tabular}
\end{minipage}
\end{figure}

While these estimates of heterogeneity are interesting in their own right, they represent the effect of changing one covariate while fixing the remaining covariates. This may not be plausible in certain situations. For instance, we would expect that precincts with a higher percentage of residents with a bachelor's degree will have lower unemployment rates. In light of this, we attempted to group precincts based on their characteristics and estimate treatment effect heterogeneity across precincts of different profiles. To do this, we used k-nearest neighbors with $k=5$ to cluster our precincts into five distinct clusters. We chose $k=5$ as this led to clusters that were very diverse socioeconomically, and because the within cluster sum of squares did not decrease substantially after 5 clusters. These clusters can be visualized in Figure \ref{fig:NYCcluster}, which shows a clear spatial structure to the clusters of precincts. This is because neighboring precincts and certain boroughs of NYC tend to have similar covariate profiles. We then estimate the treatment effect within each of these clusters separately, and the results for proactive arrests are shown in Figure \ref{fig:NYCcluster}. It is clear that the treatment effect varies greatly across the different areas of NYC. There is no treatment effect in wealthier, predominantly white areas of Manhattan (cluster 2), while the treatment effect is significantly negative in working-class neighborhoods with higher proportions of Black or Latino people (clusters 1, 3, and 5) indicating that the number of proactive arrests dropped dramatically in those precincts. This heterogeneity in the treatment effect for proactive arrests is missed in Figure \ref{fig:NYChetero}, which only looks one covariate at a time, highlighting the importance of comparing distinct, plausible covariate values. 

\section{Discussion}

In this paper, we estimated the effect of neighborhood policing on crime and arrests in New York City using a novel approach to estimating causal effects of policies with staggered adoption that additionally estimates heterogeneity of the effects by observed covariates. We showed through realistic simulations based on the observed data that our approach is able to estimate the causal effects of a precinct-level treatment with good finite sample properties. We found that neighborhood policing reduces low-level arrests and that this effect was more pronounced in working-class neighborhoods of NYC with larger proportions of Black or Latino people. Another crucial takeaway from our analysis is that neighborhood policing does not reduce violent crime,  in alignment with previous criminological research showing community policing has null to minimal impacts on crime. This suggests that neighborhood policing, and possibly other policies that reduce arrests, can be implemented without increasing violent crime.

The city officials that launched neighborhood policing hoped it would promote racial equity. A large body of research reveals that police arrest Black people at starkly disproportionate rates, underscoring the importance of this goal. We found, however, that neighborhood policing had no impact on racial disparities in discretionary arrests. Even as police made fewer arrests, the racial balance remained the same. Changing the durable disparities in criminal justice outcomes will likely require more dramatic interventions. The policy was not without its benefits, however. Mayors and city councilmembers might want to reduce the number of low-level arrests their police departments make. Research suggests this is a worthwhile goal, as low-level arrests have negative consequences for both police and the people arrested while having minimal crime control benefits \citep{natapoff2018punishment, national2018proactive}. We found that adopting neighborhood policing would be an effective way to achieve fewer misdemeanor arrests without increasing crime.

One feature of the proposed work is that it is better suited for estimating short-term causal effects rather than long-term downstream effects of a policy. Our approach relies on forecasting the potential outcomes in the absence of the policy, and these forecasts become more uncertain over time. For instance, it is possible that the policy has an effect on violent crime, but that this effect takes longer to propagate than the 10 months considered here. Another feature of the proposed approach is the generalizability of results to other populations of interest, such as other cities that may adopt this policy. While our approach focuses on sample-level estimands unique to the population being studied, by looking at heterogeneity of the causal effect we may provide better insights into how this policy would affect other populations with different covariate distributions. 

As with any causal analysis of observational data, the validity of our approach depends on certain assumptions. One of these assumptions is that there is no spillover of the treatment effect into neighboring precincts. While we believe that this is a reasonable assumption in our study of neighborhood policing (Section \ref{sec:AnalysisAssumptions}), in other contexts it may be less likely to hold. In Appendix B, we extended our approach to allow the potential outcomes to depend on the treatment status of neighboring units \citep{verbitsky2012causal,papadogeorgou2019causal}, and found little evidence of spillover of the treatment effect, which increases our confidence in the results of the NYC policing study. Further research could look to improve these extensions or allow for other interference mechanisms, such as letting the potential outcomes depend on the proportion of treated units \citep{miles2019causal}. Importantly, our approach does not rely on an unconfoundedness assumption. This is critical because we can never know if we've measured all relevant covariates in observational studies, and unmeasured confounders are always a primary concern. We showed that our approach is instead based on a time series stationarity assumption.  While this is an assumption in its own right, it is one that can be evaluated in the pre-treatment time periods to assess its plausibility. We did this in the simulation study based on the observed New York City policing data, and found that our approach attained very good credible interval coverage that was at or near the nominal rate for all estimands. This shows that our proposed procedure is well-suited to the problem at hand, and therefore the results in the New York City data are much more believable.


\section*{Acknowledgements}
The authors would like to thank Georgia Papadogeorgou, Aaron Molstad, and Rohit Patra for extremely insightful comments on the manuscript.

\bibliographystyle{apa}
\bibliography{PoliceCausal}

\appendix

\setcounter{figure}{0}

\renewcommand{\thefigure}{A.\arabic{figure}}

\section{Identification of population estimands}

As discussed in the manuscript, sample treatment effects are generally not identified solely as a function of the observed data distribution, though we will show that the population counterpart of our estimands is identified under the stationarity, consistency, and no anticipatory effects assumptions. Our sample average treatment effect is the average difference between the potential outcome at the observed treatment time and the potential outcome assuming the treatment is never initiated. The population version of this estimand that we will target is $$E(Y_{t_0 + q}(t_0) - Y_{t_0 + q}(\infty) \vert T_0 = t_0).$$ Note that we condition on $T_0$ in the population estimand, because the sample estimand looks at the observed $T_0$ only, and does not examine what would happen had everyone received treatment at a particular time point, which would be analogous to a marginal estimand that does not condition on $T_0$. The first of these terms is immediately identifiable by the consistency assumption, which implies that $E(Y_{t_0 + q}(t_0) \vert T_0 = t_0) = E(Y_{t_0 + q} \vert T_0 = t_0)$, which is a function of the observed data distribution. Now, we can identify the second term:
\begin{align*}
    E(Y_{t_0 + q}(\infty) \vert T_0 = t_0) &= \int_{y_{t_0 + q}} y_{t_0 + q} f_{Y_{t_0 + q} (\infty) | T_0 = t_0}(y_{t_0 + q}) dy_{t_0 + q} \\
    &= \int_{y_{t_0 + q}} \int_{y_{t_0 + q-1}} y_{t_0 + q} f_{Y_{t_0 + q} (\infty) | Y_{t_0 + q - 1}(\infty), T_0 = t_0}(y_{t_0 + q - 1}) \\
    &\times f_{Y_{t_0 + q - 1} (\infty) | T_0 = t_0} dy_{t_0 + q - 1} dy_{t_0 + q} \\
    &= \int_{y_{t_0 + q}} \int_{y_{t_0 + q-1}} \dots \int_{y_{t_0}}  y_{t_0 + q} \\
    &\times \bigg[ \prod_{j=0}^q f_{Y_{t_0 + q - j} (\infty) | Y_{t_0 + q - j - 1}(\infty), T_0 = t_0}(y_{t_0 + q - j}) \bigg] \\
    &\times f_{Y_{t_0 - 1} (\infty) | T_0 = t_0} dy_{t_0} dy_{t_0 + 1} \dots dy_{t_0 + q} \\
    &= \int_{y_{t_0 + q}} \int_{y_{t_0 + q-1}} \dots \int_{y_{t_0}}  y_{t_0 + q} \\
    &\times \bigg[ \prod_{j=0}^q f_{Y_{t_0 + q - j} (\infty) | Y_{t_0 + q - j - 1}(\infty), T_0 = t_0}(y_{t_0 + q - j}) \bigg] \\
    &\times f_{Y_{t_0 - 1} | T_0 = t_0} dy_{t_0} dy_{t_0 + 1} \dots dy_{t_0 + q}
\end{align*}
The last equality held because $Y_{t_0 - 1}(\infty) = Y_{t_0 - 1}$ by the consistency assumption. The only remaining component of this expression that is not a function of the observed data is the density of $Y_{t_0 + q - j} (\infty)$ given both $T_0=t_0$ and $Y_{t_0 + q - j - 1}(\infty)$, denoted by $f_{Y_{t_0 + q - j} (\infty) | Y_{t_0 + q - j - 1}(\infty), T_0 = t_0}(y_{t_0 + q - j})$. The stationarity assumption, however, makes this a function of observed data as we can assume that this conditional distribution after time $t_0$ is the same as in the periods before $t_0$ where the potential outcome is fully observed under the consistency and no anticipatory effects assumptions.  Note we have used densities throughout here, but if the outcome is a discrete random variable, then analogous expressions would hold. Throughout we have only conditioned on one time point in the past, however, the identification formula holds in the same manner if additional time points are considered. While not necessary for identification, conditioning on additional time points can 1) improve efficiency of results, and 2) make it more likely that the aforementioned stationarity assumption holds. Additionally, we could adjust for time-varying covariates, though this requires an additional assumption that the covariates are unaffected by treatment. 

\section{Alleviating the no interference assumption}

In this section, we describe two distinct extensions to our approach in order to remove the no interference assumption and estimate spillover effects of neighboring precincts adopting the policy. To the best of our knowledge, interference in panel-data settings has not been rigorously studied. In a simplified setting, interference was addressed in \cite{menchetti2020estimating}. They have pairs of data points which can interfere with each other, but the nature of the interference does not change over time. In our data example, we have a large number of data points that could in principle interfere with each other, and whether or not a unit has nearby precincts with the policy is constantly changing over time as units adopt treatment at vastly different times.

When interference is present, potential outcomes should be denoted by the time of treatment initiation of all precincts. Specifically, we can extend our potential outcomes to be $Y_{it}(t_0, \boldsymbol{t}_{-i})$ where $t_0$ is the treatment start time for unit $i$ while $\boldsymbol{t}_{-i}$ represents the starting time for the remaining units. There are far too many possibilities for possible treatment times of the remaining units, and therefore we must simplify the interference structure somewhat so that the potential outcome does not depend on the treatment status of all other units. Let $g_t(\boldsymbol{t}_{-i})$ be a function of the treatment times for all of the remaining individuals. We will make the following assumption, which is less restrictive than the no interference assumption adopted in the original manuscript: \\
\\
\textit{Neighboring interference assumption:} For two sets of treatment times for the remaining precincts given by $\boldsymbol{t}_{-i}$ and  $\boldsymbol{t}'_{-i}$, we have that $Y_{it}(t_0, \boldsymbol{t}_{-i}) = Y_{it}(t_0, \boldsymbol{t}'_{-i})$ if $g_t(\boldsymbol{t}_{-i}) = g_t(\boldsymbol{t}'_{-i})$. \\
\\
This assumption is closely related to the exposure mappings found in recent proposals for estimating spillover effects in network settings \citep{aronow2017estimating, forastiere2021identification}. For both proposals, we will make the simplifying assumption that the interference is restricted to the presence of a neighbor that has implemented the policy. Intuitively, this means that we are assuming that the treatment status of neighboring units can affect the outcomes of a particular unit, but that the treatment status of far away precincts can not affect a unit's outcome time series. Additionally, we are assuming that the number of treated neighbors does not matter, and that it is only the presence or absence of a treated neighbor that affects the potential outcome. In our setting, we let $g_t(\boldsymbol{t}_{-i})$ be an indicator of whether any of the neighbors to unit $i$ have already began treatment at time $t$. Formally, we can write this function as $g_t(\boldsymbol{t}_{-i}) = 1( \exists \ j \in \mathcal{N}_i : t_{j0} \leq t)$, where $\mathcal{N}_i$ represents the set of precincts that are geographic neighbors of precinct $i$. With this simplification, we can now define potential outcomes as $Y_{it}(t_0, g)$, where $t_0$ is the time that unit $i$ adopts treatment, and $g$ is a binary indicator of whether any of the neighbors of unit $i$ have began treatment by time $t$. Further, let $G_{it}$ be the random variable denoting whether $g_t(\boldsymbol{T}_{-i}) = 1$, i.e. whether unit $i$ has a neighbor that has already begun treatment by time $t$. We now discuss two different estimands and estimation strategies for estimating the effects of having a neighbor with the policy. 

\subsection{Assessing spillover of the realized policy on not-yet treated precincts}

\begin{figure}[htbp]
    \centering
    \includegraphics[width = 0.9\linewidth]{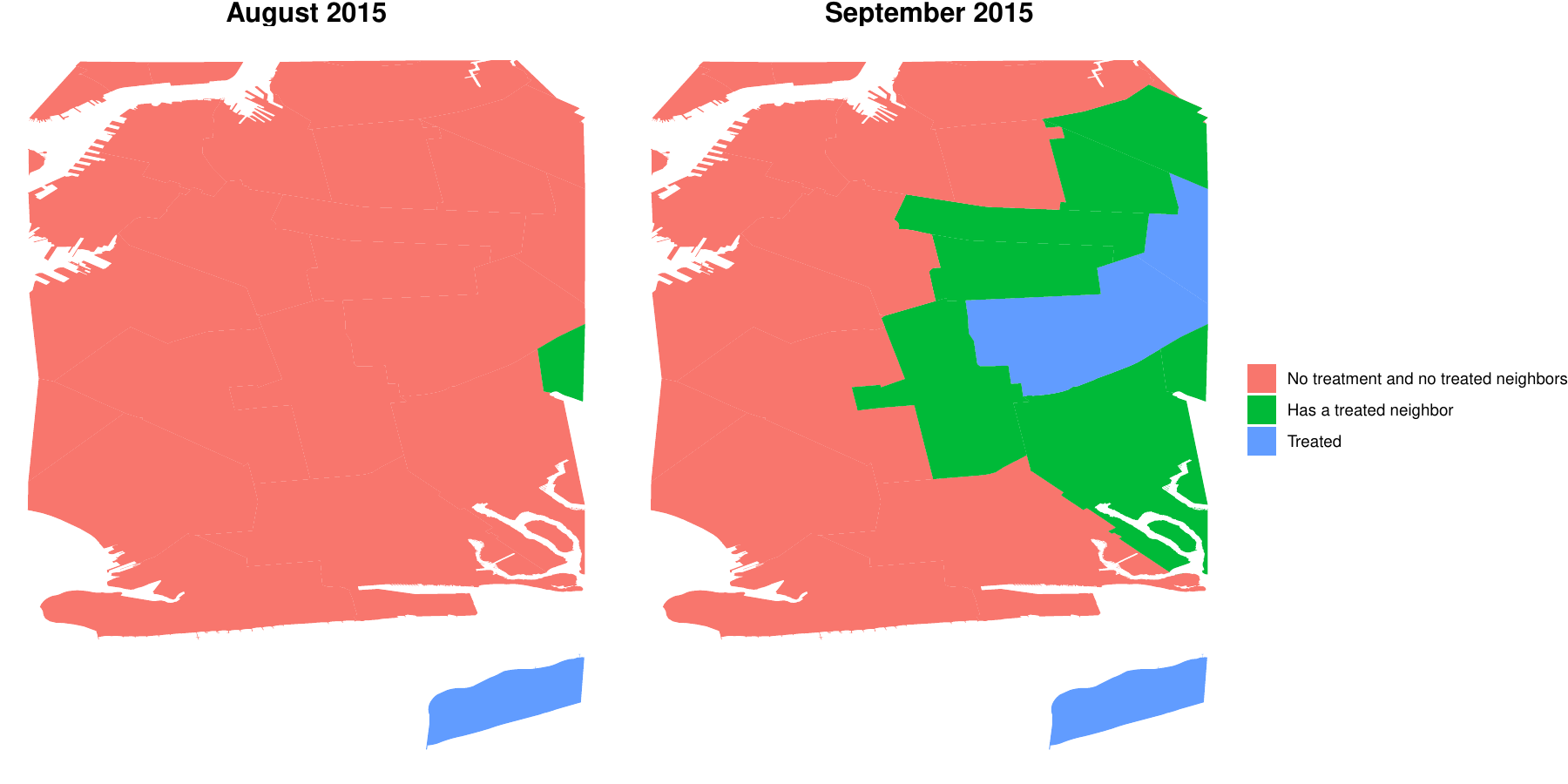}
\caption{Treatment status of precincts in two successive months. Blue denotes a unit that has already adopted policy, while green denotes units that have not yet adopted policy and have a neighboring precinct with the policy. }
\label{fig:IllustrationAugSep}
\end{figure}

One way in which interference can manifest is that untreated precincts can experience an effect of neighboring treatments receiving the policy before they themselves adopt the policy. In terms of our potential outcomes, this can be denoted by $$Y_{it}(\infty, 1) - Y_{it}(\infty, 0). $$
The first of these two quantities will be fully observed if at time $t$, unit $i$ does not yet have the policy implemented, but one of their neighbors does. On the other hand, $Y_{it}(\infty, 0)$ is fully observed if at time $t$, neither unit $i$ or their neighbors have adopted the policy yet. We take a similar strategy to estimation as in the main manuscript, which is to build a model for $\boldsymbol{Y}_{t}(\infty, 0)$. This amounts to fitting a model for all of the observed data prior to either a unit initiating treatment or a neighbor initiating treatment. At this point, we will use our model to forecast $Y_{it}(\infty, 0)$ in these time periods after either a unit or their neighbor becomes treated. We can then compare this to the fully observed value of $Y_{it}(\infty, 1)$ to see what the effect of the neighboring treatment is on unit $i$. Note that this quantity will not be observed for all $n$ precincts in the sample. Some precincts are treated before their neighbors and we will never be able to observe $Y_{it}(\infty, 1)$ for them. For this reason, we will restrict attention to the subset of the sample that is treated after one of their neighbors receives treatment. Let $\mathcal{F}_q$ be the set of indices in $1, \dots, n$ that adopt policy at least $q$ time points after their first neighbor adopts the policy. For an illustration of such as situation, see Figure \ref{fig:IllustrationAugSep}. We see that in August of 2015 most units are not yet treated, but in September 2015 two precincts on the right of the figure become treated. The green precincts are those without the policy, but also have a neighboring precinct that has already adopted the policy. Each of these green precincts will be included in $\mathcal{F}_0$ because they have a neighbor with treatment, and they are not yet treated at this time. If these precincts remain untreated in the following month, they will be included in $\mathcal{F}_1$, and this process continues until they are treated and then they are no longer included in this set.

Before we can introduce our full estimand, we require one more piece of notation. Again let $T_{i0}$ be the time at which unit $i$ begins treatment, but now let $N_{i0}$ be the first time period for which a neighbor of precinct $i$ has adopted treatment. Our estimand can be written as 
$$I(q) = \frac{1}{|\mathcal{F}_q|} \sum_{i \in \mathcal{F}_q} \bigg\{ Y_{i, N_{i0}+q}(\infty, 1) - Y_{i, N_{i0}+q}(\infty, 0) \bigg\}.$$

\begin{figure}[htbp]
    \centering
    \includegraphics[width = 0.4\linewidth]{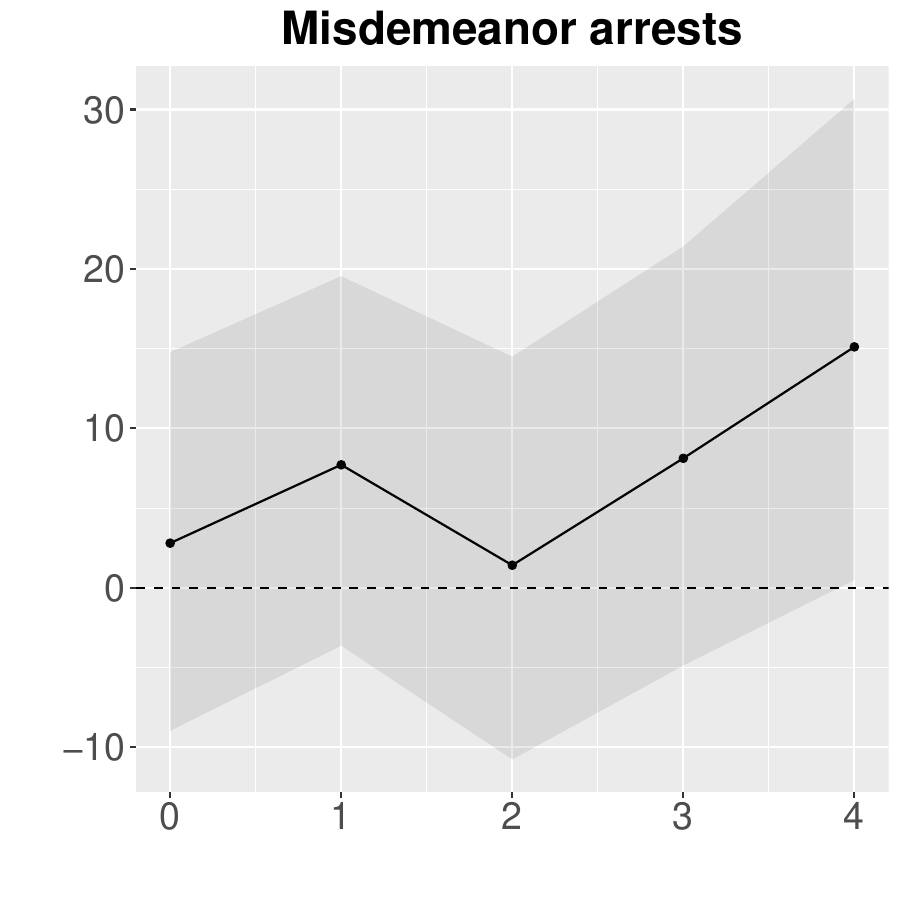}
    \includegraphics[width = 0.4\linewidth]{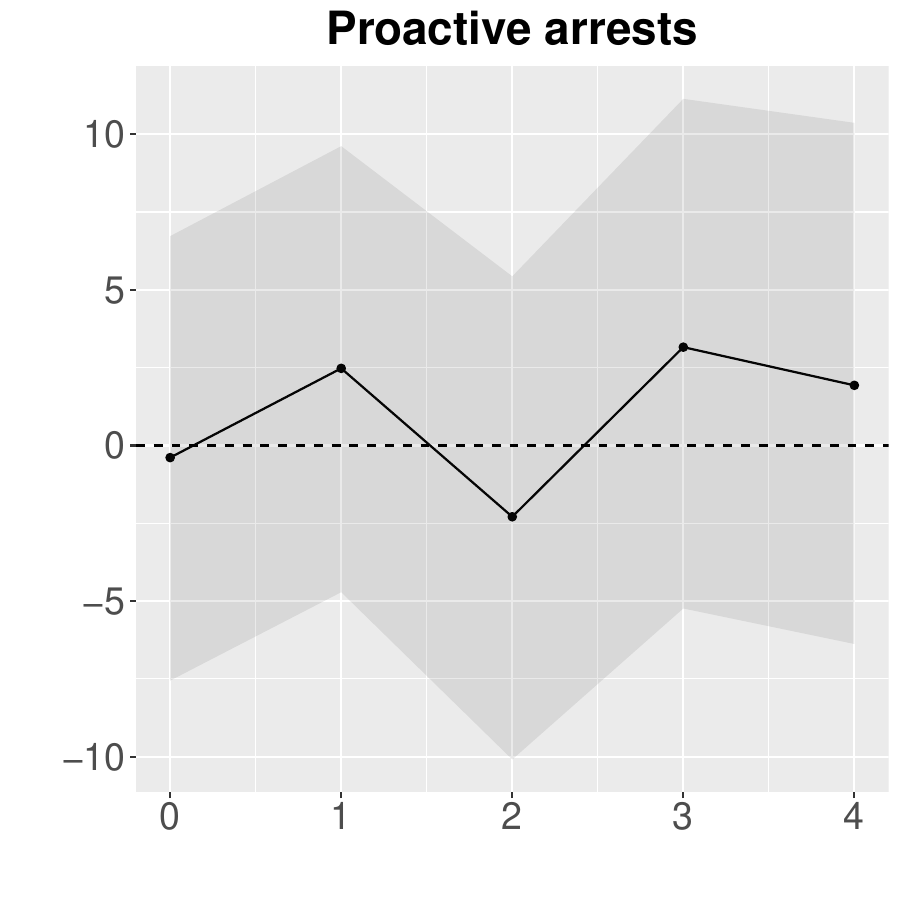} \\
    \includegraphics[width = 0.4\linewidth]{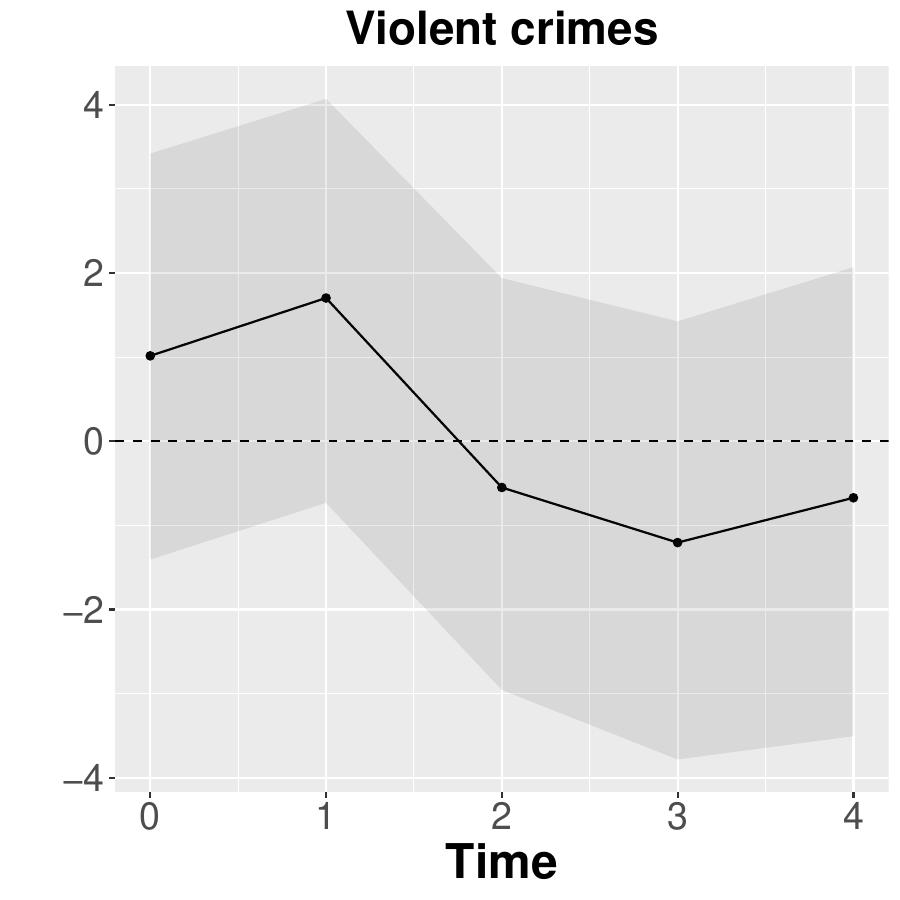}
    \includegraphics[width = 0.4\linewidth]{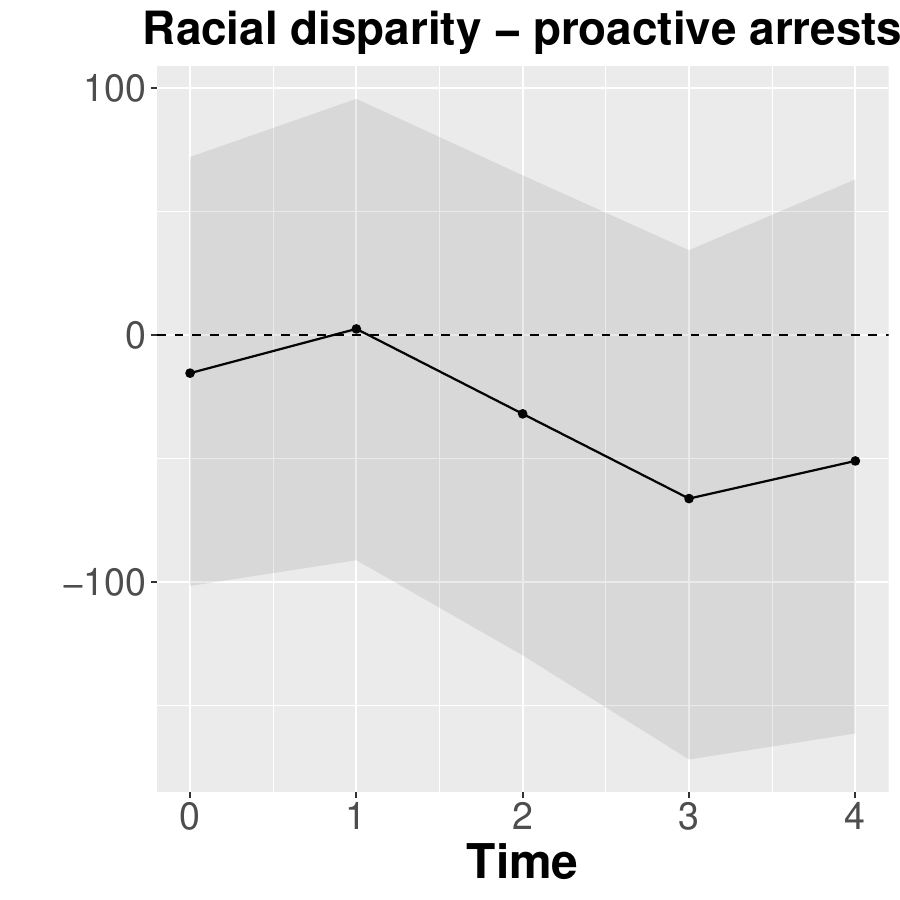}
\caption{Estimates of $I(q)$ for $q=0, 1, \dots, 4$ under each of the four outcomes considered. }
\label{fig:Iq}
\end{figure}

This can be interpreted as the average effect on untreated precincts of having a neighbor adopt policy, and this effect is allowed to vary by the time since the neighbor adopted policy. This has the added complication that the sample over which we are averaging our effects is changing at each time period $q$. Note that the set $\mathcal{F}_q$ is decreasing in size with $q$, i.e. $|\mathcal{F}_q| \geq |\mathcal{F}_{q+1}|.$ We find that 75\% of our sample is included in $\mathcal{F}_0$ and this steadily decreases as we increase the amount of time periods we look forward due to the fact that more units become treated. For this reason, we only examine $I(q)$ for $q=0,1,\dots, 4$ so that the sample being averaged over remains relatively stable. The estimates and corresponding 95\% credible intervals can be found in Figure \ref{fig:Iq}. We see that there does not appear to be any strong spillover effects on precincts that have not yet adopted the policy. Most of the estimates are close to zero with credible intervals that contain zero, with the one exception being $I(4)$ for misdemeanor arrests. The outcome with the largest estimated treatment effect in the manuscript when interference was ignored, proactive arrests, shows no spillover effect of the policy. This increases our belief in our no interference assumption of the manuscript and strengthens the findings of the manuscript when interference was assumed away. To further assess the plausibility of the no interference assumption, we explore a separate approach in the following section.

\subsection{Simple parameterization of interference effects}

\begin{figure}[htbp]
    \centering
    \includegraphics[width = 0.4\linewidth]{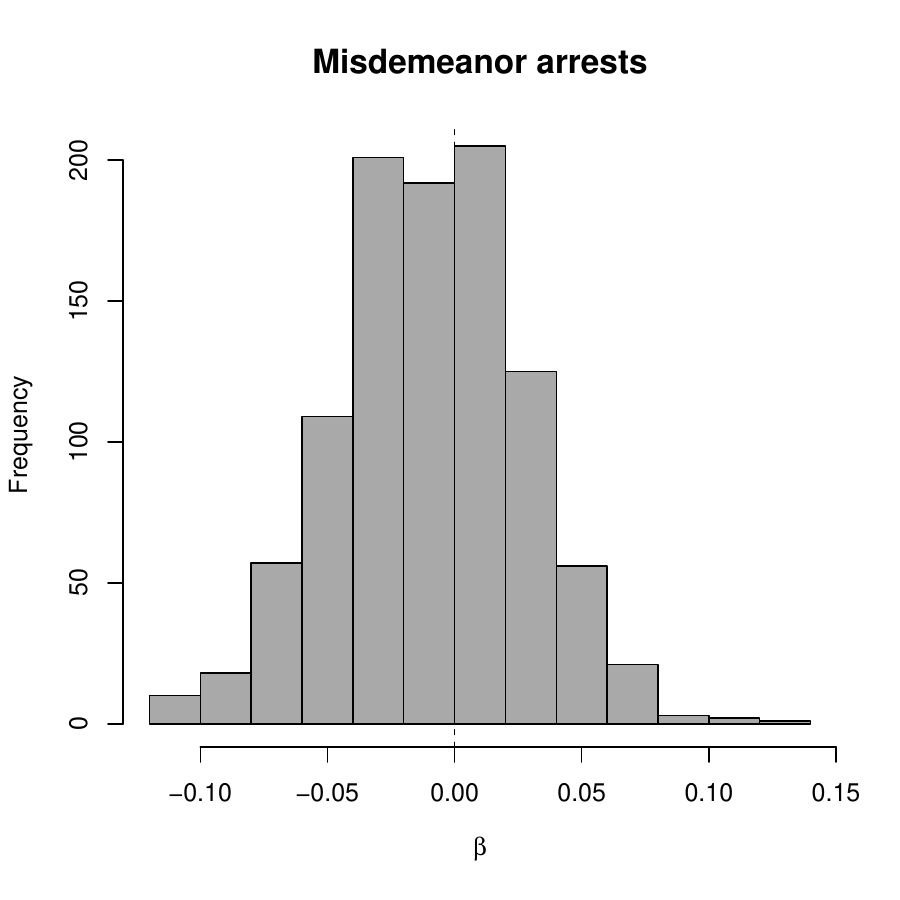} 
    \includegraphics[width = 0.4\linewidth]{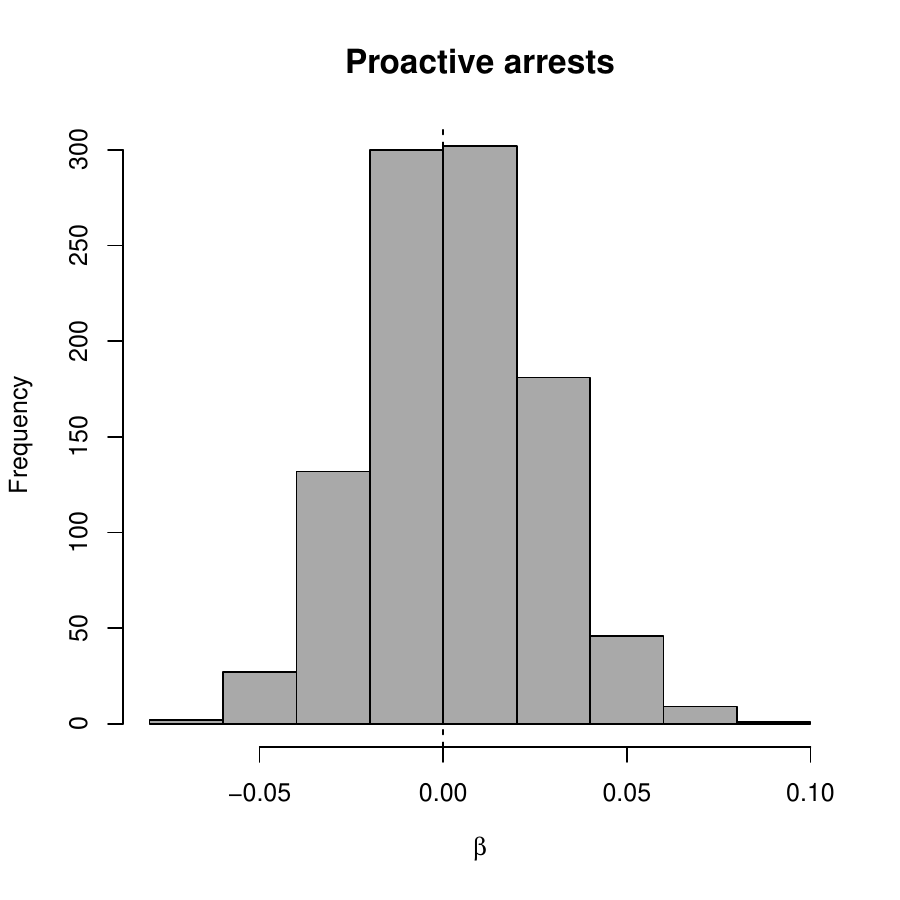} \\ 
    \includegraphics[width = 0.4\linewidth]{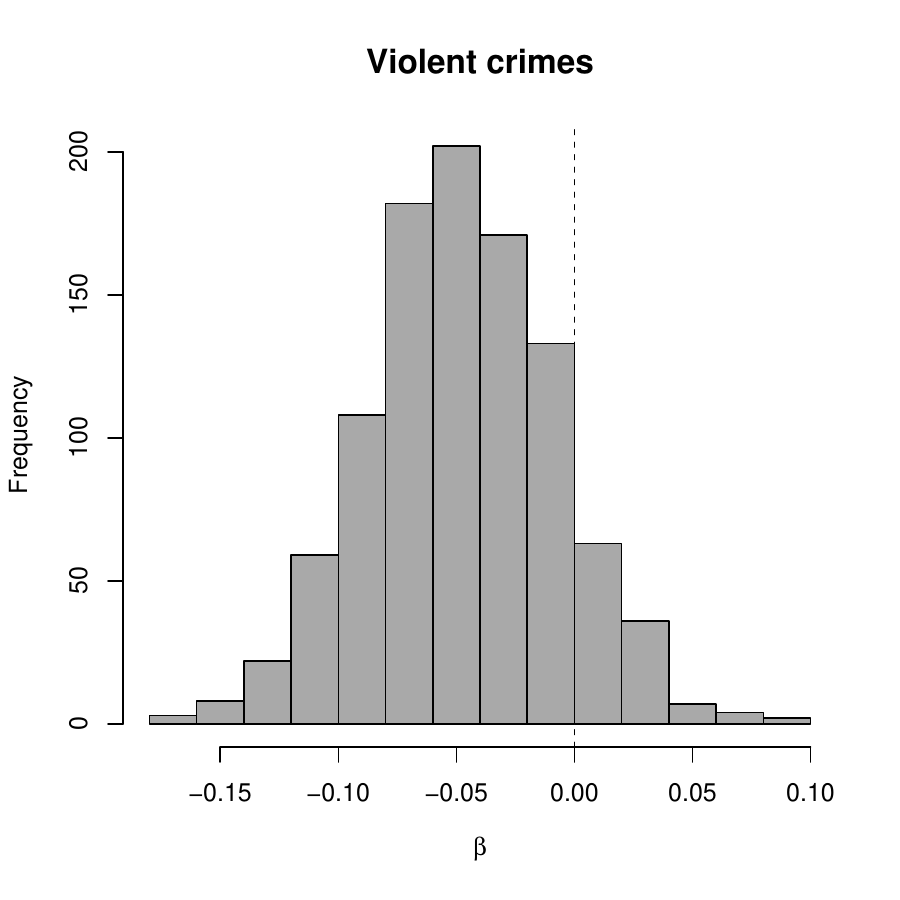} 
    \includegraphics[width = 0.4\linewidth]{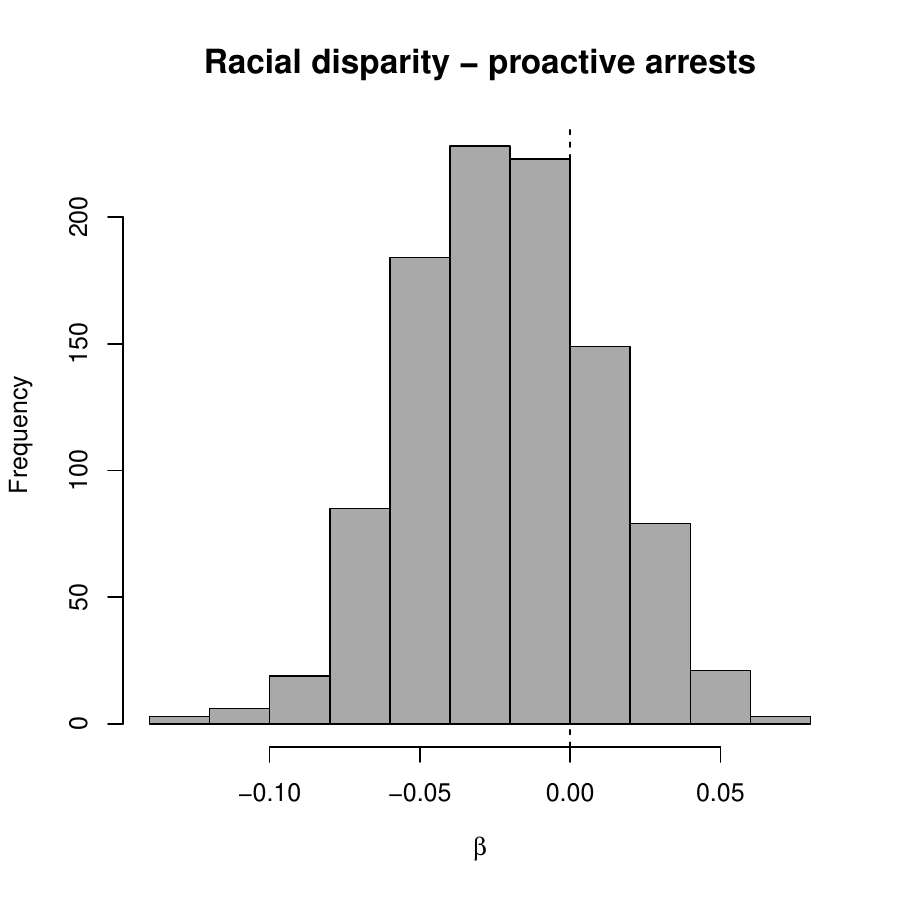}
\caption{Posterior distribution of $\beta$ for each of the four outcomes considered. A value of $\beta=0$ indicates there is no spillover of the treatment effect. }
\label{fig:InterferenceParameter}
\end{figure}

Here, we aim to separate the overall effect of the policy into two distinct effects: one that targets spillover effects of the policy, and one that targets the direct effect of the policy on the precinct that it is applied to. Specifically, we target the estimands defined as:
\begin{align*}
    \Delta^{sp}(q) &= \frac{1}{n} \sum_{i=1}^n \bigg\{ Y_{i, T_{i0} + q}(\infty, G_{i, T_{i0}+q}) - Y_{i, T_{i0} + q}(\infty, 0) \bigg\} \\
    \Delta^{dir}(q) &= \frac{1}{n} \sum_{i=1}^n \bigg\{ Y_{i, T_{i0}+q}(T_{i0}, G_{i, T_{i0}+q}) - Y_{i, T_{i0}+q}(\infty, G_{i, T_{i0}+q}) \bigg\}.
\end{align*}
The first of these two quantities targets a spillover effect of precincts that have already adopted the policy influencing the outcomes of those precincts that have not yet had the policy. One can interpret $\Delta^{sp}(q)$ as the average effect of neighboring precinct's policy decisions when a precinct does not yet itself have the policy implemented. Note that it looks at the realized treatment of neighbors given by $G_{i, T_{i0}+q}$, which is zero for some of the precincts in the study and one for others. This means that it represents the impact of the policy on untreated neighbors under the rollout of the policy that was observed in the study. On the other hand, $\Delta^{dir}(q)$ can be seen as average impact of having the policy on precincts, given the observed status of the neighbors of those precincts. Interestingly, if we sum these two quantities we obtain 
$$\Delta^{tot}(q) = \Delta^{sp}(q) + \Delta^{dir}(q) = \frac{1}{n} \sum_{i=1}^n \bigg\{ Y_{i, T_{i0}+q}(T_{i0}, G_{i, T_{i0}+q}) - Y_{i, T_{i0} + q}(\infty, 0) \bigg\},$$ the overall impact of the policy as it was adopted in New York City, which is closely related to $\Delta(q)$ of the main manuscript. We know that under a consistency assumption, we have that $Y_{i, T_{i0}+q}(T_{i0}, G_{i, T_{i0}+q})$ is a fully observed quantity. The remaining quantities are unobserved, and effectively amount to needing to estimate $Y_{i, T_{i0} + q}(\infty, g)$ for $g=0$ and $g=G_{i, T_{i0}+q}$. To estimate these missing counterfactuals, we fit the following model to the precincts and time periods prior to them adopting the intervention themselves:
\begin{align*}
\boldsymbol{Y}_t &= \boldsymbol{\mu}_t + \beta \boldsymbol{G}_t + \boldsymbol{\epsilon}_t \nonumber \\
\boldsymbol{\mu}_t &= \boldsymbol{\mu}_{t-1} + \boldsymbol{\delta}_{t-1} + \boldsymbol{\eta}_t^{\mu} \\
\boldsymbol{\delta}_t &= \boldsymbol{\delta}_{t-1} + \boldsymbol{\eta}_t^{\delta} \\
\boldsymbol{\eta}_t^{\mu} &\sim \mathcal{N}(\boldsymbol{0},  \boldsymbol{D}_{\mu}) \\
\boldsymbol{\eta}_t^{\delta} &\sim \mathcal{N}(\boldsymbol{0}, \boldsymbol{D}_{\delta})
\end{align*}
Here $\beta$ is a scalar quantity that captures the effect of neighboring precincts' treatment status on a particular unit. If $\beta=0$, then $\Delta^{sp}(q) = 0$ and there is no spillover of the policy. Also note that we are using the same $\beta$ parameter for each unit's time series, and therefore we must standardize each unit's time series before fitting this model to ensure that the magnitude of the effect of neighboring treatment status is shared across units. We fit this model to each of the four outcomes of interest in our study, and Figure \ref{fig:InterferenceParameter} shows the resulting posterior distributions for $\beta$ under each outcome. We see that for each of the four outcomes considered, zero clearly lies within the posterior distribution of $\beta$ suggesting that interference does not play a large role in any of our analyses. 

\begin{figure}[htbp]
    \centering
    \includegraphics[width = 0.4\linewidth]{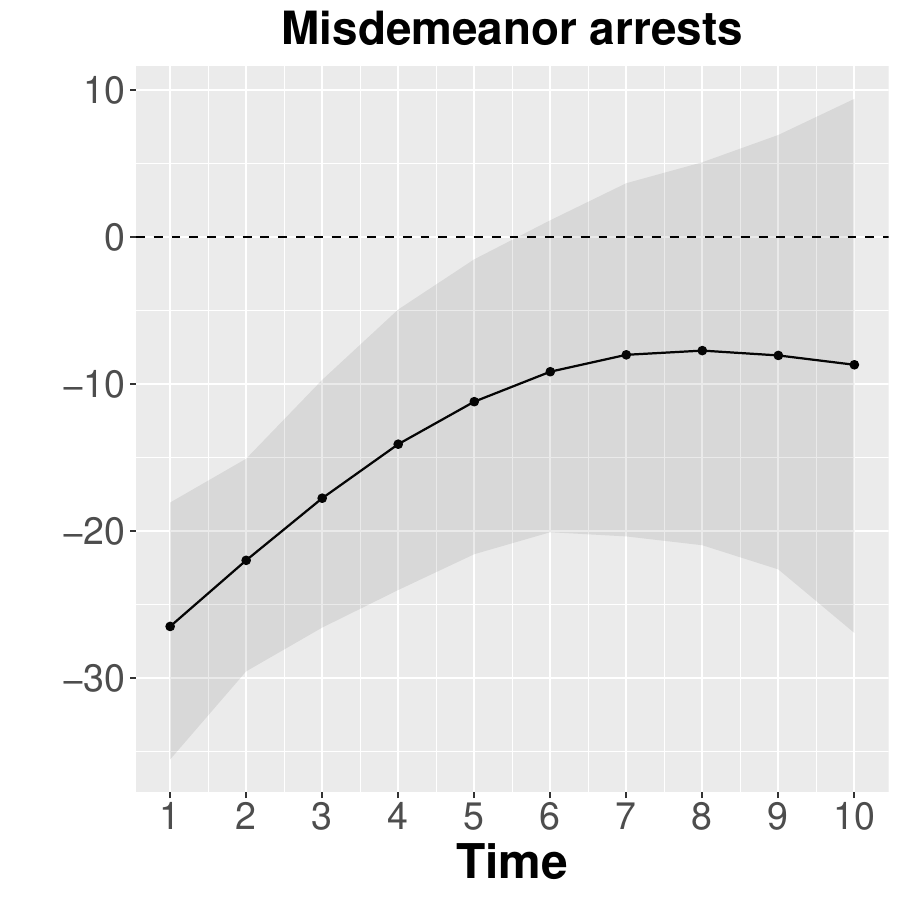}
    \includegraphics[width = 0.4\linewidth]{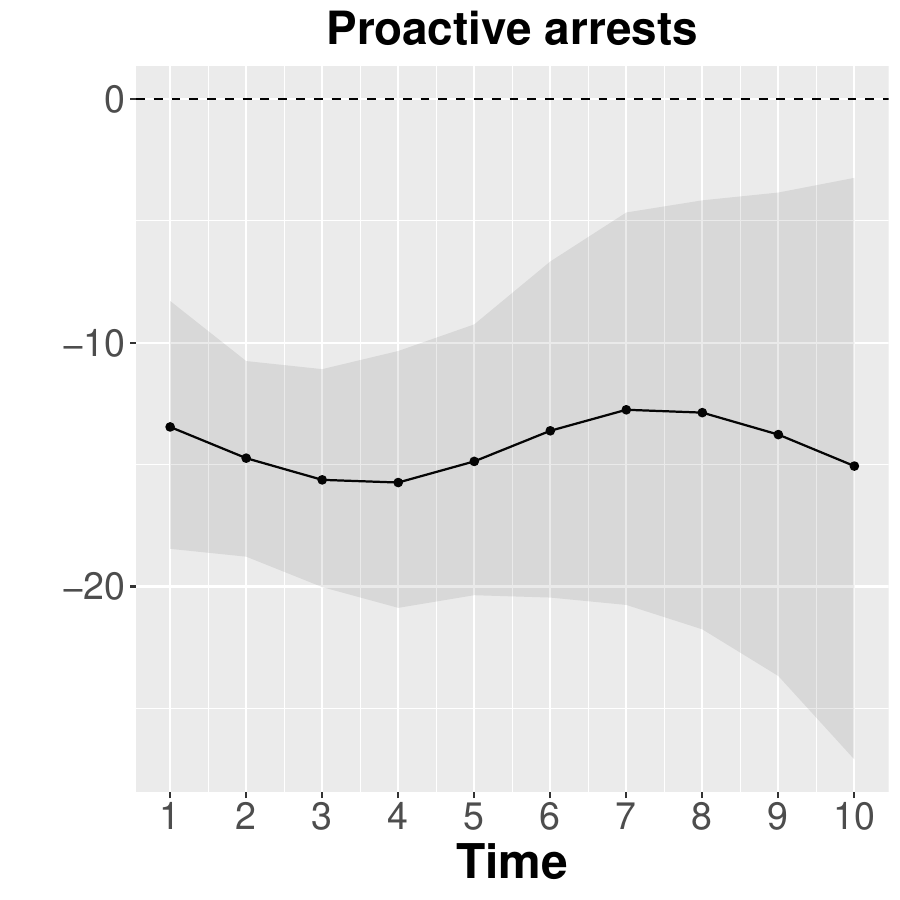} \\
    \includegraphics[width = 0.4\linewidth]{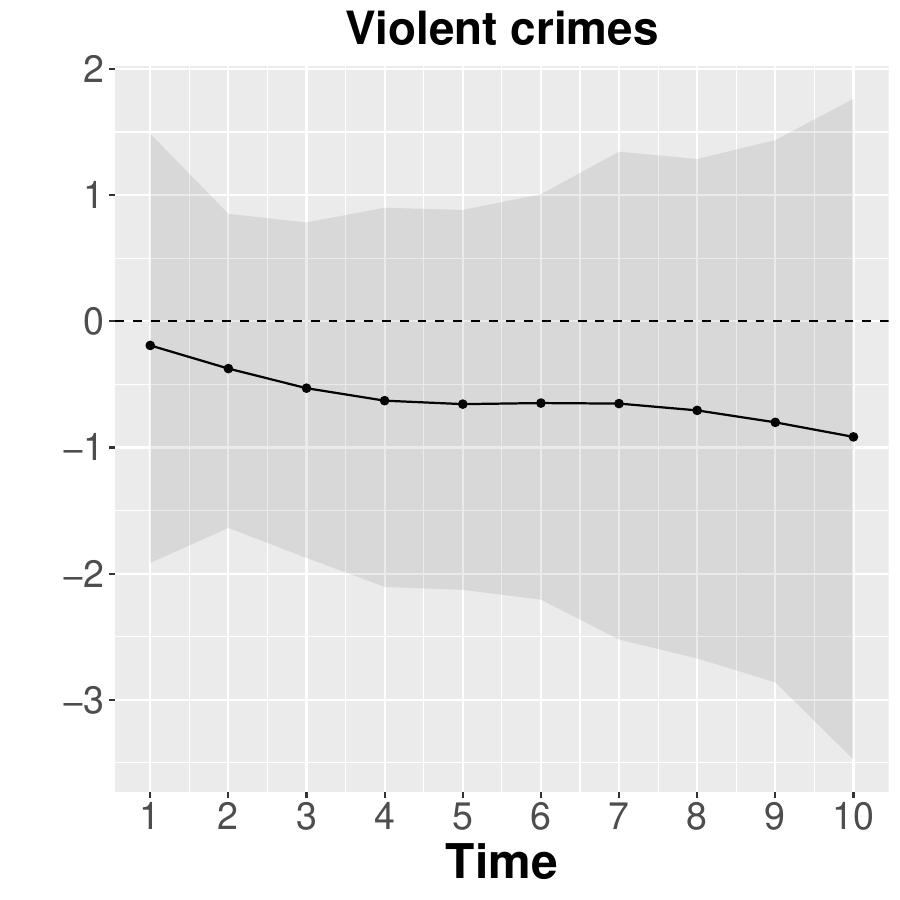}
    \includegraphics[width = 0.4\linewidth]{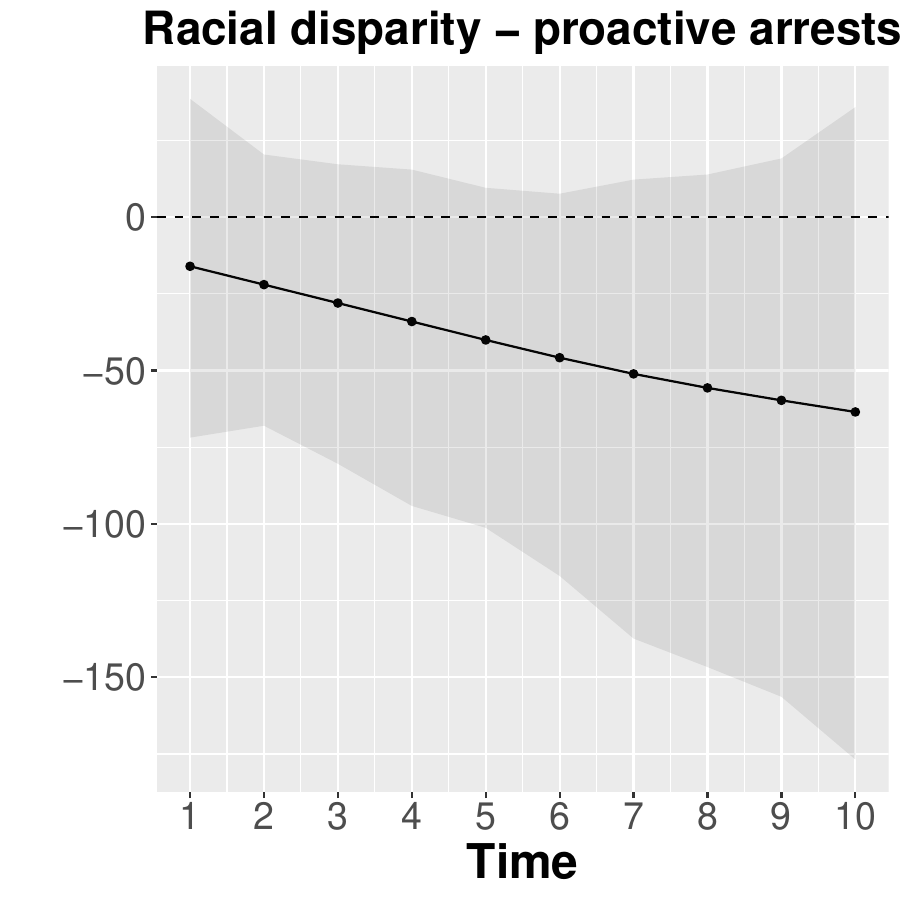}
\caption{Estimates of $\Delta^{dir}(q)$ for $q=0, 1, \dots, 9$ under each of the four outcomes considered.}
\label{fig:Delta2}
\end{figure}

We can also examine estimates of $\Delta^{dir}(q)$, which corresponds to the average direct effect of treatment on the precincts. The results can be found in Figure \ref{fig:Delta2}. We see results that closely mirror the effects of neighborhood policing that were seen in the main manuscript. There is a negative and significant effect of the policy on both proactive arrests and misdemeanor arrests, and the magnitude of these effects closely mirror the estimates of $\Delta(q)$ from the manuscript. We do not see large direct effects of the policy on either violent crimes or the racial disparity measure, which again agrees with the findings of the manuscript. 

\section{Example highlighting differences between stationarity and existing assumptions}

\label{sec:DifferentEstimators}

Here we highlight a simple example that illustrates how the approach taken in the current paper can still provide unbiased results of treatment effects even in the presence of a time-varying confounder that is not measured, which can negatively impact results under different assumptions. To allow for the use of existing approaches to panel data causal inference problems, we generate data with both treated units and units that never receive treatment. Specifically, we simulate $n=30$ independent units measured over $T=80$ time points. We generate an unmeasured, time-varying covariate $\boldsymbol{U}$ for each unit from an AR(1) process with correlation $\rho$ between successive time points. It will help to write this as $$U_{it} = \alpha U_{i, t-1} + \eta_{it}$$
where $\eta_{it}$ is random noise and $\alpha$ dictates the degree of temporal correlation. We generate the treatment times $T_{i0}$ to be a function of the unmeasured covariate. Specifically, all units are untreated until time 76 when certain units are treated. We set $A_{i,76} = 1(U_{i, 75} > 0)$, which means that all individuals who have a positive value of the unmeasured confounder at time 75 are then treated and remain treated for the remaining time points. This can also be written as
\[ T_{i0} = \begin{cases} 
      76 & U_{i,75} > 0 \\
      \infty & \text{otherwise.} 
   \end{cases}
\]
In this setting, we have an unmeasured variable that perfectly determines the treatment time of individuals. The potential outcomes in the absence of treatment are given by
$$Y_{it}(0) = \beta t + \gamma U_{it} + \epsilon_{it}, \quad \epsilon_{it} \sim \mathcal{N}(0,1).$$
The potential outcomes are therefore increasing with time, but are also a function of the unmeasured confounder. Note that this situation breaks the assumptions of a number of existing panel data methods, which assume that the distribution of the residuals for the potential outcome are the same in the treated and control groups. Clearly this is violated here, as the residuals are a function of both $\epsilon_{it}$ and $U_{it}$, but the distribution of $U_{it}$ differs in the treated and control groups since treated individuals have positive values of this unmeasured variable at time 75, while control units have negative values. We will show for the difference in differences (DID) and synthetic control (SC) estimators how this can lead to biased estimation, but how the stationarity assumption still holds in this setting and therefore our approach can provide valid inference in this setting. For simplicity, we assume that $Y_{it}(1) = Y_{it}(0) + \tau$ and therefore there is a constant treatment effect over time and across different populations. 

\subsection{Impact on existing estimators}

Here we show that the difference in difference and synthetic control estimators will not be able to provide unbiased estimates of causal effects in this setting. First we can look at the DID estimator, which relies on a parallel trends assumption. In our setting, this assumption is that $$E(Y_{i, 76}(0) - Y_{i,75}(0) | T_{i0} = 76) = E(Y_{i, 76}(0) - Y_{i,75}(0) | T_{i0} = \infty).$$
Under the simple model described above, we can easily show that this equality does not hold, since
\begin{align*}
    E(Y_{i, 76}(0) - Y_{i,75}(0) | T_{i0} = 76) &= E(Y_{i, 76}(0) - Y_{i,75}(0) | U_{i, 75} > 0) \\
    &= \beta + \gamma E(U_{i, 76} - U_{i,75} | U_{i, 75} > 0) \\
    &= \beta + \gamma E((\alpha - 1) U_{i,75} + \eta_{i, 76} | U_{i, 75} > 0) \\
    &= \beta - \gamma (1 - \alpha) E(U_{i,75} | U_{i, 75} > 0) \\
    &\neq \beta - \gamma (1 - \alpha) E(U_{i,75} | U_{i, 75} \leq 0) \\
    &= E(Y_{i, 76}(0) - Y_{i,75}(0) | T_{i0} = \infty).
\end{align*}
We see that the parallel trends assumption does not hold and the degree to which it does not hold depends on both $\gamma$ and $\alpha$. 

The synthetic control approach is a little more difficult to derive analytically what would happen in this context, but we can provide intuition for its performance here. Synthetic controls aim to find weights $w_{i}$ for treated individual $i$ such that
$Y_{it}(0) \approx \sum_{j : T_{j0} = \infty} w_{ij} Y_{jt}(0)$. This means that for each treated unit, we try to find a weighted linear combination of control units that well approximates the potential outcome of the treated unit, which we can use to impute the missing potential outcome in the absence of the treatment. These weights are assigned a sum to one constraint such that $\sum_{j : T_{j0} = \infty} w_{ij} = 1$. We point readers to \cite{ben2019synthetic} for more details on the estimation of these weights in more complex settings with multiple treated units. We can see, however that the expected outcome for a treated unit in this case is given by
$$E(Y_{i,76}(0) | U_{i, 75} > 0) = 76 \beta + \gamma E(U_{i, 76} | U_{i, 75} > 0).$$
The synthetic controls will try to approximate this with a weighted combination of control units, which have expectation
\begin{align*}
    E \Bigg(\sum_{j : T_{j0} = \infty} w_{ij} Y_{jt}(0) \ \Big| \ U_{j, 75} \leq 0 \ \forall \ j \Bigg) &= 76 \beta + \gamma \sum_{j : T_{j0} = \infty} w_{ij} E(U_{j, 76} | U_{j, 75} \leq 0) \\
    & < 76 \beta + \gamma E(U_{i, 76} | U_{i, 75} > 0) \\
    &= E(Y_{i,76}(0) | U_{i, 75} > 0)
\end{align*}
and therefore the synthetic controls will tend to underestimate the missing potential outcome in this setting, which will lead to overestimating the treatment effect. 

\subsection{Showing stationarity holds}

An alternative approach is to rely on a stationarity assumption on the control potential outcomes for the treated units. As in the manuscript, this approach will utilize the pre-treatment data (up to time period 75) to build a model for $Y_{it}(0)$ in the treated group and then use this model to forecast times $t \geq T_{i0}$. Mathematically, we will be estimating $E[Y_{it}(0) | U_{i,75} > 0]$ in the pre-treatment period, and need this model to continue to hold in the post-treatment period to obtain accurate predictions of the missing potential outcome. One can write this expectation as
\begin{align*}
    E[Y_{it}(0) | U_{i,75} > 0] &= E(\beta t + \gamma U_{it} + \epsilon_{it} | U_{i,75} > 0) \\
    &=\beta t + \gamma E[U_{it} | U_{i,75} > 0].
\end{align*}
Now we can show that this expectation is the same in the pre and post-treatment periods. First we can look at the case where $t > 75$:
\begin{align*}
\beta t + \gamma E[U_{it} | U_{i,75} > 0] &=
\beta t + \gamma E[E[U_{it} |U_{i,t-1} = u_{i,t-1}, U_{i,75} > 0]] \\
&= \beta t + \gamma \alpha E[ U_{i,t-1} | U_{i,75} > 0] \\
& \ \vdots \\
&= \beta t + \gamma \alpha^{t - 75} E[ U_{i,75} | U_{i,75} > 0].
\end{align*}
Now, we can perform similar operations for $t \leq 75:$
\begin{align*}
\beta t + \gamma E[U_{it} | U_{i,75} > 0] &=
\beta t + \gamma E[E[U_{it} |U_{i,t+1} = u_{i,t+1}, U_{i,75} > 0]] \\
&= \beta t + \frac{\gamma}{\alpha} E[ U_{i,t+1} | U_{i,75} > 0] \\
& \ \vdots \\
&= \beta t + \frac{\gamma}{\alpha^{75 - t}} E[ U_{i,75} | U_{i,75} > 0] \\
&= \beta t + \gamma \alpha^{t - 75} E[ U_{i,75} | U_{i,75} > 0].
\end{align*}
We see that these mean functions are the same in the pre and post-treatment periods and therefore we can use the pre-treatment outcomes to estimate this model and use it to forecast into time periods after treatment initiation. Note that even though the true outcome model is a linear function of time and the unmeasured confounder, after integrating over possible values of the unmeasured confounder, we now have a nonlinear function of time $t$. 

\subsection{Simulation results}

\begin{figure}[H]
		\centering
		\includegraphics[width=0.85\linewidth]{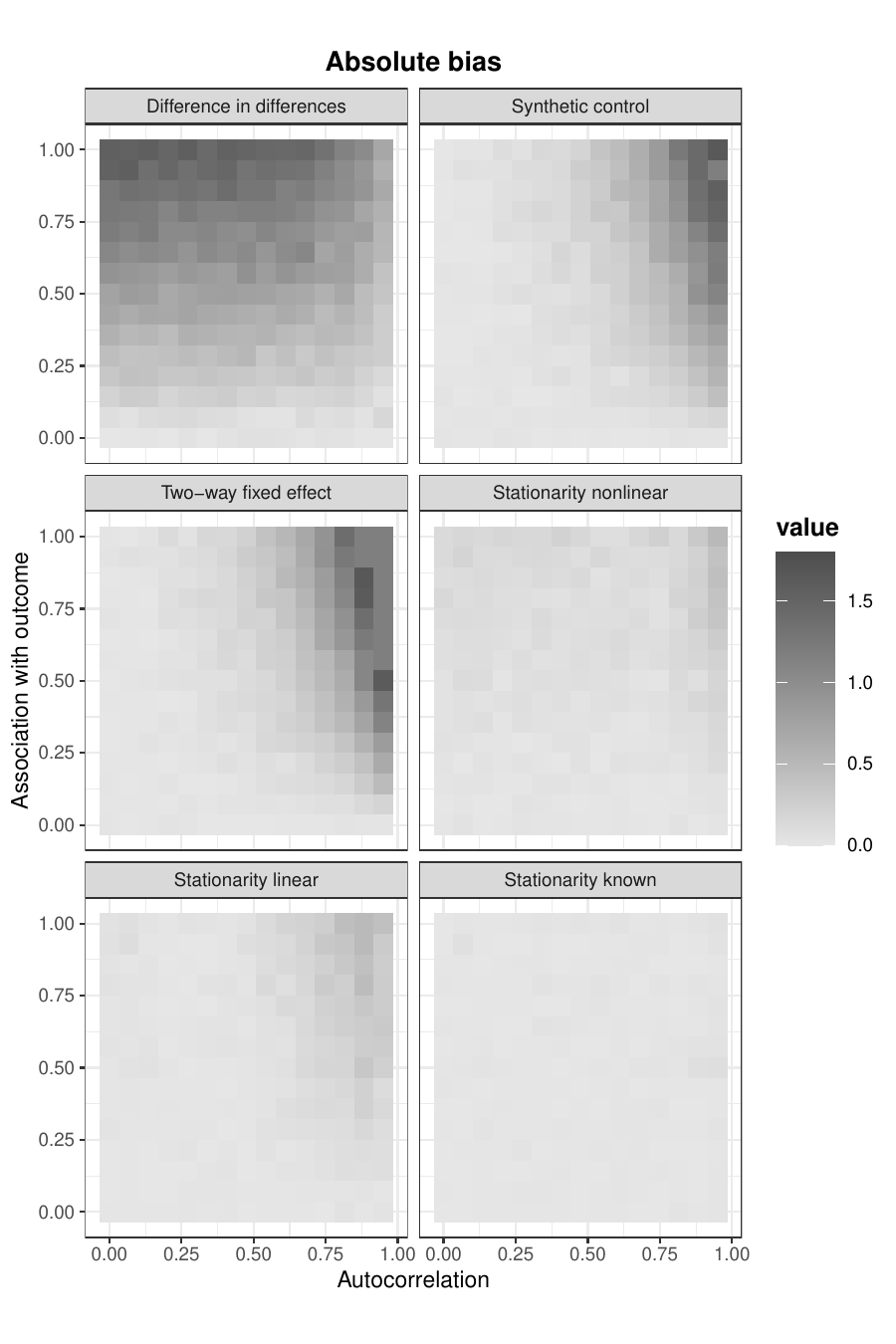}
		\caption{Absolute value of the bias of the various estimators under the time-varying unmeasured confounder simulation across a range of values of $\rho$ and $\gamma$. }
		\label{fig:StationaritySim}
\end{figure}

Now, we repeat this simulation study 100 times to evaluate the bias of a variety of estimators in this situation. We focus on the following list of estimators:
\begin{enumerate}
    \item A difference in differences estimator using the R package \texttt{did} \citep{callaway2021package, callaway2019difference}.
    \item A pooled synthetic control estimator from \cite{ben2019synthetic}.
    \item A two-way fixed effects estimator of the form $$E(Y_{it}) = \beta_0 + \beta_i + \beta_t + \tau A_{it}.$$
    \item The proposed approach where we model $E(Y_{it}) = f(t)$ during the pre-treatment period for the treated subjects and use it to forecast the potential outcomes in the post-treatment period. We estimate $f(t)$ using 3 degree of freedom natural cubic splines.
    \item The proposed approach where we estimate $f(t)$ using a linear function of time.
    \item The proposed approach where we use the \textit{known} function of time $f(t) = \beta t + \gamma \alpha^{t - 75} E[ U_{i,75} | U_{i,75} > 0]$ to forecast future time points. This estimator is not feasible in practice, but we use it here to compare with our approaches that estimate this unknown function.
\end{enumerate}
The results from this simulation can be found in Figure \ref{fig:StationaritySim}. We see that the bias results for both the DID and synthetic control estimators are as expected given the results above. The DID estimator is biased in nearly every situation, except for when there is no association between the unmeasured confounder and the outcome ($\gamma = 0$). The synthetic control estimator has increasing bias as we increase both $\gamma$ and the autocorrelation. The direction of the bias, which is not shown here, is as expected given the results above in that the synthetic control estimator tends to overestimate the treatment effect. The DID estimator overestimates the unknown trend given by $E(Y_{i, 76}(0) - Y_{i,75}(0) | T_{i0} = 76)$, which leads to underestimation of the causal effect. The two-way fixed effects estimator also has substantial amounts of bias for larger values of autocorrelation and $\gamma$. The stationarity approaches do not suffer from this bias. When we estimate a nonlinear function of time, which is seen in the middle-right panel of Figure \ref{fig:StationaritySim}, we obtain relatively small amounts of bias in all situations. The bias is not as small as the situation when we know the true underlying $f(t)$ function (bottom-right panel), but it approximates this ideal situation relatively well. The model assuming stationarity with a linear function of time (bottom-left panel) also does reasonably well, but is more biased than the nonlinear function of time, since the true underlying function is nonlinear. 

\subsection{Simulation study with seasonal unmeasured confounder}

\begin{figure}[htbp]
		\centering
		\includegraphics[width=0.65\linewidth]{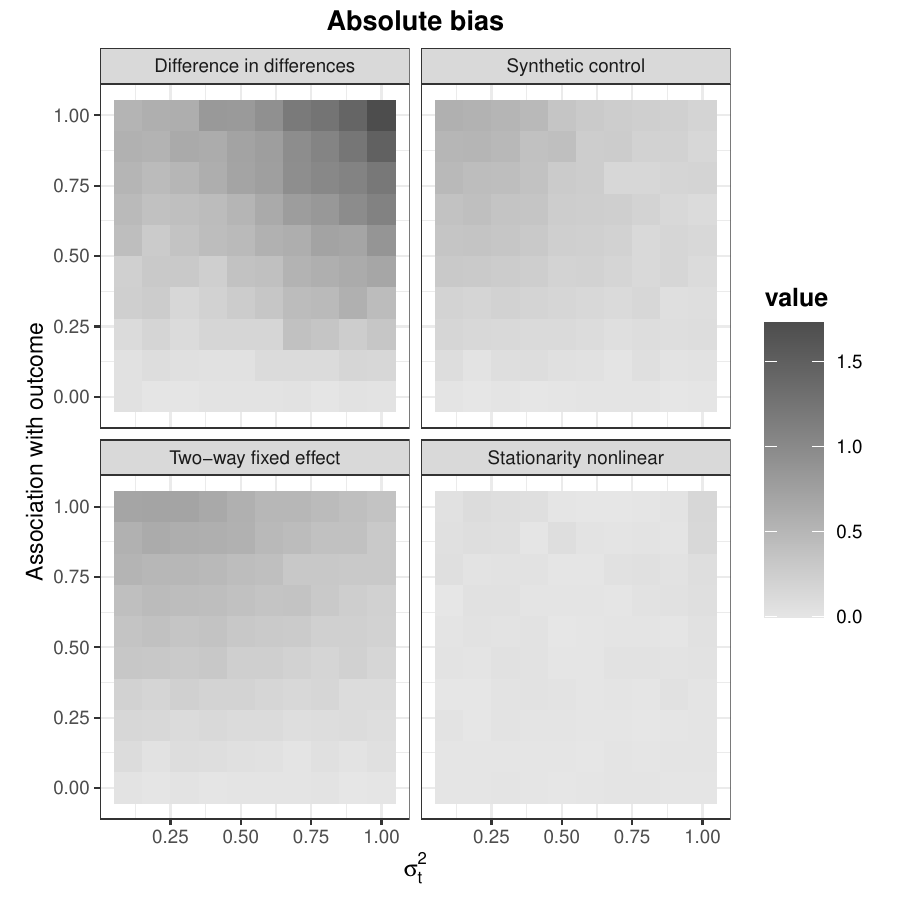}
		\caption{Absolute value of the bias of the various estimators under the time-varying unmeasured confounder simulation across a range of values of $\sigma_t^2$ and $\gamma$.}
		\label{fig:SeasonalSim}
\end{figure}

We now explore a related situation where the unmeasured confounder has a seasonal trend instead of following an AR(1) process. We use the same simulation structure as above where we have $n=30$ units and $T = 80$ time periods. We use the same structure for defining the start time of treatment adoption and simulating the potential outcomes. Specifically, we simulate treatment start time as
\[ T_{i0} = \begin{cases} 
      76 & U_{i,75} > 0 \\
      \infty & \text{otherwise,} 
   \end{cases}
\]
and generate the potential outcomes according to the following:
$$Y_{it}(0) = \beta t + \gamma U_{it} + \epsilon_{it}, \quad \epsilon_{it} \sim \mathcal{N}(0,1).$$
We now generate the unmeasured variable as $U_{it} = \sin(t/3 + a_i) + \eta_{it}$ where $\eta_{it} \sim \mathcal{N}(0, \sigma_t^2)$ and $a_i$ is a randomly sampled integer between 1 and 10. We utilize the same estimators as before, but we exclude the linear stationarity estimator that clearly won't work in this nonlinear setting and the known stationarity estimator that is not feasible in practice. The results can be found in Figure \ref{fig:SeasonalSim}, which shows the bias of the four estimators considered across a range of values for $\sigma_t^2$ and $\gamma$. We see that the DID estimator again has the most bias of all of the estimators, while the synthetic control and two-way fixed effect estimators are biased for large values of $\gamma$ and small values of $\sigma_t^2$. The proposed approach based on stationarity is relatively unbiased for all parameter values considered. 

\section{Computational details for MCMC sampling}

Here we discuss the details of the MCMC algorithm used to sample from the multivariate Bayesian structural time series model that was utilized in the paper. We also present computational details for a vector autoregressive model that can be used for the same forecasts in Section \ref{sec:AppendixAltModel}. We will discuss sampling for a particular structural time series model, though adding additional complexities, such as seasonal terms, is relatively straightforward to add into the proposed Gibbs sampler. Specifically, we will be sampling from the following model:
\begin{align*}
\boldsymbol{Y}_t &= \boldsymbol{\mu}_t + \boldsymbol{\epsilon}_t \nonumber \\
\boldsymbol{\mu}_t &= \boldsymbol{\mu}_{t-1} + \boldsymbol{\delta}_{t-1} + \boldsymbol{\eta}_t^{\mu} \\
\boldsymbol{\delta}_t &= \boldsymbol{\delta}_{t-1} + \boldsymbol{\eta}_t^{\delta} \\
\boldsymbol{\eta}_t^{\mu} &\sim \mathcal{N}(\boldsymbol{0},  \boldsymbol{D}_{\mu}) \\
\boldsymbol{\eta}_t^{\delta} &\sim \mathcal{N}(\boldsymbol{0}, \boldsymbol{D}_{\delta}) \\
\boldsymbol{\epsilon}_t &\sim \mathcal{N}(\boldsymbol{0}, \boldsymbol{\Sigma}),
\end{align*}
where $\boldsymbol{D}_{\mu}$ is a diagonal matrix with elements given by $\sigma_{\mu,i}^2$ for $i = 1, \dots, n$. $\boldsymbol{D}_{\delta}$ is defined analogously, but with variance parameters given by $\sigma_{\delta,i}^2$ for $i = 1, \dots, n$. Given this model specification, one can show the conditional updates for the vector of mean and trend states at time period 1 to be given by
\begin{itemize}
    \item $\boldsymbol{\delta}_1 \vert \cdot \sim \mathcal{N} \Bigg( \Big(\boldsymbol{D}_{\mu}^{-1} + 2 \boldsymbol{D}_{\delta}^{-1} \Big)^{-1} \Big((\boldsymbol{\mu}_2 - \boldsymbol{\mu}_1)^T\boldsymbol{D}_{\mu}^{-1} + \boldsymbol{\delta}_2^T \boldsymbol{D}_{\delta}^{-1}  \Big), \Big(\boldsymbol{D}_{\mu}^{-1} + 2 \boldsymbol{D}_{\delta}^{-1} \Big)^{-1} \Bigg)$
     \item $\boldsymbol{\mu}_1 \vert \cdot \sim \mathcal{N} \Bigg( \Big(\boldsymbol{\Sigma}^{-1} + 2 \boldsymbol{D}_{\mu}^{-1} \Big)^{-1} \Big( \boldsymbol{Y}_1^T \boldsymbol{\Sigma}^{-1} +   (\boldsymbol{\mu}_2 - \boldsymbol{\delta}_1)^T \boldsymbol{D}_{\mu}^{-1} \Big), \Big(\boldsymbol{\Sigma}^{-1} + 2 \boldsymbol{D}_{\mu}^{-1} \Big)^{-1} \Bigg).$
\end{itemize}
Now we can show the updates for a time $1 < t < T$ where all units are still fully observed. This means that $T_{i0} > t$ for all $i=1, \dots, n$. These updates are given by
\begin{itemize}
    \item $\boldsymbol{\delta}_t \vert \cdot \sim \mathcal{N} \Bigg( \Big(\boldsymbol{D}_{\mu}^{-1} + 2 \boldsymbol{D}_{\delta}^{-1} \Big)^{-1} \Big((\boldsymbol{\mu}_{t+1} - \boldsymbol{\mu}_{t})^T \boldsymbol{D}_{\mu}^{-1} + \boldsymbol{\delta}_{t-1}^T \boldsymbol{D}_{\delta}^{-1} + \boldsymbol{\delta}_{t+1}^T \boldsymbol{D}_{\delta}^{-1}  \Big), \Big(\boldsymbol{D}_{\mu}^{-1} + 2 \boldsymbol{D}_{\delta}^{-1} \Big)^{-1} \Bigg)$
     \item $\boldsymbol{\mu}_t \vert \cdot \sim \mathcal{N} \Bigg( \Big(\boldsymbol{\Sigma}^{-1} + 2 \boldsymbol{D}_{\mu}^{-1} \Big)^{-1} \Big( \boldsymbol{Y}_t^T \boldsymbol{\Sigma}^{-1} +   (\boldsymbol{\mu}_{t-1} + \boldsymbol{\delta}_{t-1})^T \boldsymbol{D}_{\mu}^{-1} + (\boldsymbol{\mu}_{t+1} - \boldsymbol{\delta}_{t})^T \boldsymbol{D}_{\mu}^{-1} \Big), \Big(\boldsymbol{\Sigma}^{-1} + 2 \boldsymbol{D}_{\mu}^{-1} \Big)^{-1} \Bigg).$
\end{itemize}

Lastly, we can show the updates for time periods for which some units may be on their final time period before treatment initiation, and other units may have already adopted treatment and therefore do not contribute to the likelihood of our models anymore. To do this, we must first introduce some additional notation. Let the $*$ superscript denote vectors and matrices that only contain the data from the units who have yet to receive treatment. For instance, $\boldsymbol{Y}_t^*$ is a vector of outcomes at time $t$ for individuals with $T_{i0} > t$. Similarly, $\boldsymbol{\Sigma}^*$ is a $k$ by $k$ matrix, where $k$ is the number of individuals for whom $T_{i0} > t$, and it corresponds to the submatrix of $\boldsymbol{\Sigma}$ that only has the rows and columns corresponding to the untreated units at time $t$. Lastly, let the 0 subscript correspond to vectors and matrices that have zeroes for indices of individuals with $T_{i0} = t+1$, i.e. receive treatment in the next time period. For instance $\boldsymbol{\mu}_{t,0}^*$ is a vector of length $k$ with values of $\boldsymbol{\mu}_t$ for whom $T_{i0} > t$, with zeroes for individuals with $T_{i0} = t+1$. The updates for the mean and trend parameters at time $t$ for individuals with $T_{i0} > t$ are given by
\begin{itemize}
    \item $\boldsymbol{\delta}_t^* \vert \cdot \sim \mathcal{N} \Bigg( \Big({\boldsymbol{D}_{\mu, 0}^*}^{-1} + {\boldsymbol{D}_{\delta}^*}^{-1} + {\boldsymbol{D}_{\delta, 0}^*}^{-1} \Big)^{-1} \Big((\boldsymbol{\mu}_{t+1, 0}^* - \boldsymbol{\mu}_{t,0}^*)^T{\boldsymbol{D}_{\mu, 0}^*}^{-1} + {\boldsymbol{\delta}_{t-1}^*}^T {\boldsymbol{D}_{\delta}^*}^{-1} + {\boldsymbol{\delta}_{t+1, 0}^*}^T {\boldsymbol{D}_{\delta}^*}^{-1}  \Big), \\
    \Big({\boldsymbol{D}_{\mu, 0}^*}^{-1} + {\boldsymbol{D}_{\delta}^*}^{-1} + {\boldsymbol{D}_{\delta, 0}^*}^{-1} \Big)^{-1} \Bigg)$
     \item $\boldsymbol{\mu}_t^* \vert \cdot \sim \mathcal{N} \Bigg( \Big({\boldsymbol{\Sigma}^*}^{-1} + {\boldsymbol{D}_{\mu}^*}^{-1} + {\boldsymbol{D}_{\mu, 0}^*}^{-1} \Big)^{-1} \Big( {\boldsymbol{Y}_t^*}^T {\boldsymbol{\Sigma}^*}^{-1} +   (\boldsymbol{\mu}_{t-1}^* + \boldsymbol{\delta}_{t-1}^*)^T {\boldsymbol{D}_{\mu}^*}^{-1} + (\boldsymbol{\mu}_{t+1, 0}^* - \boldsymbol{\delta}_{t, 0}^*)^T {\boldsymbol{D}_{\mu}^*}^{-1} \Big), \\ \quad \Big({\boldsymbol{\Sigma}^*}^{-1} + {\boldsymbol{D}_{\mu}^*}^{-1} + {\boldsymbol{D}_{\mu, 0}^*}^{-1} \Big)^{-1} \Bigg).
     $
\end{itemize}
That concludes the updates for the mean and trend parameters, however, we also have to update the variance parameters $(\sigma_{\delta,i}^2, \sigma_{\mu,i}^2)$ for $i = 1, \dots, n$. We assign these independent inverse gamma priors with hyper parameters $a_{\sigma}$ and $b_{\sigma}$. This leads to conjugate updates within the Gibbs sampler that are given by
\begin{itemize}
    \item $\sigma_{\delta, i}^2 \vert \cdot \sim \mathcal{IG}\bigg( a_{\sigma} + \frac{T_{i0} - 1}{2}, b_{\sigma} + \delta_{1i}^2/2 + \frac{1}{2} \sum_{t=2}^{T_{i0} - 1} (\delta_{ti} - \delta_{t-1,i})^2 \bigg)$
    \item $\sigma_{\mu, i}^2 \vert \cdot \sim \mathcal{IG}\bigg( a_{\sigma} + \frac{T_{i0} - 1}{2}, b_{\sigma} + \mu_{1i}^2/2 + \frac{1}{2} \sum_{t=2}^{T_{i0} - 1} (\mu_{ti} - \mu_{t-1,i} - \delta_{t-1,i})^2 \bigg).$
\end{itemize}
One can iterate through each of the steps described above to implement a Gibbs sampler that updates all parameters in our model. 

\section{Alternative model specification}
\label{sec:AppendixAltModel}

In addition to the structural time series model used throughout the manuscript, we explored a vector autoregressive model as an alternative model for the outcome process over time. Specifically, to account for both spatial and temporal dependencies, we specify a local-mean first-order vector autoregressive model of the form 
\begin{align}\boldsymbol{Y}_t = \boldsymbol{f}(t) + \boldsymbol{A}  \big(\boldsymbol{Y}_{t-1} -  \boldsymbol{f}(t-1) \big) + \boldsymbol{\epsilon}_t, \label{eqn:MainModel}
\end{align}
where $\boldsymbol{\epsilon}_t \sim \mathcal{N}(\boldsymbol{0}_n, \boldsymbol{\Sigma})$. For a more general discussion of these models, see \cite{banbura2018forecasting}. The vector of functions $\boldsymbol{f}(t)$ accounts for unit-specific intercepts and trends over time. These will be estimated using basis functions by allowing $f_i(t) = \sum_{k=1}^K \beta_{ik} \phi_k(t)$, where $\{ \phi_k(t) \}_{k=1}^K$ are pre-specified basis functions that include an intercept. $\boldsymbol{A}$ is an $n \times n$ matrix of parameters that control the extent of temporal dependence over time. Diagonal elements of $\boldsymbol{A}$ allow for dependence across time within each subject, while non-diagonal elements of $\boldsymbol{A}$ dictate the amount of dependence across subjects over time. As in the manuscript, the error term $\boldsymbol{\epsilon}_t$ allows for spatial dependence across units in the study at a particular time period through the covariance matrix $\boldsymbol{\Sigma}$. 

\subsection{Sparsity of $A$ matrix}

In high-dimensional time series settings, we do not have sufficient data to estimate all $n^2$ parameters in the $\boldsymbol{A}$ matrix, and some form of dimension reduction or shrinkage is required. Both shrinkage priors \citep{banbura2010large,kastner2017sparse,ghosh2019high} and point mass priors \citep{korobilis2013var} have been used to improve estimation of $\boldsymbol{A}$. One complicating factor unique to our setting is that the observed time series lengths in the pre-treatment period may vary drastically across subjects. This complicates computation as nearly all algorithms have been developed for situations with equal numbers of time periods. Additionally, this can cause a problem for forecasting as it is difficult to use values from one precinct to predict the values of another if they are observed for drastically different lengths of time. For this reason, we force $A_{ij} = 0$ for any subjects $i$ and $j$ that initiate treatment at substantially different times. This greatly increases the sparsity in $\boldsymbol{A}$ leading to more stable estimation, and avoids problems due to time series being observed at very different time points. Further dimension reduction can be achieved by setting any $A_{ij} = 0$ for subjects that are not geographic neighbors. This is based on the fact that geographic neighbors should be more correlated than geographically distant subjects, and is a strategy we will utilize for the New York City policing data. The nonzero entries of $\boldsymbol{A}$, along with the parameters $\boldsymbol{\beta}$ for the time trends, are given independent normal prior distributions with diffuse variances, as the dimension of the parameter space has been sufficiently reduced to alleviate the need for shrinkage priors. 

\subsection{Computational details for MCMC sampling}

All unknown parameters have full conditional distributions that are from known distributions and therefore the model can be implemented using a standard Gibbs sampling algorithm. First we will detail the Gibbs sampling update for $\boldsymbol{\beta}_k$ for $k=1, \dots, K$. Let us define 
$$\boldsymbol{R}_{t} = \boldsymbol{Y}_{t} - \sum_{j \neq k} \boldsymbol{X}_j(t) \boldsymbol{\beta}_j - \boldsymbol{A} \boldsymbol{Y}_{t-1} + \boldsymbol{A} \sum_{j \neq k} \boldsymbol{X}_j(t-1) \boldsymbol{\beta}_j,$$ 
for $t=2, \dots T$, and the first time point is defined as
$$\boldsymbol{R}_{1} = \boldsymbol{Y}_{1} - \sum_{j \neq k} \boldsymbol{X}_j(1) \boldsymbol{\beta}_j.$$
Further, we must define the following:
$$\widetilde{\boldsymbol{X}}_k(t) = \boldsymbol{X}_k(t) - \boldsymbol{A} \boldsymbol{X}_k(t-1),$$
for $t = 2, \dots, T$, and $\widetilde{\boldsymbol{X}}_k(1) = \boldsymbol{X}_k(1)$. Next, define $O_t = \{ j: t_{j0} > t \}$ to be the set of subjects who have not yet been exposed to the treatment. We will use the $*$ superscript to denote versions of all relevant vectors and matrices that have elements corresponding to indices not in $O_t$ set to zero. Specifically, let $\boldsymbol{Y}_t^*$ be defined such that
\[ Y_{it}^* =  \begin{cases} 
      Y_{it} & i \in O_t \\
      0 & i \notin O_t.
   \end{cases}
\]
identical notation will be used for $\boldsymbol{R}_t$. For matrices, we will let $\boldsymbol{X}_j^*(t)$ be equal to $\boldsymbol{X}_j^*(t)$ except all values in row $i$ and column $i$ will be set to zero if $i \notin O_t$. Lastly, we will let ${\boldsymbol{\Sigma}_t^*}^{-1}$ be an $n \times n$ matrix with any elements in row and column $i$ set to zero for all $i \notin O_t$. The remaining elements of ${\boldsymbol{\Sigma}_t^*}^{-1}$ will be set to the inverse of the submatrix of $\boldsymbol{\Sigma}$ defined by indices in $O_t$. 

The update for $\boldsymbol{\beta}_k$ then proceeds as follows: for $k =1, \dots, K$ sample $\boldsymbol{\beta}_k$ from a multivariate normal distribution with mean $\boldsymbol{M}$ and variance $\boldsymbol{V}$ defined by 
\begin{align*}
    \boldsymbol{V} &= \bigg( \sum_{t=1}^T \widetilde{\boldsymbol{X}}_k^*(t)' {\boldsymbol{\Sigma}_t^*}^{-1} \widetilde{\boldsymbol{X}}_k^*(t)  \bigg)^{-1} \\
    \boldsymbol{M} &= \boldsymbol{V} \bigg( \sum_{t=1}^T \widetilde{\boldsymbol{X}}_k^*(t)' {\boldsymbol{\Sigma}_t^*}^{-1} \boldsymbol{R}_t \bigg)
\end{align*}

Now we will detail the update for $\boldsymbol{A}$, which can be done in a very similar manner to $\boldsymbol{\beta}_k$ if structured properly. To simplify the updates, assume that we are trying to update one element in each row of $\boldsymbol{A}$ simultaneously, implying that we are updating $n$ values at a time, one for each subject in the data. This is nearly identical to the situation posed when updating the $n-$vector $\boldsymbol{\beta}_k$, with some slight modifications needed. We again need to define a residual value as
$$\boldsymbol{E}_{t} = \boldsymbol{Y}_{t} - \sum_{k=1}^K \boldsymbol{X}_k(t) \boldsymbol{\beta}_k - \boldsymbol{A}_0 \bigg( \boldsymbol{Y}_{t-1} - \sum_{k=1}^K \boldsymbol{X}_k(t-1) \boldsymbol{\beta}_k \bigg),$$ 
where $\boldsymbol{A}_0$ is equal to $\boldsymbol{A}$ with the $n$ elements we are updating all set to zero. If we let $j_i$ be the index of $\boldsymbol{A}_{i}$ that we are updating, we can define $\boldsymbol{W}_t$ to be a diagonal matrix with the $(i,i)$ element equal to the $j_i$ element of $\bigg( \boldsymbol{Y}_{t-1} - \sum_{k=1}^K \boldsymbol{X}_k(t-1) \boldsymbol{\beta}_k \bigg)$. We will again use the $*$ superscript to zero the relevant elements of all vectors and matrices as we did for the update of $\boldsymbol{\beta}_k$. We can update the vector of $n$ values from $\boldsymbol{A}$ from a multivariate normal distribution with mean $\boldsymbol{M}$ and variance $\boldsymbol{V}$ defined by 
\begin{align*}
    \boldsymbol{V} &= \bigg( \sum_{t=1}^T {\boldsymbol{W}_t^*}' {\boldsymbol{\Sigma}_t^*}^{-1} \boldsymbol{W}_t^*  \bigg)^{-1} \\
    \boldsymbol{M} &= \boldsymbol{V} \bigg(\sum_{t=1}^T {\boldsymbol{W}_t^*}' {\boldsymbol{\Sigma}_t^*}^{-1} \boldsymbol{E}_t \bigg)
\end{align*}
This process is then iterated until all elements of $\boldsymbol{A}$ have been updated. If each row in $\boldsymbol{A}$ has exactly $q$ nonzero elements, then this process can simply be iterated $q$ times. If there is an unequal number of nonzero terms in the rows of $\boldsymbol{A}$ then this process will be repeated $q_{max}$ times where $q_{max}$ is the maximum number of nonzero elements in a row of $\boldsymbol{A}$. In this setting, rows of $\boldsymbol{A}$ that have less than $q_{max}$ nonzero elements can either recycle their nonzero elements and therefore they get updated more than once per MCMC iteration, or the relevant matrices and vectors in the construction of $\boldsymbol{M}$ and $\boldsymbol{V}$ can have the indices corresponding to these rows set to zero. In the latter setup we will update the parameters of $\boldsymbol{A}$ from a multivariate normal with mean and variance given by only the elements of $\boldsymbol{M}$ and $\boldsymbol{V}$ that correspond to the indices being updated. 

\subsubsection{Posterior predictive distribution and causal effects}

Once we have posterior samples of $\boldsymbol{\beta}_k$ and $\boldsymbol{A}$, we can now produce posterior samples of outcome values in the absence of treatment, denoted by $\widetilde{\boldsymbol{Y}}(\infty).$ For all time periods before the treatment is initiated, this value is observed and known. We will use time series forecasting from our model to predict these values for values post-treatment initiation. Let $t_{min}$ be the first time point that treatment is initiated throughout the study. The following algorithm will generate the $m^{th}$ posterior draw from $P(\boldsymbol{\widetilde{Y}}(\infty) \vert \boldsymbol{Y}, \boldsymbol{X})$: for $t = t_{min}, \dots, T$ perform the following steps:
\begin{enumerate}
    \item Draw values $\boldsymbol{\beta}_k^{(m)}$ for $k=1, \dots, K$ and $\boldsymbol{A}^{(m)}$ from the posterior distribution of both parameters.
    \item Obtain residuals from the previous time point as 
    $$\boldsymbol{r}_{t-1} = \boldsymbol{\widetilde{Y}}_{t-1}^{(m)}(\infty) - \sum_{k=1}^K \boldsymbol{X}_k(t-1) \boldsymbol{\beta}_k^{(m)} $$
    \item For the current time point, calculate
    $$\boldsymbol{M}_t = \sum_{k=1}^K \boldsymbol{X}_k(t) \boldsymbol{\beta}_k^{(m)} + \boldsymbol{A}^{(m)} \boldsymbol{r}_{t-1}$$
    \item If $t_{i0} \leq t \ \forall \ i$ then draw $\boldsymbol{\widetilde{Y}}_{t}^{(m)}(\infty)$ from a multivariate normal distribution with mean $\boldsymbol{M}_t$ and variance $\boldsymbol{\Sigma}$. If there exists an $i$ such that $t_{i0} > t$, then split the mean and covariance matrices as
    $$\boldsymbol{M}_t = 
\begin{pmatrix}
\boldsymbol{M}_{t1} \\
\boldsymbol{M}_{t2}
\end{pmatrix} \quad \boldsymbol{\Sigma} = \begin{pmatrix}
\boldsymbol{\Sigma}_{11} & \boldsymbol{\Sigma}_{12} \\
\boldsymbol{\Sigma}_{21} & \boldsymbol{\Sigma}_{22}
\end{pmatrix},$$
where $\boldsymbol{M}_{t1}$ represents the components of $\boldsymbol{M}_{t}$ corresponding to subjects $i$ with $t_{i0} \leq t$. Further $\boldsymbol{\Sigma}_{11}$ corresponds to the covariance matrix of residual errors for the same set of subjects. We can also let $\boldsymbol{\widetilde{Y}}_{t1}^{(m)}(\infty)$ and $\boldsymbol{\widetilde{Y}}_{t2}^{(m)}(\infty)$ be defined similarly. We are able to observe $\boldsymbol{\widetilde{Y}}_{t2}^{(m)}(\infty) = \boldsymbol{Y}_{t2}(\infty)$ whereas we will draw $\boldsymbol{\widetilde{Y}}_{t1}^{(m)}(\infty)$ from a multivariate normal distribution with mean
$$\boldsymbol{M}_{t1} + \boldsymbol{\Sigma}_{12} \boldsymbol{\Sigma}_{22}^{-1} (\boldsymbol{Y}_{t2}(\infty) - \boldsymbol{M}_{t2}),$$
and variance
$$\boldsymbol{\Sigma}_{11} + \boldsymbol{\Sigma}_{12}\boldsymbol{\Sigma}_{22}^{-1}\boldsymbol{\Sigma}_{21}.$$
Now that we have posterior draws of $\boldsymbol{\widetilde{Y}}_{t}(\infty)$ we automatically have posterior draws of the treatment effect at any time point as $\boldsymbol{Y}_{t,obs} - \boldsymbol{\widetilde{Y}}_{t}(\infty)$, the difference between the observed data (under treatment) and the prediction of what would have happened in the absence of treatment. 
\end{enumerate}

\subsection{Alternating least squares estimate of $\Sigma$}

Here we detail how we find initial estimates of ($\widehat{\boldsymbol{f}}(t), \widehat{\boldsymbol{A}}$), which can be used to construct an estimate of the residual covariance matrix. We will be constructing $\boldsymbol{f}(t)$ as $f_i(t) = \sum_{k=1}^K \beta_{ik} \phi_k(t)$, and therefore our model can be expressed as follows:

\begin{align*}
\boldsymbol{Y}_t &= \boldsymbol{f}(t) + \boldsymbol{A}  \big(\boldsymbol{Y}_{t-1} -  \boldsymbol{f}(t-1) \big) + \boldsymbol{\epsilon}_t \\
&= \sum_{k=1}^K \boldsymbol{X}_k(t) \boldsymbol{\beta}_k + \boldsymbol{A}  \bigg(\boldsymbol{Y}_{t-1} -  \sum_{k=1}^K \boldsymbol{X}_k(t-1) \boldsymbol{\beta}_k \bigg),
\end{align*}
where $\boldsymbol{X}_k(t)$ is an $n \times n$ diagonal matrix with each element of the diagonal equal to $\phi_k(t)$. $\boldsymbol{\beta}_k$ is a vector of length $n$ representing coefficients $\beta_{ik}$ for $i=1, \dots, n.$ We will construct an algorithm to estimate all unknown parameters, which are given by $(\boldsymbol{\beta}_1, \dots, \boldsymbol{\beta}_k, \boldsymbol{A})$. Note that we will also adopt the convention that $\boldsymbol{A}_{i} = [A_{i1}, \dots, A_{in}]$. Our algorithm will iterate across all unknown parameters using least squares at each step while conditioning on the current estimates of the remaining parameters. We will do this until the estimates have converged, which we will assess using the l2 norm between iterative updates to see if the difference has dropped below a pre-chosen $\delta$ level. Below are the specific updates for each parameters involved in the iterative algorithm. 

\subsubsection{Update of $\beta_k$}

Let us first describe how to estimate $\boldsymbol{\beta}_k$ given current estimates of the other parameters, $\widehat{\boldsymbol{\beta}}_j$ for $j \neq k$, and $\widehat{\boldsymbol{A}}$. Let us define 
$$\boldsymbol{R}_{t} = \boldsymbol{Y}_{t} - \sum_{j \neq k} \boldsymbol{X}_j(t) \widehat{\boldsymbol{\beta}}_j - \widehat{\boldsymbol{A}} \boldsymbol{Y}_{t-1} + \widehat{\boldsymbol{A}} \sum_{j \neq k} \boldsymbol{X}_j(t-1) \widehat{\boldsymbol{\beta}}_j,$$ 
for $t=2, \dots T$, and the first time point is defined as
$$\boldsymbol{R}_{1} = \boldsymbol{Y}_{1} - \sum_{j \neq k} \boldsymbol{X}_j(1) \widehat{\boldsymbol{\beta}}_j.$$
Lastly, we must define the following:
$$\widetilde{\boldsymbol{X}}_k(t) = \boldsymbol{X}_k(t) - \widehat{\boldsymbol{A}} \boldsymbol{X}_k(t-1),$$
for $t = 2, \dots, T$, and $\widetilde{\boldsymbol{X}}_k(1) = \boldsymbol{X}_k(1)$. As we are estimating the parameters with least squares, our goal is to minimize the following quantity:
\begin{align*}
    \sum_{t=1}^T ||\boldsymbol{R}_t - \widetilde{\boldsymbol{X}}_k(t) \boldsymbol{\beta}_k||^2.
\end{align*}
Taking the derivative of this expression with respect to $\boldsymbol{\beta}_k$, setting the derivative equal to zero, and solving for $\boldsymbol{\beta}_k$, we can see that our estimate is 
$$\widehat{\boldsymbol{\beta}}_k = \bigg( \sum_{t=1}^T \widetilde{\boldsymbol{X}}_k(t)^T \widetilde{\boldsymbol{X}}_k(t)  \bigg)^{-1} \bigg( \sum_{t=1}^T \widetilde{\boldsymbol{X}}_k(t)^T \boldsymbol{R}_t \bigg)$$. 

\subsubsection{Update of $A$}

To update $\boldsymbol{A}$ we can separately estimate $\boldsymbol{A}_i$ for $i=1, \dots, n$. As described in the manuscript, many elements of $\boldsymbol{A}_i$ will be zero by construction. To simplify notation, let $\boldsymbol{A}_i^*$ be a vector containing only the nonzero elements of $\boldsymbol{A}_i$. For instance, if $S = \{s: A_{is} = 1 \}$ and $|S| = q$, then $\boldsymbol{A}_i^* = (A_{iS_1}, \dots, A_{i S_q})$. We will adopt the same convention for the vectors $\boldsymbol{\widehat{\beta}}_k$ and $\boldsymbol{Y}_t$, and we will let $\boldsymbol{X}_k^*(t)$ be the rows of $\boldsymbol{X}_k(t)$ corresponding to the indices in $S$. We can now write our model as follows
\begin{align*}
    Y_{it} &= \sum_{k=1}^K \widehat{\beta}_{ik} \phi_k(t) + \boldsymbol{A}_i \bigg(\boldsymbol{Y}_{t-1} -  \sum_{k=1}^K \boldsymbol{X}_k(t-1) \widehat{\boldsymbol{\beta}}_k \bigg) \\
    &= \sum_{k=1}^K \widehat{\beta}_{ik} \phi_k(t) + \boldsymbol{A}_i^* \bigg(\boldsymbol{Y}_{t-1}^* -  \sum_{k=1}^K \boldsymbol{X}_k^*(t-1) \widehat{\boldsymbol{\beta}}_k^* \bigg)
\end{align*}
Now we can define $\boldsymbol{E}_i = (E_{i2}, \dots, E_{iT})$, where $E_{it} = Y_{it} - \sum_{k=1}^K \widehat{\beta}_{ik} \phi_k(t)$. Further, define $\widetilde{\boldsymbol{W}}_i$ to be a $(T-1) \times q$ matrix where row $t$ is given by
$$\widetilde{\boldsymbol{W}}_{it} = \boldsymbol{Y}_{t-1}^* -  \sum_{k=1}^K \boldsymbol{X}_k^*(t-1) \boldsymbol{\widehat{\beta}}_k^*.$$
Our least squares estimate of $\boldsymbol{A}_i$ is then equal to 
$$\widehat{A}_i = (\widetilde{\boldsymbol{W}}_i^T \widetilde{\boldsymbol{W}}_i)^{-1} \widetilde{\boldsymbol{W}}_i^T \boldsymbol{E}_i$$

\subsubsection{Update of $\Sigma$}

Once estimates of all parameters are obtained from the above algorithm, then estimates of $\widehat{Y}_t$ can be obtained using the following
\begin{align*}
    \widehat{Y}_1 &= \sum_{k=1}^K \boldsymbol{X}_k(t) \widehat{\boldsymbol{\beta}}_k \\
    \widehat{Y}_t &= \sum_{k=1}^K \boldsymbol{X}_k(t) \widehat{\boldsymbol{\beta}}_k + \widehat{\boldsymbol{A}}  \bigg(\boldsymbol{Y}_{t-1} -  \sum_{k=1}^K \boldsymbol{X}_k(t-1) \widehat{\boldsymbol{\beta}}_k \bigg) \quad \text{for } t=2, \dots, T
\end{align*}
Once these have been obtained, then we can construct an estimate of the covariance matrix by first using the sample covariance matrix defined by
$$\widehat{\boldsymbol{S}} = \frac{1}{T - K - 1}\sum_{t=1}^T (\boldsymbol{Y}_t - \widehat{\boldsymbol{Y}}_t) (\boldsymbol{Y}_t - \widehat{\boldsymbol{Y}}_t)'.$$
Lastly, this covariance matrix is unstable unless $T$ is large relative to $n$, which is not the case in the policing example described in the manuscript. To improve estimation, we will enforce sparsity on the inverse of $\boldsymbol{\Sigma}$ by solving the following constrained optimization problem:
\begin{align*} 
\widehat{\boldsymbol{\Omega}} = \underset{\Omega}{\operatorname{argmin}} \text{ tr}(\Omega \widehat{S}) - \log \det \Omega \\
\text{such that } \Omega \in \mathcal{Q},
\end{align*}
where $\mathcal{Q}$ is the space of all positive semi-definite matrices whose $(i,j)$ element is zero for any $i$ and $j$ that are not neighbors in the data. $\widehat{\boldsymbol{\Omega}}$ can be inverted to provide our final estimate of $\widehat{\boldsymbol{\Sigma}}$.

\subsection{Simulation results}

\begin{figure}[htbp]
		\centering
		\includegraphics[width=0.315\linewidth]{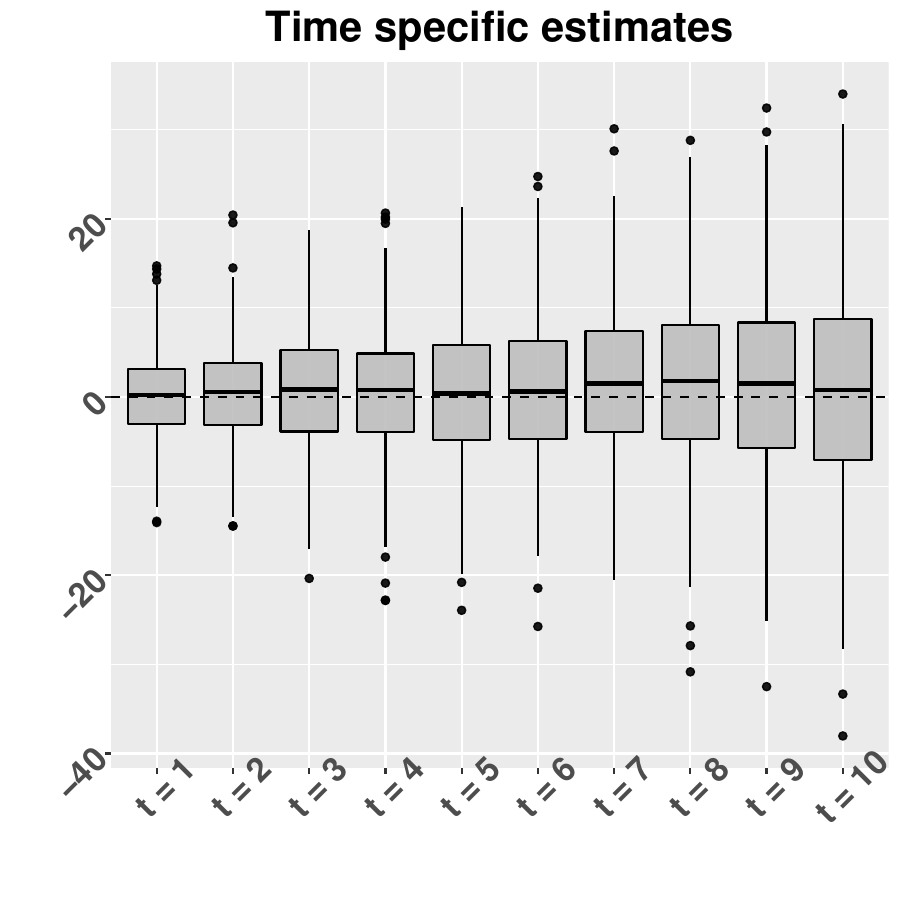}
		\includegraphics[width=0.33\linewidth]{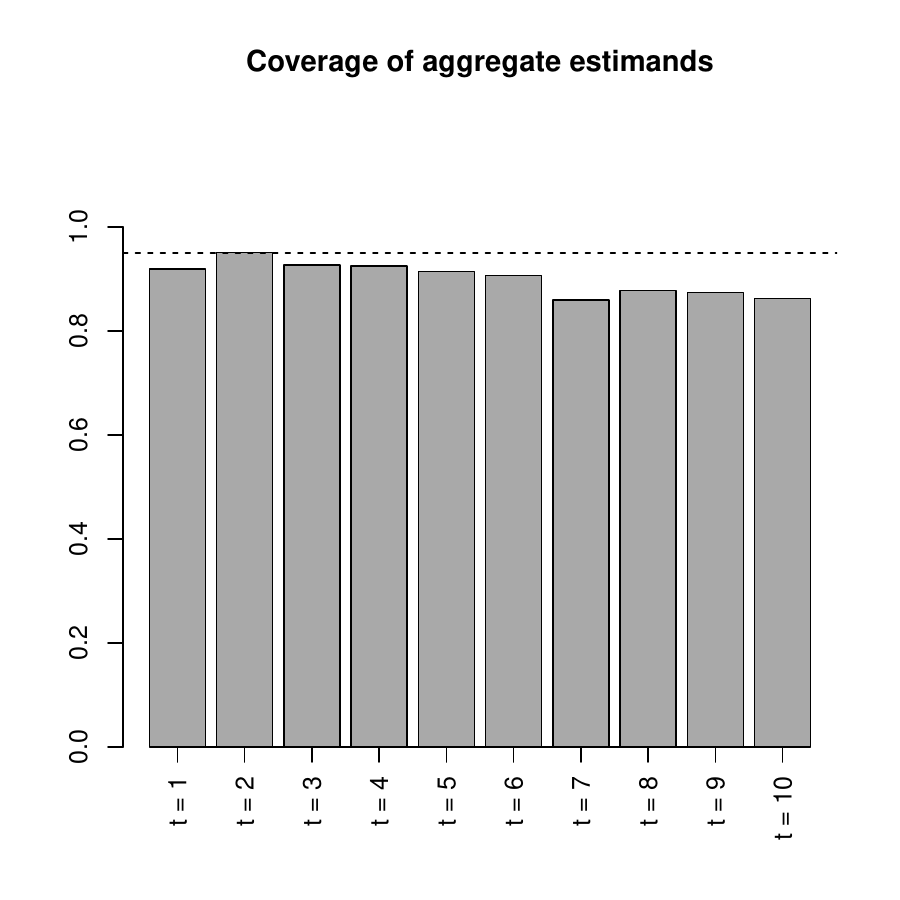}
		\includegraphics[width=0.333\linewidth]{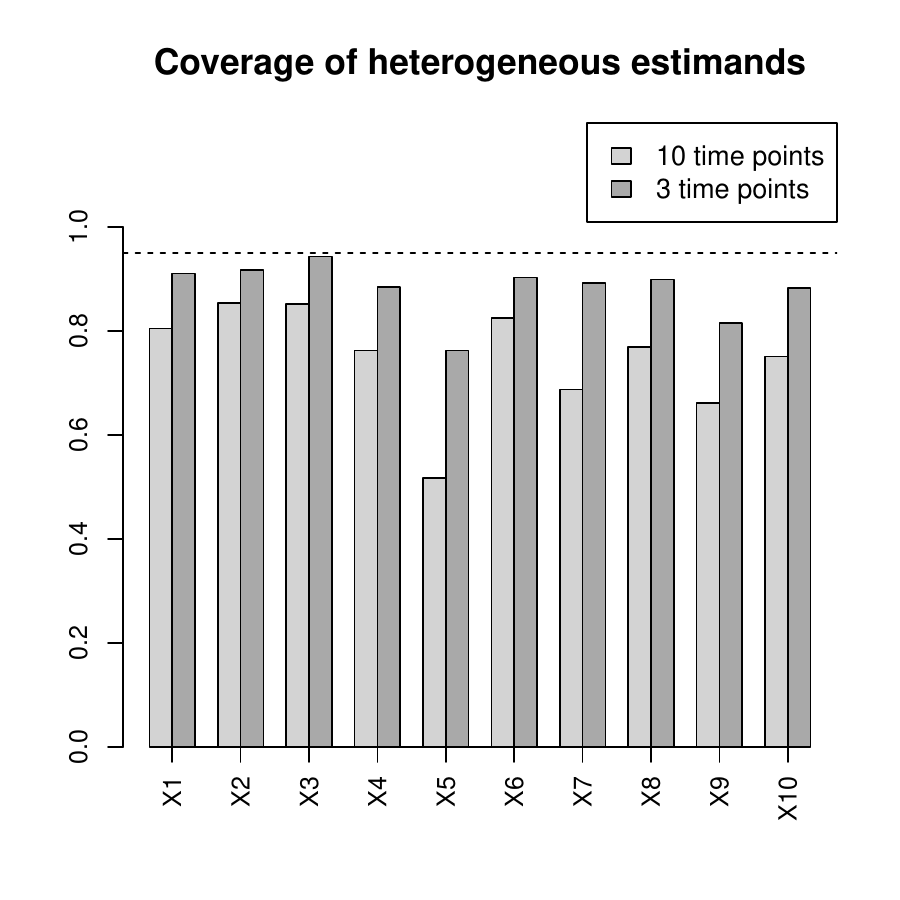}
		\caption{Results from the simulation study for misdemeanor arrests when using vector autoregressive models.}
		\label{fig:VARsim43}
\end{figure}

Here we present simulation results analogous to those used from the manuscript, but for the VAR model described above. The simulation setup is identical to the one presented in the manuscript, and the results can be found in Figure \ref{fig:VARsim43}. The results are fairly similar with respect to the marginal estimands $\Delta(q)$ as this approach is able to achieve nearly the nominal 95\% coverage rate. With the VAR model, coverage does dip slightly lower into the 86\% range for longer-term causal effects. Another key difference arises in the estimation of heterogeneous treatment effects, where the VAR model leads to worse coverage of heterogeneous estimands as seen in the right panel of Figure \ref{fig:VARsim43}. We see that if instead of trying to forecast 10 time periods into the future, we only forecast 3 time periods, then the coverage of the heterogeneous estimands improves and is closer to the nominal level. While it is intuitive that forecasting farther into the future is more difficult, the state space model used in the main manuscript did not suffer from this same limitation. For brevity we do not show the results here, but we applied this VAR model to simulations from all four outcomes of interest and it generally performs slightly worse than the model used in the manuscript. 

\subsection{Results on NYC policing study}

\begin{figure}[!t]
		\centering
		\includegraphics[width=0.48\linewidth]{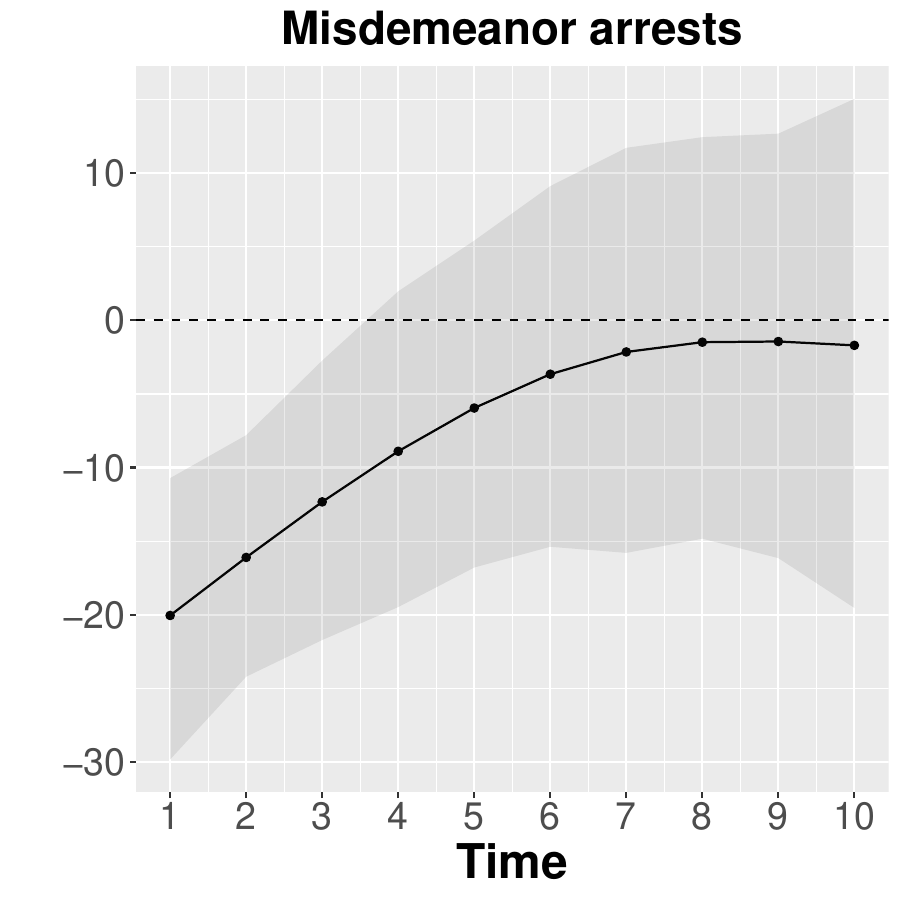}
		\includegraphics[width=0.48\linewidth]{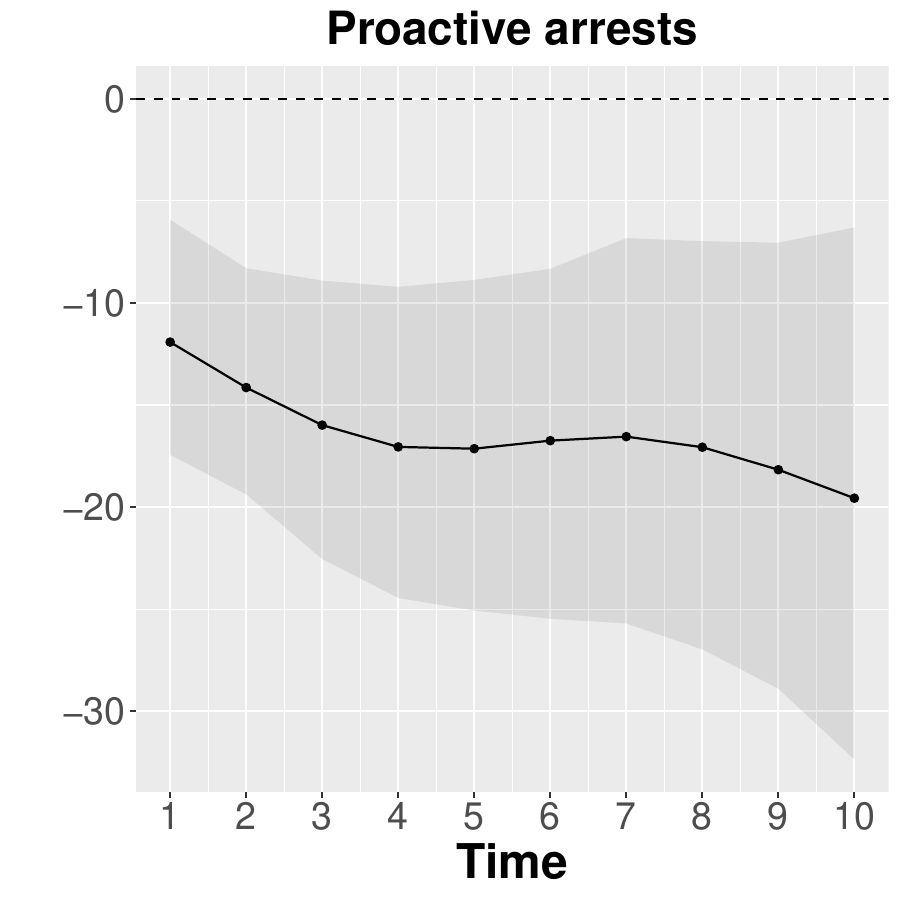} \\
		\includegraphics[width=0.48\linewidth]{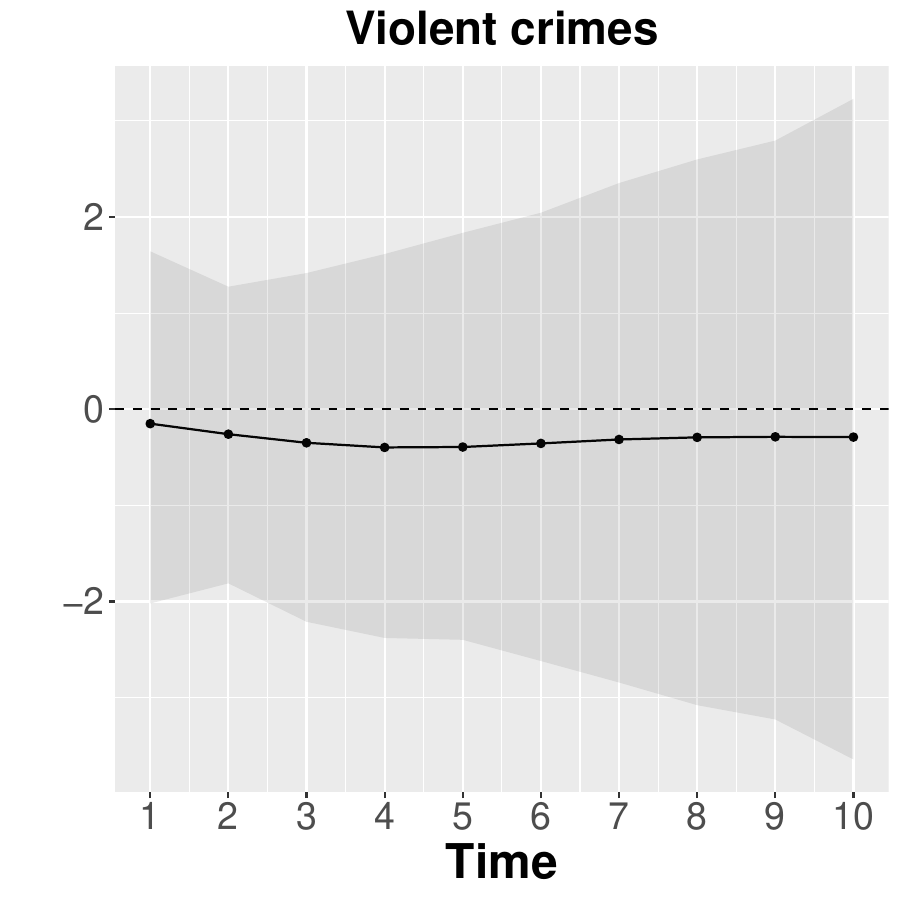}
		\includegraphics[width=0.48\linewidth]{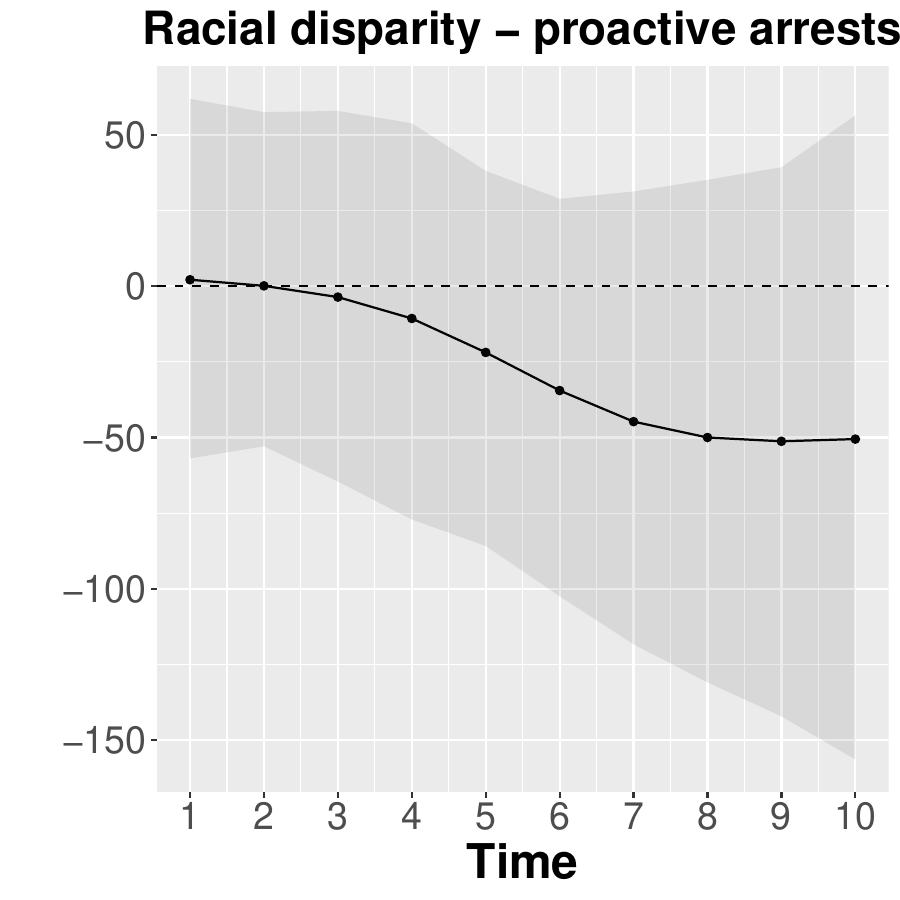}
		\caption{Estimates and 95\% credible intervals for time specific effects $\Delta(q)$ of neighborhood policing on misdemeanor arrests (first panel), proactive arrests (second panel), Violent crimes (third panel), and the difference of black and white proactive arrest rates (fourth panel) when using a VAR outcome model.}
		\label{fig:VAR_NYCmarginal}
\end{figure}

Here we present the findings of our NYC neighborhood policing analysis when using VAR models to forecast the potential outcomes in the absence of the policy. The estimates of $\Delta(q)$ for each of the four outcomes as a function of $q$ can be found in Figure \ref{fig:VAR_NYCmarginal}. We see a very similar story as with the models in the main manuscript, which is that there is a strong effect of the policy on both misdemeanor and proactive arrests that seems more sustained for proactive arrests. The estimated effects for both violent crimes and the racial disparity of proactive arrests are very close to zero indicating little to no effect of the policy on these outcomes.

\begin{figure}[htbp]
		\centering
		\includegraphics[width=0.98\linewidth]{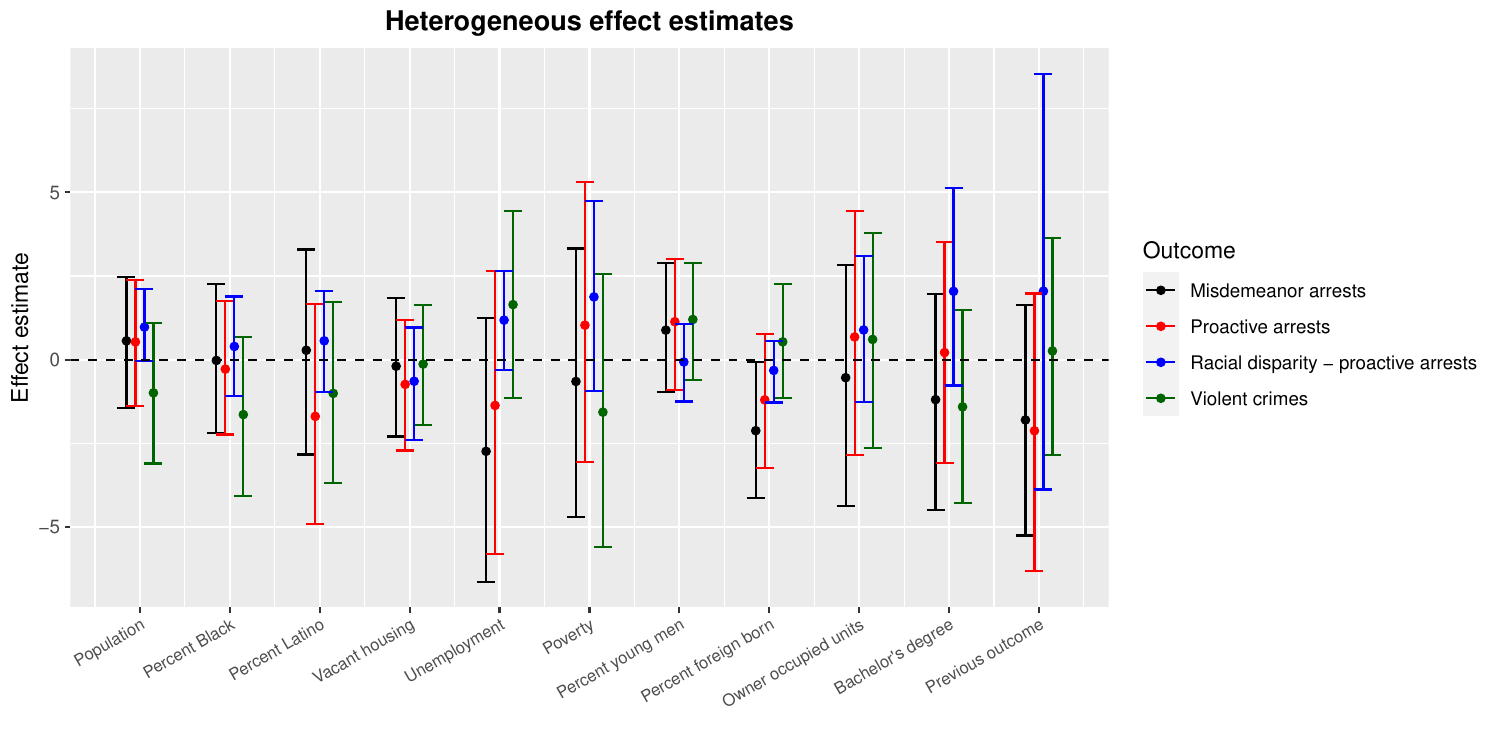}
		\caption{Estimates of coefficients from the heterogeneous treatment effect functions when using VAR models for time series forecasting.}
		\label{fig:VAR_NYChetero}
\end{figure}

We can also examine whether the effects of neighborhood policing varied by observed characteristics or locations of New York City. First, looking at whether the covariates modify the treatment effect in Figure \ref{fig:VAR_NYChetero}, we see a similar story as in the main manuscript. None of the covariates strongly affect the magnitude or direction of the effect of neighborhood policing. Lastly, we can investigate whether the treatment effect differs in the five regions defined by the clustering algorithm of the manuscript. Previously, we had seen very negative effects in certain areas of New York City, and very little effect of the policy in other neighborhoods. We see very similar estimates here, with significant and negative effects on clusters 1,3, and 5, while there is very little evidence of a treatment effect in clusters 2 and 4. Overall, the results are fairly similar across the VAR and state space modeling approaches, which gives us increased beliefs in the main findings of the manuscript. 

\begin{table}[htbp]
\caption{Estimates of the treatment effect on proactive arrests for each cluster when using a VAR outcome model.}
\label{tab:VARNYCcluster}
\begin{tabular}{lr}
  \hline
Cluster & Treatment Effect \\ 
  \hline
  1 & -10.58 (-18.11, -3.5) \\ 
    2 & 1.38 (-7.33, 10.37) \\ 
    3 & -25.05 (-39.39, -10.16)  \\ 
    4 & -1.7 (-11.49, 8.01) \\ 
    5 & -34.34 (-53.42, -15.56) \\ 
   \hline
\end{tabular}
\end{table}

\section{Results on NYC policing study using existing estimators}

Here we apply both a difference in differences and synthetic control estimator to estimate the marginal estimands in the neighborhood policing study in NYC. We utilize the same DID and synthetic control estimators described in Section \ref{sec:DifferentEstimators} that are applicable to the staggered adoption setting. One difference between these estimators and the one proposed in the manuscript is that these estimators require at least one control observation that is never treated during the study, while every precinct in NYC eventually adopts neighborhood policing. To avoid this issue, we will estimate the effect of the policy on a subset of the data that is treated earlier, while using precincts that are treated at later times as control units. Specifically, we have 6 precincts that become treated at time point 153 in our study, 6 that become treated at time point 150, and all other precincts are treated on or before time period 147. For this reason, we drop the precincts that are treated at time point 150 and we include those treated at time point 153 as control precincts. Due to this restriction, we also only estimate treatment effects for the first five time periods after treatment adoption, so that even precincts treated at time 147 can utilize these control precincts. It is important to note that while we are performing these analyses to confirm that we obtain similar results as those from our approach in the manuscript, these are slightly different estimands that are averaging over a subset of the precincts. Any differences seen between our approach and these approaches could be due to the statistical approach taken, the underlying assumptions associated with each approach, or the fact that we are looking at a slightly different estimand. Also note that we do not consider estimands highlighting heterogeneity by covariates here as these approaches are currently not developed for this purpose.  

\begin{figure}[htbp]
		\centering
		\includegraphics[width=0.85\linewidth]{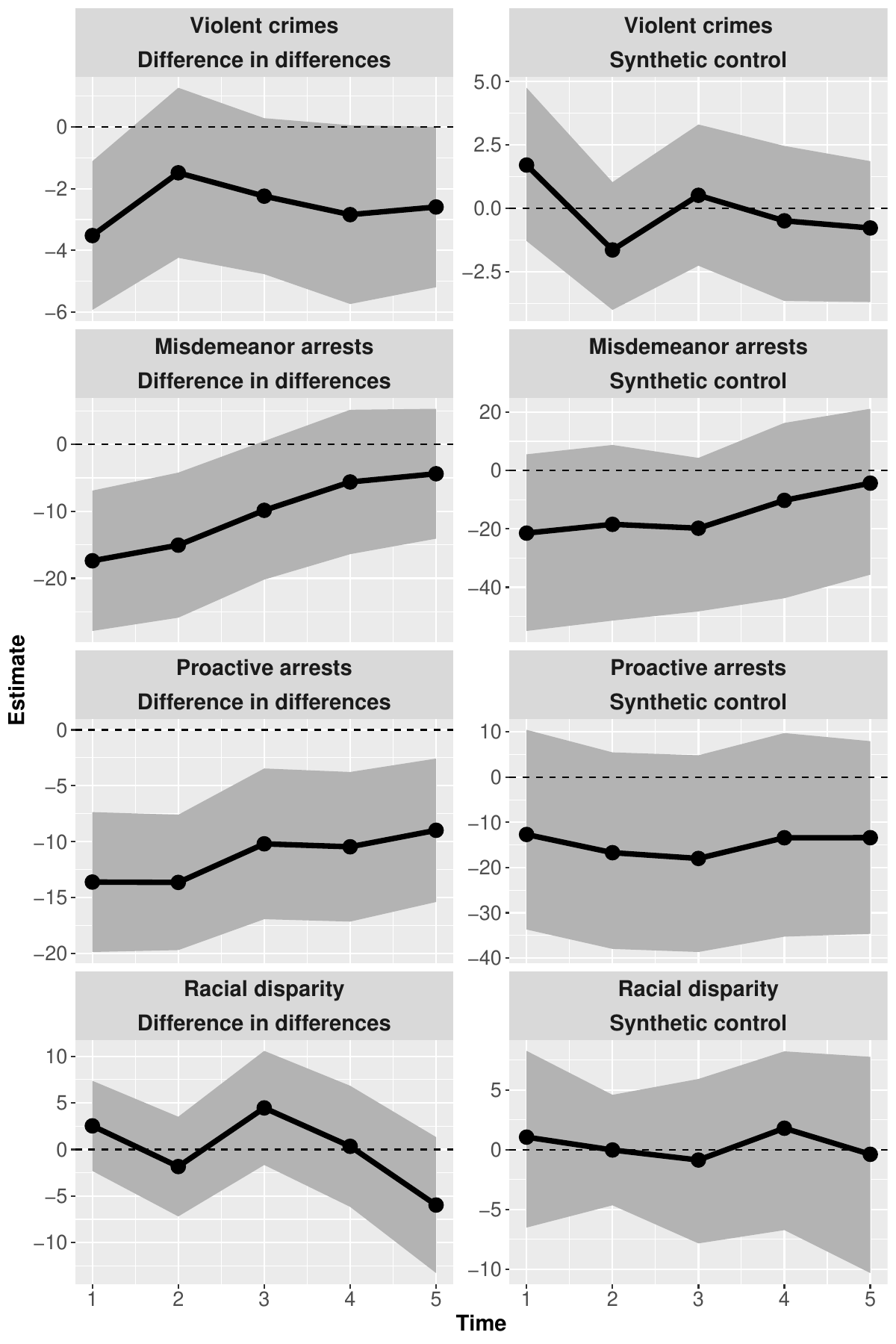}
		\caption{Estimates of the marginal effect of neighborhood policing in the first five time periods after treatment initiation for both a DID and synthetic control estimator. }
		\label{fig:ExistingNYC}
\end{figure}

The results for all four outcomes considered in the manuscript can be seen in Figure \ref{fig:ExistingNYC}. We see relatively similar results to those seen in the manuscript, which provides additional evidence for the overall findings of our approach. Neither approach finds any effect of neighborhood policing on the racial disparity in proactive arrests. In terms of both misdemeanor and proactive arrests, the DID approach provides extremely similar results to those seen in the manuscript. The DID estimator finds a strong, negative effect of neighborhood policing on proactive arrests, and finds a moderate effect on misdemeanor arrests that slightly decreases in strength over time. The synthetic control estimator also finds highly similar results to the proposed approach in terms of point estimates for both proactive and misdemeanor arrests. They have slightly wider 95\% confidence intervals, however, which leads to the intervals generally covering 0. The increased width in the confidence intervals is at least partially caused by the fact that this analysis uses a subset of the precincts for estimation of the treatment effect, and this decreased sample size would be expected to lead to increased uncertainty. In terms of violent crimes, the synthetic control estimator finds no effect of neighborhood policing, which closely aligns with the results seen in the manuscript, while the DID estimator finds a moderately negative effect on violent crimes. Overall, these results paint a relatively similar picture as those presented in the manuscript, which is that the most pronounced effects of neighborhood policing are on both proactive and misdemeanor arrests. 

\section{Simulations on other outcomes of interest}

In the manuscript, we focused our simulations on misdemeanor arrests, though we examine proactive arrests, violent crimes, and the racial disparity in proactive arrests as well. Here, we present identical simulation studies to the homogeneous treatment effect simulation of the main manuscript, though we use the three additional outcomes for the data in the simulation. The stationarity assumption, which is critical to our approach, is unique to each outcome and therefore we must run this simulation and model checking across all outcomes to ensure validity of our results.  

\begin{figure}[htbp]
		\centering
		\includegraphics[width=0.315\linewidth]{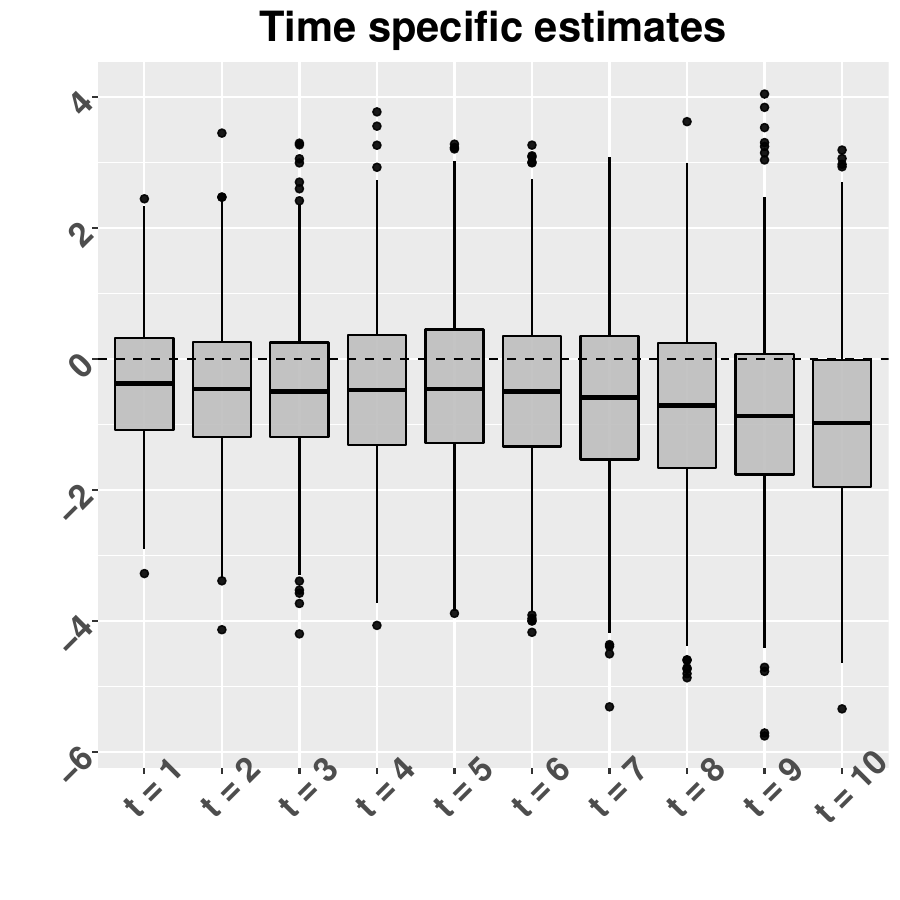}
		\includegraphics[width=0.33\linewidth]{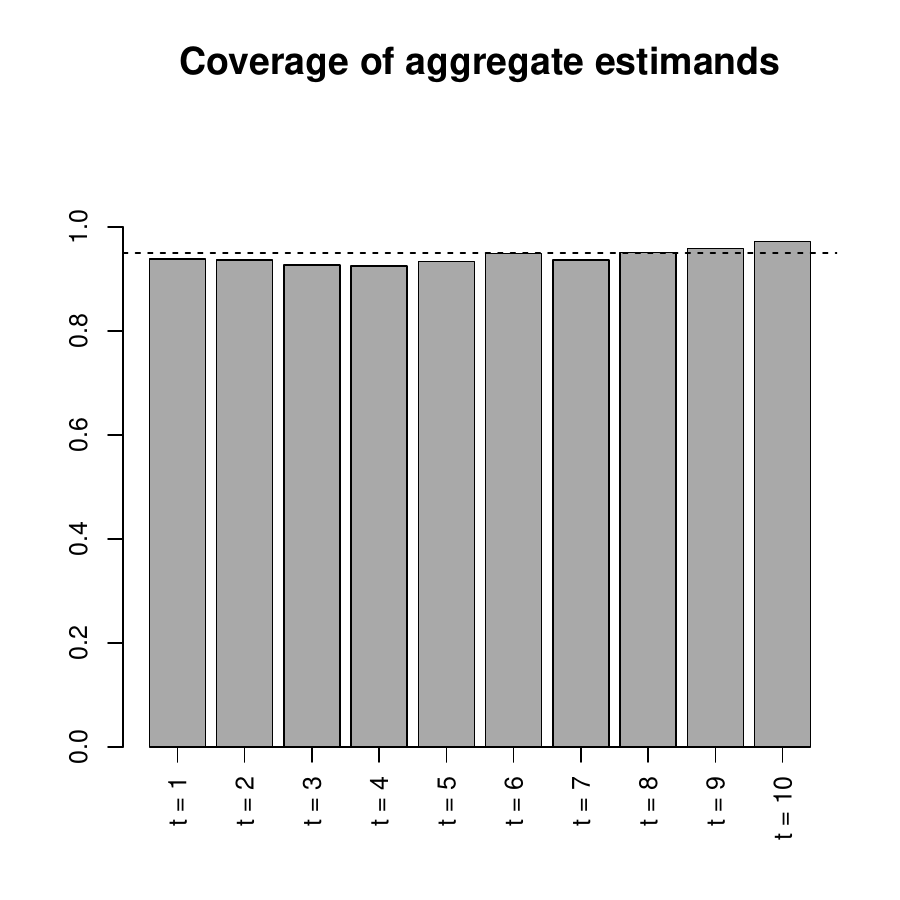}
		\includegraphics[width=0.333\linewidth]{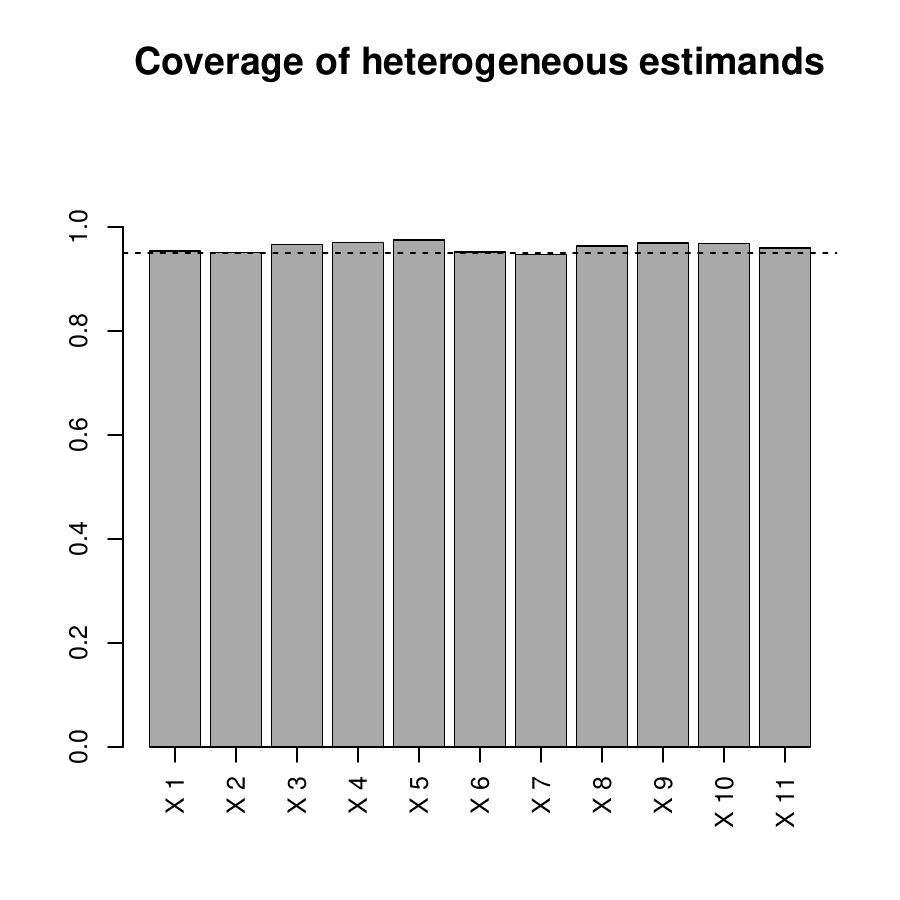}
		\caption{Results from the homogeneous treatment effect simulation study when applied to violent crimes instead of misdemeanor arrests.}
		\label{fig:33Sim}
\end{figure}

\begin{figure}[htbp]
		\centering
		\includegraphics[width=0.315\linewidth]{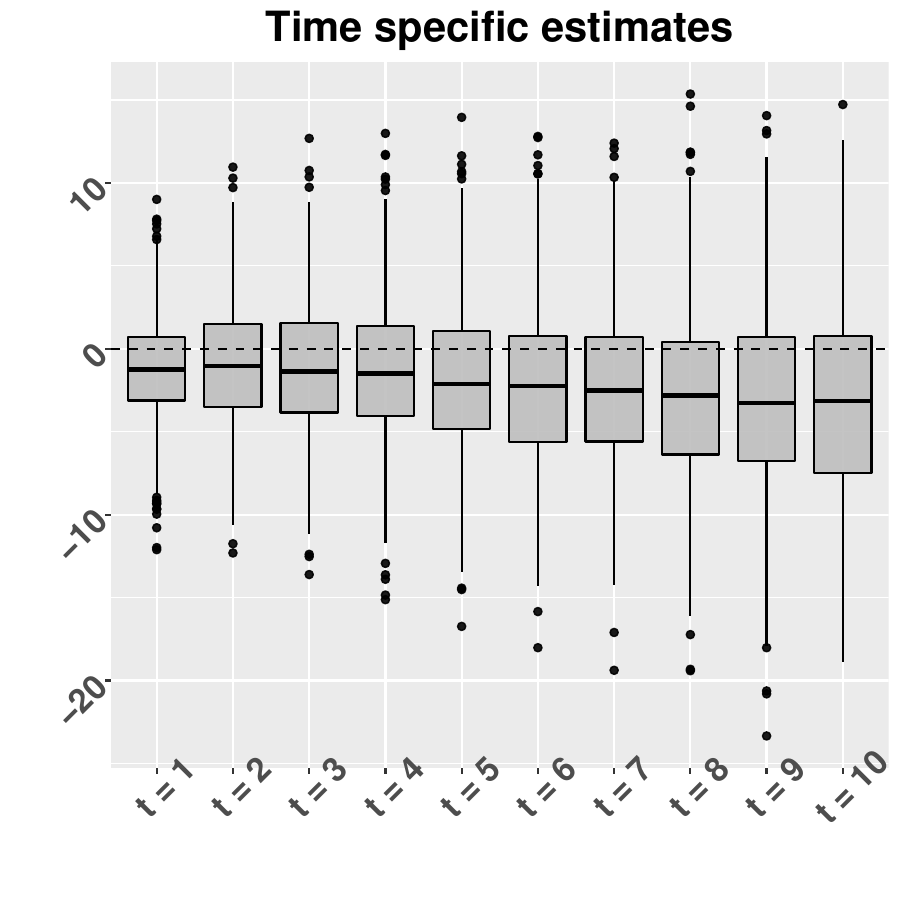}
		\includegraphics[width=0.33\linewidth]{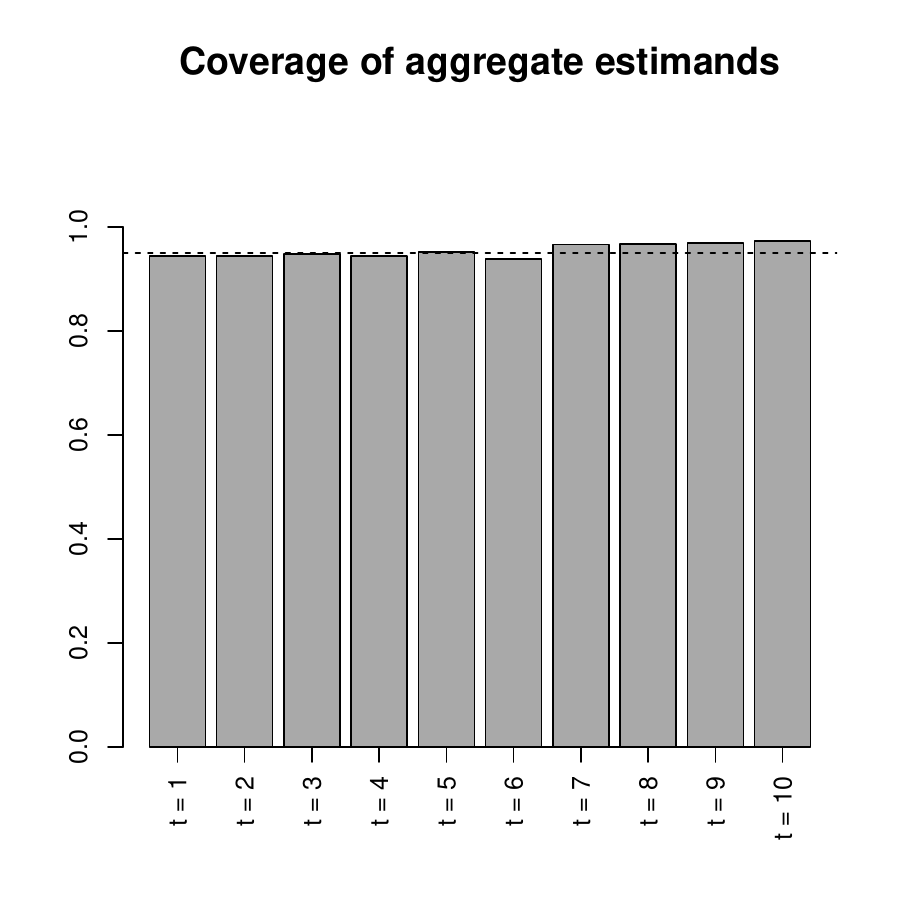}
		\includegraphics[width=0.333\linewidth]{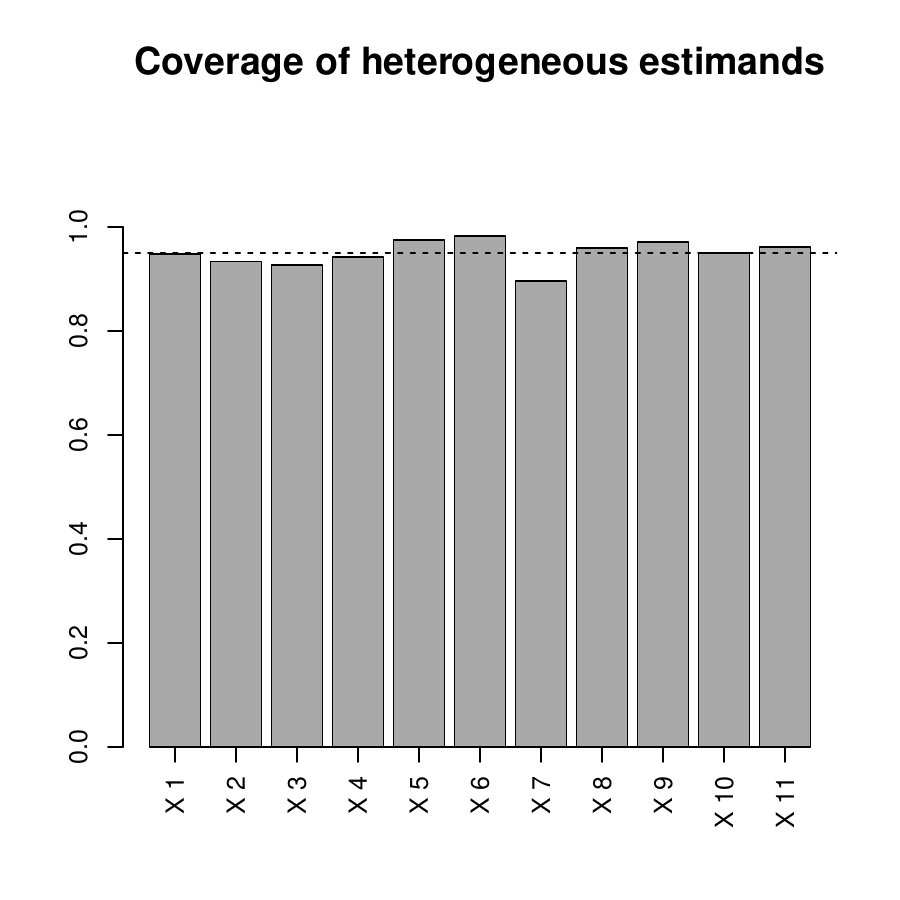}
		\caption{Results from the homogeneous treatment effect simulation study when applied to proactive arrests instead of misdemeanor arrests.}
		\label{fig:23Sim}
\end{figure}

\begin{figure}[htbp]
		\centering
		\includegraphics[width=0.315\linewidth]{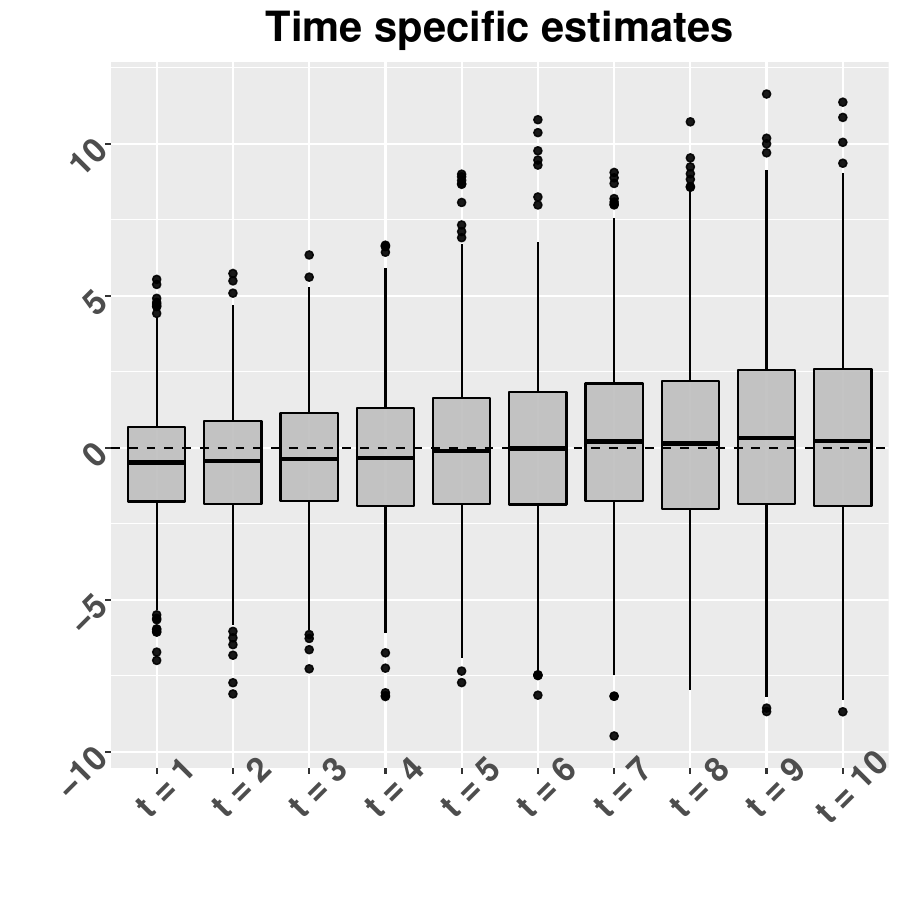}
		\includegraphics[width=0.33\linewidth]{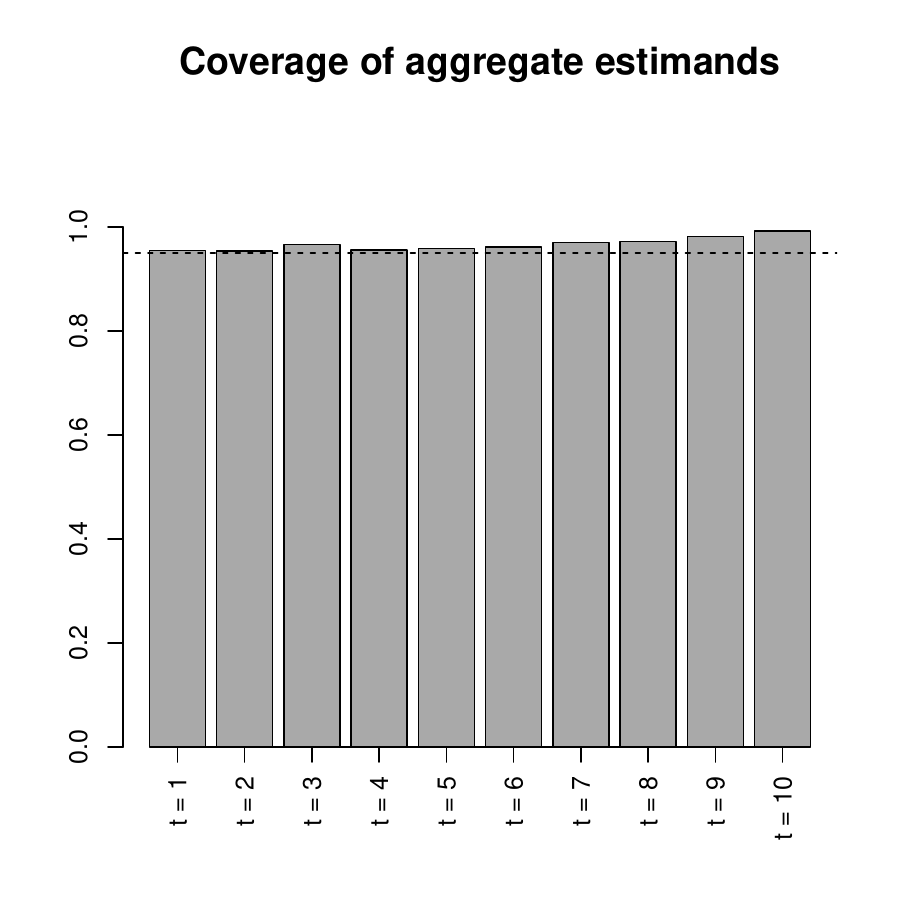}
		\includegraphics[width=0.333\linewidth]{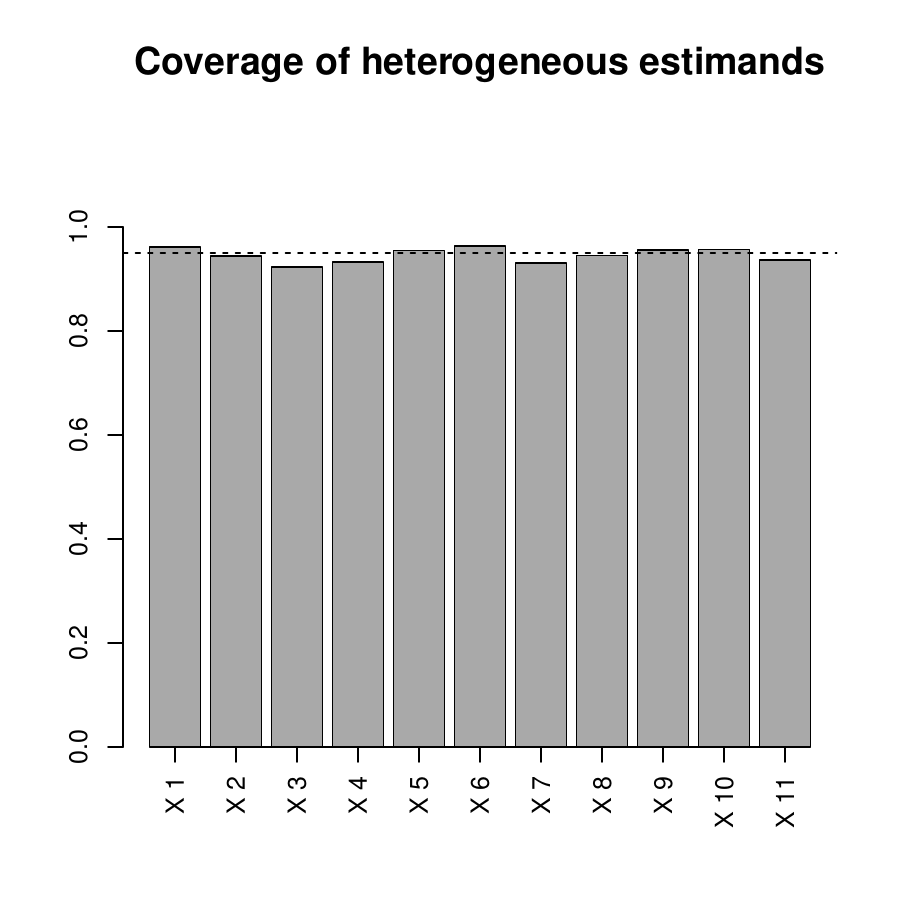}
		\caption{Results from the homogeneous treatment effect simulation study when applied to the racial disparities of proactive arrests instead of misdemeanor arrests.}
		\label{fig:63Sim}
\end{figure}

The full results for violent crimes, proactive arrests, and racial disparities in proactive arrests can all be found in Figures \ref{fig:33Sim}, \ref{fig:23Sim}, and \ref{fig:63Sim}, respectively. Here, we briefly summarize these results and how they compare with those seen in the manuscript for misdemeanor arrests. Importantly, the credible interval coverage for all marginal estimands remains very close to 0.95 for all three simulations, which echoes the results of the simulations for misdemeanor arrests in the main manuscript. Heterogeneous estimands also depict a similar story, as we are also able to achieve coverage very close to the nominal rate. The bias of the marginal treatment effects is relatively similar to the simulation from the main manuscript as biases are relatively low, and they tend to increase as we estimate causal effects farther into the future. This bias is not substantial enough to greatly affect the coverage probabilities for these estimands, which are close to 95\%. Overall, these results suggest that our approach is able to estimate treatment effects for all four outcomes of interest with a reasonably high amount of statistical validity. It appears, at least in the pre-treatment period, that the stationarity assumption is reasonable and that we can obtain accurate estimates of treatment effects of interest. 

\section{Additional simulation studies}

Here we present additional simulation results that are again based on the NYC policing data. We will present results from simulation studies that show our approach with smoothed estimates of $\Delta(q)$, and in a simulation with heterogeneous treatment effects.

\subsection{Smooth estimates of $\Delta(q)$}

\begin{figure}[!b]
		\centering
		\includegraphics[width=0.35\linewidth]{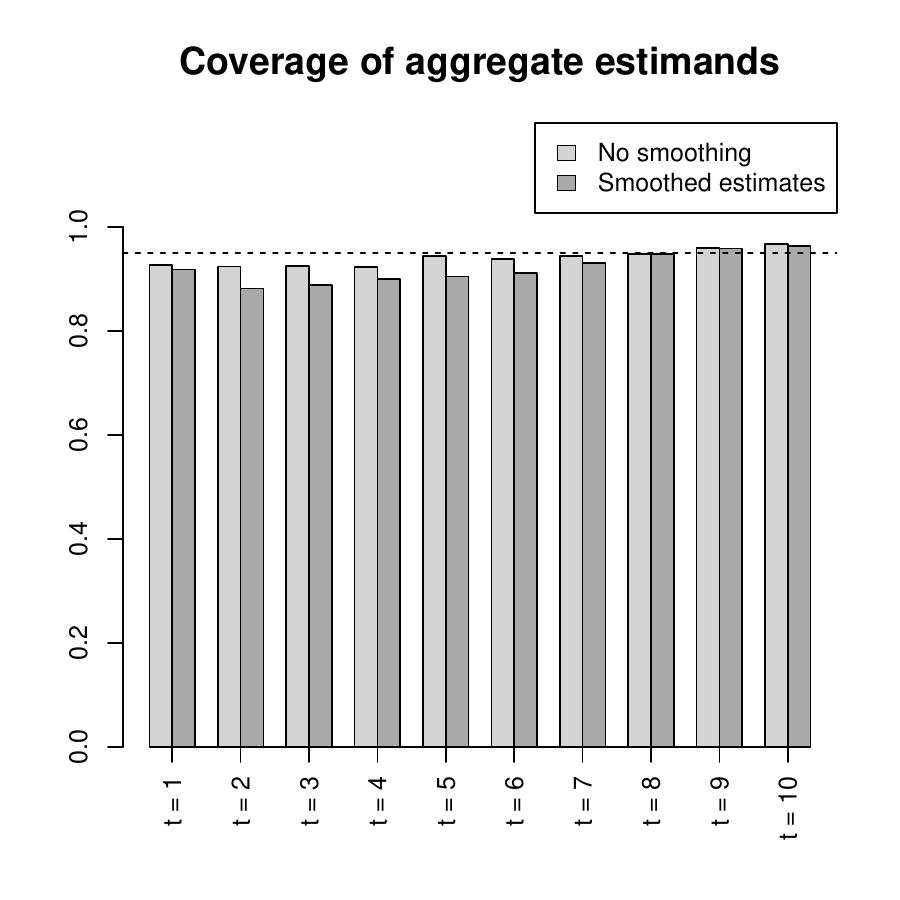}
		\includegraphics[width=0.35\linewidth]{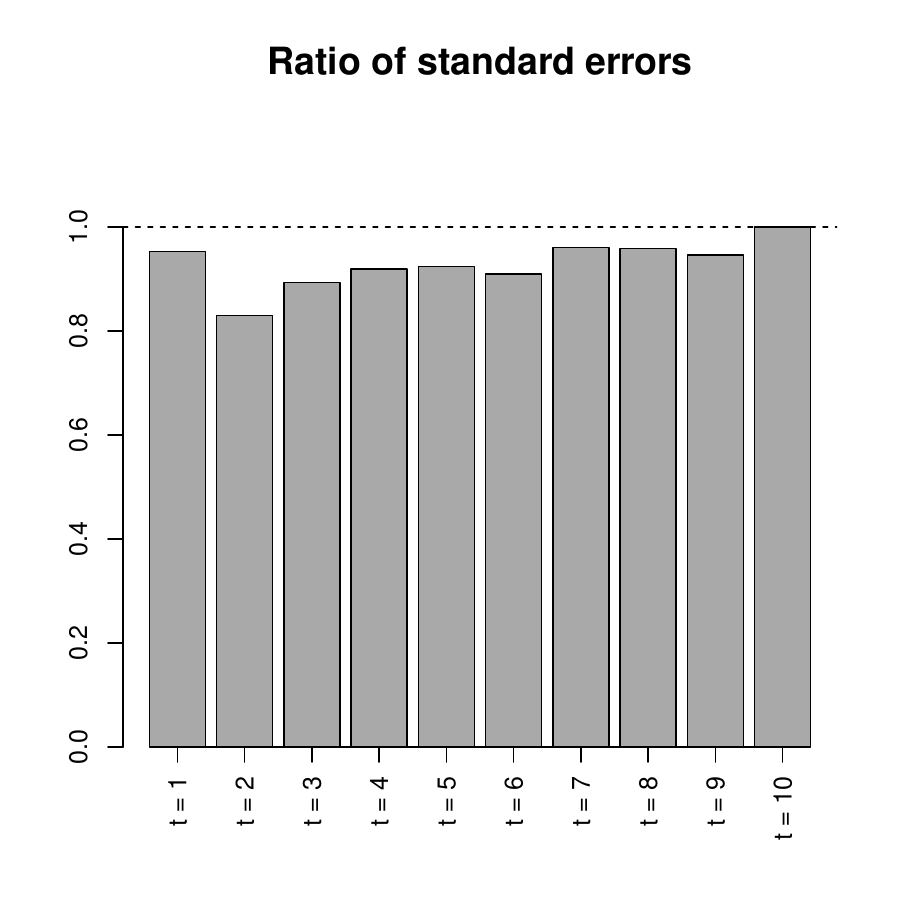}
		\caption{Results from the simulation study with smooth $\Delta(q)$ values. The left panel shows the coverage of $\Delta(q)$ both with and without smoothness, while the right panel shows the ratio of empirical standard errors between the model assuming smoothness and the one that does not assume smoothness of $\Delta(q)$.}
		\label{fig:93SimSmooth}
\end{figure}

Here we run the same simulation study as in the homogeneous simulation study of the main manuscript, except we now let the true $\Delta(q)$ be a smooth function of $q$. In particular, we let $\Delta(q) = 10 + \widetilde{\boldsymbol{Z}}_q \boldsymbol{\beta}_q$ where $\widetilde{\boldsymbol{Z}}_q$ are three degrees of freedom natural splines evaluated at $q$ and $\boldsymbol{\beta}_q = (-2,-4,-6)$. We use the proposed approach in two ways: one that assumes smoothness of $\Delta(q)$ and one that does not. The model that does not assume smoothness simply takes the posterior distribution of $\Delta_{i,T_{i0} + q, T_{i0}}$ and directly calculates the posterior distribution of the sample average treatment effect by averaging over the units in the sample. The approach that assumes smoothness takes every posterior sample of $\Delta_{i,T_{i0} + q, T_{i0}}$ and regresses these individual treatment effects against a three degrees of freedom spline representation for $q$. The fitted values from this model are then used as posterior draws of $\Delta_{i,T_{i0} + q, T_{i0}}$ and calculating sample average treatment effects proceeds analogously. We can see in the left panel of Figure \ref{fig:93SimSmooth} that both estimators provide credible interval coverages for $\Delta(q)$ that are at or near the nominal 95\% rate, with the smoothed estimates having slightly lower coverage. One key difference can be seen in the right panel of Figure \ref{fig:93SimSmooth} as the standard deviation of the estimates coming from the model assuming smoothness are generally smaller than the model that does not assume smoothness of $\Delta(q)$.

\subsection{Heterogeneous treatment effects}

Here we run the same simulation study as in the homogeneous simulation study of the main manuscript, except we now let the treatment effect for each precinct be proportional to $\boldsymbol{X}_i \boldsymbol{\beta}$.
The results are nearly identical to those from the homogeneous treatment effect setting in the manuscript. We see effectively no bias of the marginal treatment effects, and interval coverages that are close to the nominal 95\% rate. We see the same story for the heterogeneous estimands as well, as our approach is able to achieve nearly the nominal coverage rate for each covariate in the study, which shows the ability of our approach to estimate heterogeneous treatment effects in the NYC policing example.

\begin{figure}[!t]
		\centering
		\includegraphics[width=0.315\linewidth]{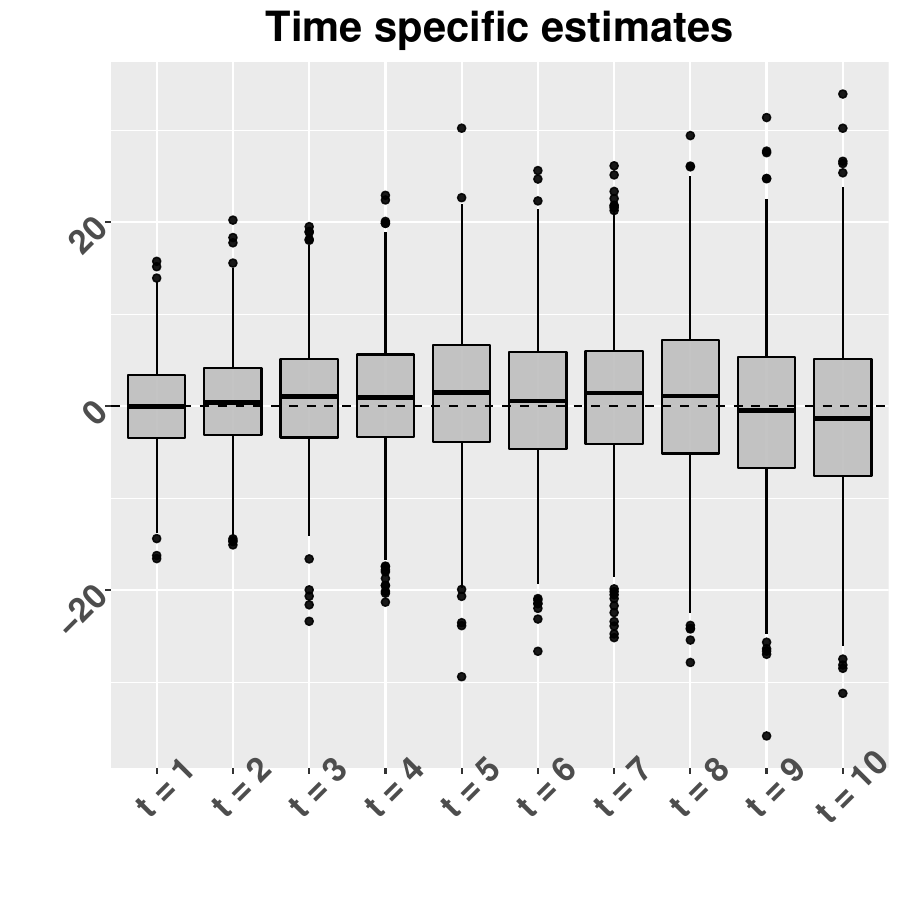}
		\includegraphics[width=0.33\linewidth]{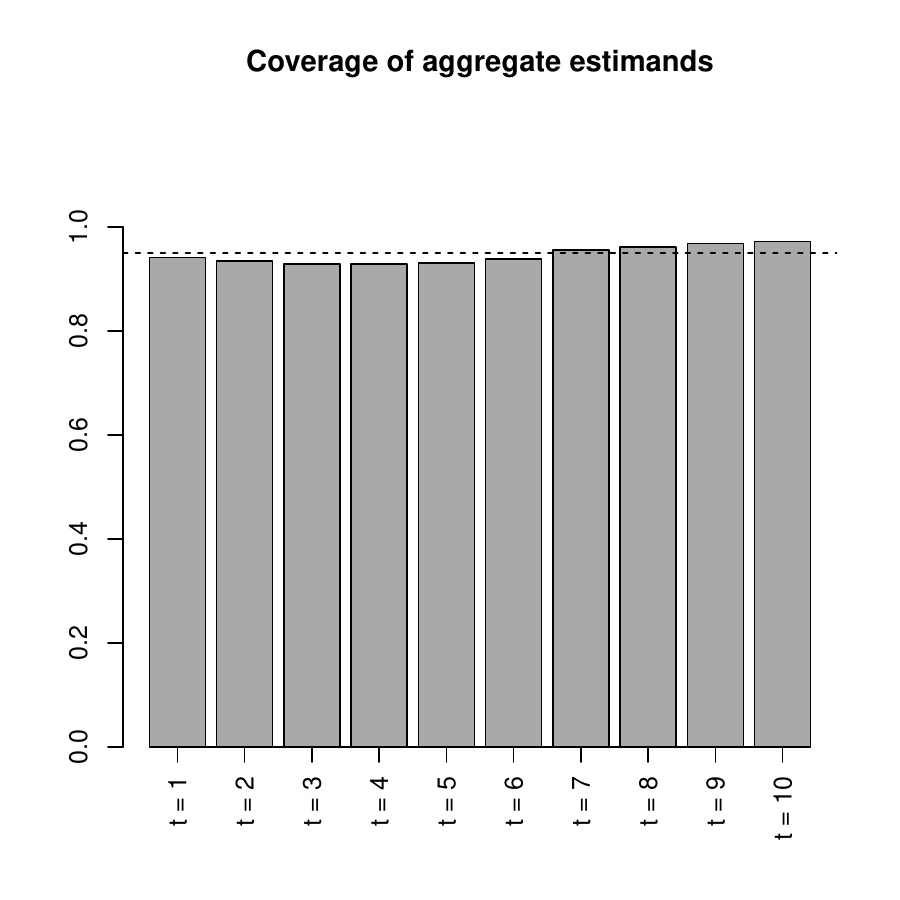}
		\includegraphics[width=0.333\linewidth]{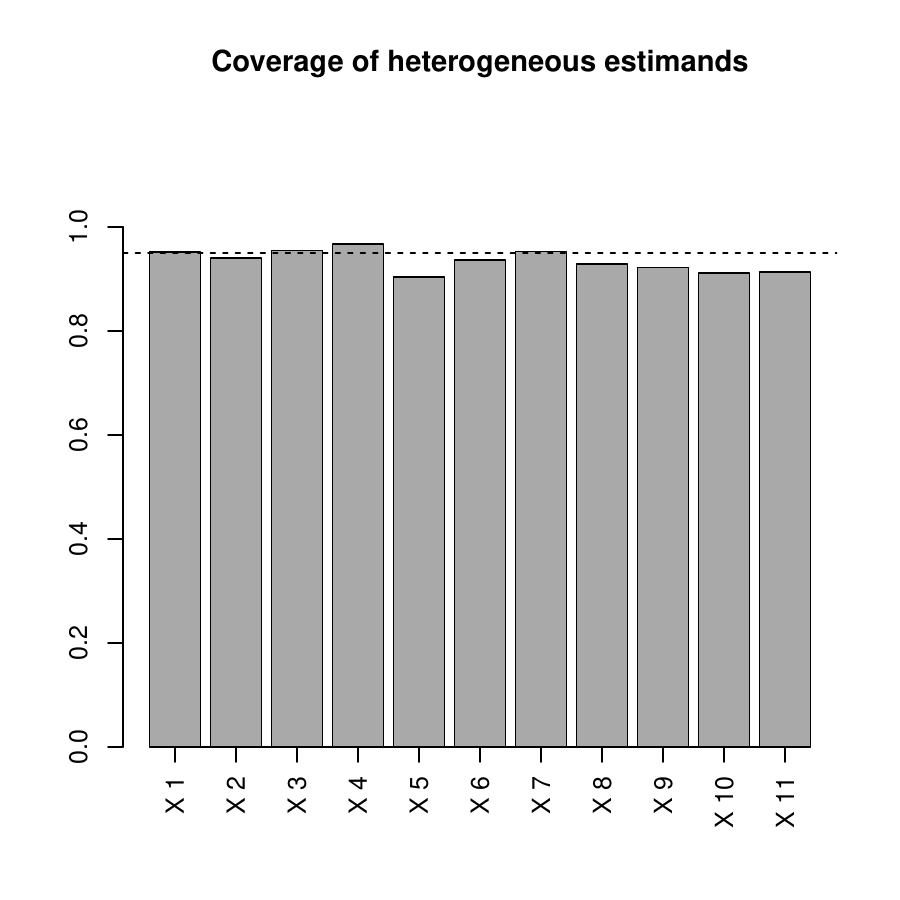}
		\caption{Results from the heterogeneous treatment effect simulation study. The left panel shows the estimates of $\Delta(q)$ for $q=0, \dots, 9$. Estimates are mean shifted so that an unbiased estimator will be centered at zero. The middle panel shows the coverage of $\Delta(q)$, while the right panel shows coverage of the heterogeneous estimands. }
		\label{fig:43SimHetero}
\end{figure}

\end{document}